\documentclass[twocolumn]{aastex63}

\usepackage{color}
\usepackage{amsmath}
\usepackage{subcaption}
\usepackage{wrapfig}

\newcommand{\rg}{r_{\rm g}}
\newcommand{\tunit}{\rg/c}

\received{14-Sep-2023}
\revised{20-Oct-2023}
\accepted{24-Oct-2023}
\submitjournal{ApJL}

\shorttitle{Surface waves in supermassive black hole jets}
\shortauthors{Davelaar et al.}

\graphicspath{{./}{}}
\begin{document}

\title{Synchrotron polarization signatures of surface waves in supermassive black hole jets}

\correspondingauthor{J. Davelaar}
\email{jdavelaar@flatironinstitute.org}

\author[0000-0002-2685-2434]{J. Davelaar}
\affiliation{Center for Computational Astrophysics, Flatiron Institute, 162 Fifth Avenue, New York, NY 10010, USA}
\affiliation{Department of Astronomy and Columbia Astrophysics Laboratory, Columbia University, 550 W 120th St, New York, NY 10027, USA}

\author[0000-0002-7301-3908]{ B. Ripperda}\affiliation{Canadian Institute for Theoretical Astrophysics, 60 St. George St, Toronto, Ontario M5S 3H8}
\affiliation{Department of Physics, University of Toronto, 60 St. George Street, Toronto, ON M5S 1A7}
\affiliation{David A. Dunlap Department of Astronomy, University of Toronto, 50 St. George Street, Toronto, ON M5S 3H4}
\affiliation{Perimeter Institute for Theoretical Physics, 31 Caroline St. North, Waterloo, ON, Canada N2L 2Y5}

\author[0000-0002-1227-2754]{L. Sironi}\affiliation{Department of Astronomy and Columbia Astrophysics Laboratory, Columbia University, 550 W 120th St, New York, NY 10027, USA}

\author[0000-0001-7801-0362]{A. A. Philippov}\affiliation{Department of Physics, University of Maryland, College Park, MD, USA}

\author[0000-0001-6833-7580]{H. Olivares}\affiliation{Department of Astrophysics/IMAPP, Radboud University, P.O. Box 9010, 6500 GL  Nijmegen, The Netherlands}

\author{O. Porth}\affiliation{Anton Pannekoek Instituut, Universiteit van Amsterdam P.O. Box 94249, 1090 GE Amsterdam, The Netherlands}

\author{B. van den Berg}\affiliation{Department of Astrophysics/IMAPP, Radboud University, P.O. Box 9010, 6500 GL  Nijmegen, The Netherlands}

\author[0000-0003-1151-3971]{T. Bronzwaer}\affiliation{Department of Astrophysics/IMAPP, Radboud University, P.O. Box 9010, 6500 GL  Nijmegen, The Netherlands}

\author[0000-0002-2825-3590]{K. Chatterjee}\affiliation{Black Hole Initiative at Harvard University, 20 Garden Street, Cambridge, MA 02138, USA}\affiliation{Harvard-Smithsonian Center for Astrophysics, 60 Garden Street, Cambridge, MA 02138, USA}

\author{M. Liska}\affiliation{Institute for Theory and Computation, Harvard University, 60 Garden Street, Cambridge, MA 02138, USA}

\begin{abstract}
Supermassive black holes in active galactic nuclei (AGN) are known to launch relativistic jets, which are observed across the entire electromagnetic spectrum and are thought to be efficient particle accelerators. Their primary radiation mechanism for radio emission is polarized synchrotron emission produced by a population of non-thermal electrons. In this Letter, we present a global general relativistic magnetohydrodynamical (GRMHD) simulation of a magnetically arrested disk (MAD). After the simulation reaches the MAD state, we show that waves are continuously launched from the vicinity of the black hole and propagate along the interface between the jet and the wind. At this interface, a steep gradient in velocity is present between the mildly relativistic wind and the highly relativistic jet. The interface is, therefore, a shear layer, and due to the shear, the waves generate roll-ups that alter the magnetic field configuration and the shear layer geometry. We then perform polarized radiation transfer calculations of our GRMHD simulation and find signatures of the waves in both total intensity and linear polarization, effectively lowering the fully resolved polarization fraction. The tell-tale polarization signatures of the waves could be observable by future Very Long Baseline Interferometric observations, e.g., by the next-generation Event Horizon Telescope.
\end{abstract}


\section{Introduction}
Accreting supermassive black holes can produce highly relativistic electromagnetically collimated outflows called jets, observed across the electromagnetic spectrum. These jets can be observed up to kilo-parsec scale in the case of active galactic nuclei (AGN). At radio and mm frequencies, the primary emission mechanism is synchrotron emission. Very Long Baseline Interferometric (VLBI) observations probe the jet substructure and reveal edge-brightened morphology, often referred to as limb brightening, see, e.g., \citep{walker2018a,kim2018a, giovannini2018,janssen2021}. However, the mechanism responsible for energizing the radiating electrons along the jet surface remains an active debate. The upcoming next-generation VLBI facilities will bring higher resolving power and dynamic range, allowing for better-resolved AGN jet images and polarization maps. In this work, we investigate the imprint of instabilities in the jet boundary and the effect of non-thermal tails in electron distribution functions on polarized emission features of AGN jets.

Since synchrotron emission is intrinsically linearly polarized \citep{rybicki1979}, where the polarization vector is perpendicular to the magnetic field lines, the observed polarization from AGN can be used to study the magnetic field structure of jets. These systems generally show very low linear polarization fractions across the entire radio band, see, e.g., \cite{zavala2003,hada2016a,walker2018a,park2019,ehtmwlscienceworkinggroup2021}. These low fractions indicate that an external Faraday screen depolarizes the jet's emission before it reaches us or that the emission from the jet is not generated in large-scale coherent magnetic field geometries. In the case of M87 at sub-mm wavelengths, observations by \cite{hada2016a} show linear polarization fractions of up to 20\%, revealing regions of coherent field geometry when observed at higher spatial resolution.

The highest resolution polarized observations of a low-luminosity AGN (LLAGN) to date are by the EHT collaboration \citep{eventhorizontelescopecollaboration2021}. The EHT showed that in M87$^*$ at horizon scales, the polarization vector shows a helical pattern, which is typically reproduced by simulations of accretion flows in the magnetically arrested disk (MAD) state (\citealt{igumenshchev2003,narayan2003,tchekhovskoy2011}). MAD accretion flows can be studied with general relativistic magnetohydrodynamical (GRMHD) simulations. To study the emission generated by the GRMHD simulations, they can be post-processed with general-relativistic radiative transfer codes \citep{dexter2012,moscibrodzka2017,davelaar2018a,wong2021,cruz-osorio2022,fromm2022} after postulating an electron temperature prescription. A magnetically arrested flow typically reaches a limit where the magnetic pressure balances the gas pressure due to accumulated magnetic flux on the event horizon. This limit is often identified with $\phi_{\rm mad} = \Phi_{\rm B}/\sqrt{\dot{M}} \approx 15$ \citep{tchekhovskoy2011}, here $\Phi_{\rm B}$ is the magnetic flux threading the horizon and $\dot{M}$ the mass accretion rate. If this threshold is reached, the antiparallel magnetic field lines in the northern and southern jet are compressed to form a thin current sheet that reconnects and expels the magnetic flux  \citep{dexter2020,ripperda2020,ripperda2022}. During such magnetic flux eruptions, the magnetic field undergoes no-guide-field reconnection, resulting in the expulsion of a flux tube consisting of a vertical (poloidal) magnetic field. This flux tube can push the accretion disk away, effectively arresting a part of the incoming flow. The flux tubes can orbit at sub-Keplerian velocities in the disk, where they can propagate to a few tens of gravitational radii. Within one orbit, the low-density fluid in the flux tube gets mixed into the higher-density disk through magnetic Rayleigh-Taylor instabilities (RTI) \citep{ripperda2022,zhdankin2023}. The magnetic flux eruptions are conjectured to power high energy flares through reconnection near the horizon \citep{ripperda2020,dexter2020,ripperda2022,hakobyan2023} and through reconnection induced by RTI at the boundary of the orbiting flux tube \citep{porth2021,zhdankin2023}.

In this Letter, we will use GRMHD simulations in Cartesian Kerr-Schild coordinates with adaptive mesh refinement to study the large-scale properties of the jets. Using Cartesian coordinates in combination with adaptive mesh refinement allows us to better resolve the shear layer separating the highly-relativistic bulk velocity jet and the mildly-relativistic disk. We will refer to this region as the jet-wind shear layer. Our simulation shows that the magnetic flux eruptions associated with MAD flows can drive waves along the jet-wind shear layer. At larger radii, the waves show roll-ups that mix high-density wind material with the low-density magnetized jet. In this non-linear phase, the waves may trigger magnetic reconnection and turbulence, as was found in local-box simulations \citep{sironi2021}. To quantify the imprint of the waves on observables, we ray-trace our simulation with our polarized radiative transfer code {\tt RAPTOR} \citep{Bronzwaer2018,Bronzwaer2020}. We find that the waves depolarize the observed synchrotron emission. 

This letter is structured as follows. Section 2 describes our numerical setup and summarizes how we compute synthetic polarized images. Section 3 explains our GRMHD and radiative transfer results, highlights linear and circular polarization properties and provides evidence that the existence of waves results in depolarization. Finally, in section 4, we discuss and summarize our main conclusions.

\section{Numerical setup}

In this section, we will describe our GRMHD simulation setup, our radiative transfer code, and our electron thermodynamics model.

\subsection{GRMHD}

To model the dynamics of the accretion flow around a Kerr black hole, we use the Black Hole Accretion Code ({\tt BHAC}) \citep{porth2017,olivares2019}, which solves the ideal GRMHD equations in curved spacetime. The equation of state is assumed to be an ideal gas law, described via the specific enthalpy 
\begin{equation}
     h(\rho, P_{\rm gas})= 1 +\frac{\gamma_{\rm ad}}{{\gamma_{\rm ad}}-1}\frac{P_{\rm gas}}{\rho},
\end{equation}
with gas pressure $P_{\rm gas}$, mass density $\rho$ in the fluid frame, and the adiabatic index is set to $\gamma_{\rm ad}=13/9$.

We initialize our simulation with a \cite{fishbone1976} torus in spherical Boyer-Lindquist coordinates, $(t,r,\theta,\phi)$, here $\theta$ is the angle with respect to the spin axis of the black hole and $\phi$ the azimuthal angle around the black hole spin axis. The initial conditions are then transformed to Cartesian Kerr-Schild coordinates, $(t,x,y,z)$. The initial disk has an inner radius of $r_{\rm in} = 20 ~\rg$, with $\rg$ the gravitational radius given by $\rg = {GM}/{c^2}$, with $G$ Newtons constant, $M$ the black hole mass, and $c$ the speed of light. We set the pressure maximum of the disk at $r_{\rm max}=40~\rg$. The initial magnetic field is given by the vector potential $A_\phi = \rho (r/r_{\rm in} \sin \theta )^3 e^{-r/400} - 0.2$, where $r$ is the radial coordinate, and $\theta$ the polar angle. The vector potential follows iso-contours of density, $\rho$. These initial conditions are chosen such that the simulation will reach the saturated state of a MAD accretion flow.

We set the dimensionless black hole spin parameter to $a=15/16$, which results in a black hole horizon size of $r_{\rm h} \approx 1.34 ~\rg$. The domain size is $(-4000~\rg,4000~\rg)$ in all three spatial Cartesian directions. Our base resolution is $192^3$ cells. We employ nine additional levels of static mesh refinement, resulting in an effective uniform Cartesian resolution of $98,304^3$. The highest resolution grid is centered at the horizon and has a resolution of $\Delta x_i = 0.08 ~\rg$. The simulation is run for $10,000 \rg/c$. We introduce a maximum for the cold magnetization parameter $\sigma=\frac{b^2}{\rho}$, where $b$ is the magnetic field strength, and we inject mass to maintain $\sigma\leq100$. Additionally, we ensure that $\beta_{\rm p}^{-1} \leq 10^3$, where $\beta_{\rm p} =  8\pi P_{\rm gas}/b^2 $ is the plasma beta parameter. We use floor profiles for density as well as pressure, given by $\rho_{\rm floor} = 10^{-4} r^{-2}$, and $P_{\rm floor} = 10^{-6} \rho_{\rm floor}^{\gamma_{\rm ad}}$. Due to our Cartesian grid, we do not have an inner radius where we can employ outflowing boundary conditions, we, therefore, introduce an artificial treatment of the fluid variables inside the black hole event horizon, known as “excision” in numerical relativity, to limit the accumulation of energy and density, which otherwise could numerically diffuse out of the event horizon when accumulated to too high values. In our case, we introduce a ceiling on density ($\rho_{\rm max}=6$) as well as pressure ($P_{\rm max}=2$) when $r<r_{\rm crit}$ where our critical radius is set to be $r_{\rm crit} = 1 ~\rg$. Given the location of the critical radius, we have four cells between the event horizon and the critical radius.
 
\subsection{Synthetic polarized images}

To generate synthetic polarized images of the accretion flow, we use the general-relativistic ray tracing code {\tt RAPTOR}. {\tt RAPTOR} solves the polarized radiative transport equations along null geodesics. The geodesic equation is solved starting from a virtual camera outside the simulation domain ($r_{\rm cam}=10^4 ~\rg$). We employ an adaptive camera as described in \cite{davelaar2022}. The adaptive camera allows a varying resolution over the image plane, resulting in a computational benefit since most of the resolution can be focused on the event horizon scale, which shows small-scale emission varying on short timescales, while the larger scale can be fully resolved with a lower resolution. We use a base resolution of $625^2$ pixels and double the resolution four times, within 60~$\rg$, 40~$\rg$, 20~$\rg$, and 10 $\rg$, respectively, bringing the effective resolution to $10,000^2$ pixels \footnote{This significantly reduces the computation cost, since the effective resolution uses a factor 50 fewer pixels compared to a standard uniform camera.}. We use an adaptive Runga-Kutta-Fehlberg method to integrate the geodesic equation and a fourth-order finite difference method to compute the metric derivatives needed for the Christoffel coefficients. The stepsize in {\tt RAPTOR} is estimated based on the wavevector in the previous step; see \cite{davelaar2019}. For this work, we developed an additional stepsize criterion based on a Courant–Friedrichs–Lewy (CFL) condition, where we require a minimum of eight steps per cell to ensure convergence of all Stokes parameters. We compute synthetic images between $80$ GHz and $100$ GHz with a frequency spacing of $3$ GHz. We also compute time-averaged spectra from the image-integrated total intensity at 25 frequencies with a logarithmic spacing between $10^{10}-10^{15}$ Hz.

The ideal GRMHD equations are dimensionless and do not describe the evolution and thermodynamics of electrons. To this end, we need to introduce a mass scaling and a prescription for the electron temperature. To scale our simulation to a specific black hole, we use a unit of length $\mathcal{L}=\rg$, a unit of time $\mathcal{T} = \rg/c$, and a unit of mass $\mathcal{M}$. The unit of mass is related to the mass accretion rate via $\dot{M} = \dot{M}_{\rm sim} \mathcal{M}/\mathcal{T}$, where $\dot{M}_{\rm sim}$ is the accretion rate in simulation units. Combinations of these units are then used to scale all relevant fluid quantities. The density is scaled by $\rho_0 = \mathcal{M}/\mathcal{L}^3$, the internal energy by $u_{0}=\rho_0 c^2$, and the magnetic fields by $B_0=c\sqrt{4\pi \rho_0}$ (where $B_0$ is expressed in Gaussian units). Since the black hole mass is often constrained observationally, the only free parameter in our system is $\mathcal{M}$, which can be used to set the energy budget of the simulation such that the total emission produced equals a user-set target. We follow \cite{moscibrodzka2016}, to parameterize the ratio between electron temperature $T_{\rm e}$ and proton temperatures $T_{\rm p}$ via

\begin{subequations} \label{eq-temp-rat}
\begin{align}
T_{\rm ratio} &= \frac{1}{1+\beta_{\rm p}^2} + R \frac{\beta_{\rm p}^2}{1+\beta_{\rm p}^2},\\
\Theta_{\rm e} &= k_{\rm b} T_{\rm e}/m_{\rm e} c^2 =  \frac{U(\gamma_{\rm ad} - 1) m_{\rm p}/m_{\rm e}}{\rho \left(1 + T_{\rm ratio} \right)},
\end{align}
\end{subequations}

where $U$ the internal energy, $m_{\rm p}$ the proton mass, $m_{\rm e}$ the electron mass, and $\Theta_{\rm e}$ the dimensionless electron temperature. The variable $R$ is a free parameter that sets the temperature ratio in regions where $\beta_{\rm p} \gg 1$, allowing for different emission morphology depending on the choice of $R$, e.g., disk dominated if $R=1$ or jet dominated if $R\gg 1$. Here, we will limit ourselves to $R=20$, e.g., a more jet-dominated model. Note that MADs are relatively insensitive to the exact value of $R$ given that most of the emission originates from a region where $\beta_{\rm p} \lesssim 1$. Finally, we must choose the electron distribution function's shape and spatial variation. We consider two models, one where the distribution function (DF) is limited to a thermal relativistic Maxwell-J\"uttner DF (MJ-DF), and one where we combine the MJ-DF with a $\kappa$-DF. The $\kappa$-DF deviates from an MJ-DF by having a thermal core and a power law at high Lorentz factors. The $\kappa$-DF \citep{xiao2006} is given by,

\begin{equation}
\frac{dn_e}{d\gamma} = n_e N \gamma \sqrt{\gamma^2 - 1 }\left(1+\frac{\gamma-1}{\kappa w}\right)^{-(\kappa +1)},
\end{equation}

where the free parameters are $w$, which sets the width of the distribution, and $\kappa$, which sets the slope of the power law, and $N$ is a normalization constant such that $\int_{1}^{\infty} \frac{dn_e}{d\gamma} d\gamma = n_{\rm e}$. The $\kappa$ parameter is related to the power-law index $p$ of the non-thermal tail of the DF via $\kappa=p+1$, such that for $\gamma \gg 1$, $\frac{dn_e}{d\gamma} \propto \gamma^{1-\kappa}$. For the width $w$, we follow \cite{davelaar2019} by enforcing that the energy in the DF equals the total available internal energy of the electrons,

\begin{equation}
    w = \frac{(\kappa-3)}{\kappa} \Theta_e,
\end{equation}

where $\Theta_e$ is computed with equation \ref{eq-temp-rat}. Note that this formula requires $\kappa>3$ ($p>2$).

We then compute emission coefficients $c_{\nu}$ (emission, absorption, and rotation coefficients), using a prescription introduced in \cite{eventhorizontelescopecollaboration2022b} that combines thermal and $\kappa$ coefficients via,

\begin{eqnarray}
    c_{\nu} = (1-\eta(\beta_{\rm p},\sigma)) c_{\rm thermal} + \eta(\beta_{\rm p},\sigma) c_{\kappa} \\
    \eta(\beta_{\rm p},\sigma) = \left(1-e^{-\beta_{\rm p}^{-2}}\right)\left(1-e^{-(\sigma/\sigma_0)^2}\right),
\end{eqnarray}

here $\sigma_0$ sets the transition point for the magnetization, which we set to $\sigma_0=0.5$. The function $\eta(\beta_{\rm p},\sigma)$ smoothly transitions from thermal to non-thermal component if $\beta_{\rm p} < 1$ and $\sigma/\sigma_{0} > 1$, representing sites with a large reservoir of magnetic energy that can dissipate to accelerate particles, e.g., in the jet's shear layer. The polarized radiation transfer coefficients are computed via fit formula. For the thermal distribution function, we use the emission and absorption coefficients presented in \cite{Dexter2016} and the rotativities from \cite{Shcherbakov2008}, for the $\kappa$ distribution function, we follow \cite{pandya2016,Marszewski2021}.

As an archetype LLAGN, we use model parameters consistent with M87*, meaning a black hole mass of $M=6.5
\times10^9 M_\odot$, and a distance of $d=16.8$ Mpc \citep{eventhorizontelescopecollaboration2019}. We set the angle between the BH spin axis and the observer to $i=160^{\circ}$, following \cite{walker2018a}. We set $\mathcal{M}$ such that the average flux at $86$ GHz is $F_{86 {\rm GHz}}= 1$ Jy, as observed by \citep{ehtmwlscienceworkinggroup2021}.  The spectrum obtained by \cite{ehtmwlscienceworkinggroup2021} also shows a spectral slope in the optically thin part of the spectrum in the near-infrared (NIR) of $F_{\nu} \propto \nu^{-1}$. Given that for optically thin synchrotron emission, the flux follows $F \propto \nu^{-(p-1)/2} = \nu^{-(\kappa-2)/2}$, to match the observed spectral slope in the NIR, we need to use $\kappa=4$.

To exclude the emission from the spine (interior of the jet, which in GRMHD simulations is typically dominated by artificial density floors), we exclude all the emission if $\sigma>5$. We expect our results to be insensitive to this choice for larger $\sigma$ values. However, lower $\sigma$ values would result in a smaller emission region at the jet-wind interface. We also exclude the larger scale disk, $\sqrt{x^2+y^2}>150 \rg$, which has not settled in a steady state for the runtime of our simulation. However, given the viewing angle of $i=160^{\circ}$, this choice does not strongly affect our results.
\section{Results}

In this section, we summarize our results. In subsection \ref{grmhd}, we find that surface waves are continuously present along the jet-wind shear layer after our simulation reaches the MAD state. In subsection \ref{spectra}, we fit our GRMHD model to the spectrum of M87 and show that the $\kappa$-jet model recovers the low-frequency radio and the NIR. In subsection \ref{linpol}, we show that the magnetic flux eruptions and waves imprint themselves in various linear polarization quantities, serving as potential tell-tale signatures that could be used to test M87's putative MAD accretion flow state. In subsection \ref{circpol}, we highlight that circular polarization plays a minor role. However, we see sign reversals in circular polarization maps that could indicate magnetic field reversals caused by the surface waves. We measure 
the Faraday rotation measure of our models in \ref{RM}. Lastly, in subsection \ref{rayprop}, we provide evidence that the polarization signatures are connected to the waves seen in the GRMHD simulation. 

\subsection{GRMHD}\label{grmhd}

\begin{figure}
  \centering
    \includegraphics[width=0.47\textwidth]{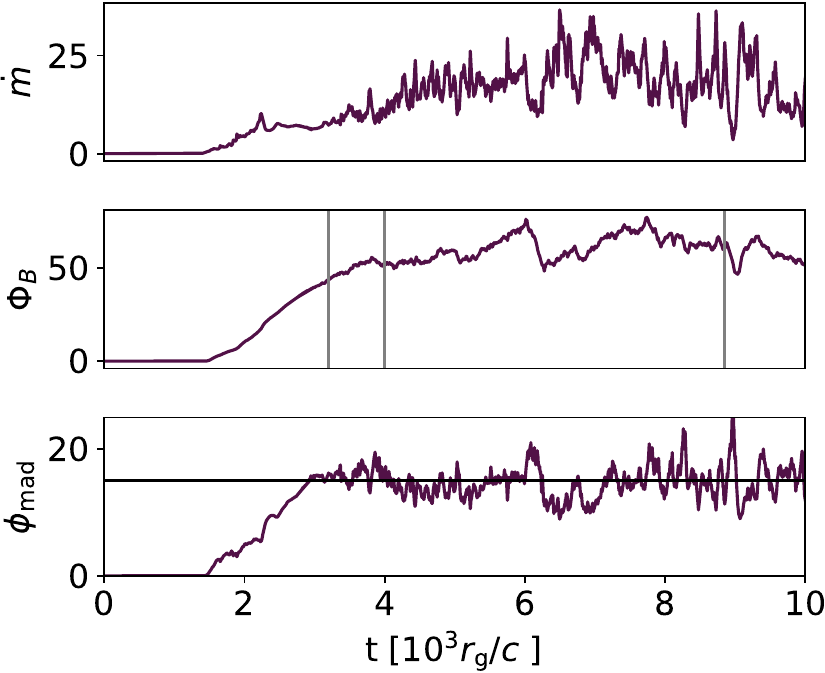}
    \caption{Time series in dimensionless units of horizon integrated mass accretion rate $\dot{m}$ (top panel), magnetic flux $\Phi_{\rm B}$ (middle panel, gray lines correspond to the slices shown in Figure \ref{fig-panel}), and the MAD parameter $\phi_{\rm mad}=\Phi_{\rm B}/\sqrt{\dot{m}}$ (bottom panel). The MAD parameter saturates at $\phi_{\rm mad}\approx15$, corresponding to the horizontal black line.}
   \label{fig-grmhd}
\end{figure}

To assess when the simulation reaches the MAD state, we compute at the black hole's event horizon: the mass accretion rate, $\dot{m}$, and magnetic flux threading the horizon, $\Phi_B$, both in dimensionless units. The mass accretion rate is defined as $\dot{m}= \int_{r=r_{\rm h}} \rho u^r \sqrt{g} d\theta d\phi$, where $u^{\rm r}$ is the radial component of the velocity and $\sqrt{g}$ the determinant of the metric. The magnetic flux is defined as $\Phi_B = 0.5 \int_{r=r_{\rm h}} \mid B^r \mid \sqrt{\gamma} d\theta d\phi$, here $B^{\rm r}$ is the radial component of the magnetic field and $\sqrt{\gamma}$ the determinant of the spatial part of the metric \footnote{Note that we use the spatial part of the metric here which differs from the mass accretion since {\tt BHAC} uses 3+1 formalism which results in the Lapse function being contracted in the definition of the magnetic fields.}. We also define the MAD parameter $\phi_{\rm mad}=\Phi_{\rm B}/\sqrt{\dot{m}}$, which was introduced by \cite{tchekhovskoy2011} to quantify that a simulation reaches the MAD state when $\phi_{\rm mad}\sim 15$. All three quantities, $\dot{m}$, $\Phi_B$ and $\phi_{\rm mad}$ are shown in Figure \ref{fig-grmhd}.
The accretion flow reaches for the first time the MAD state at $\approx3000 ~\tunit$, when the magnetic flux on the horizon saturates as $\phi_{\rm mad} \sim 15$. Our MAD simulation shows globally similar properties to standard simulations on spherical grids; see Appendix \ref{app-A} for a comparison. 

\begin{figure*}[ht]
  \centering
  \includegraphics[width=\textwidth]{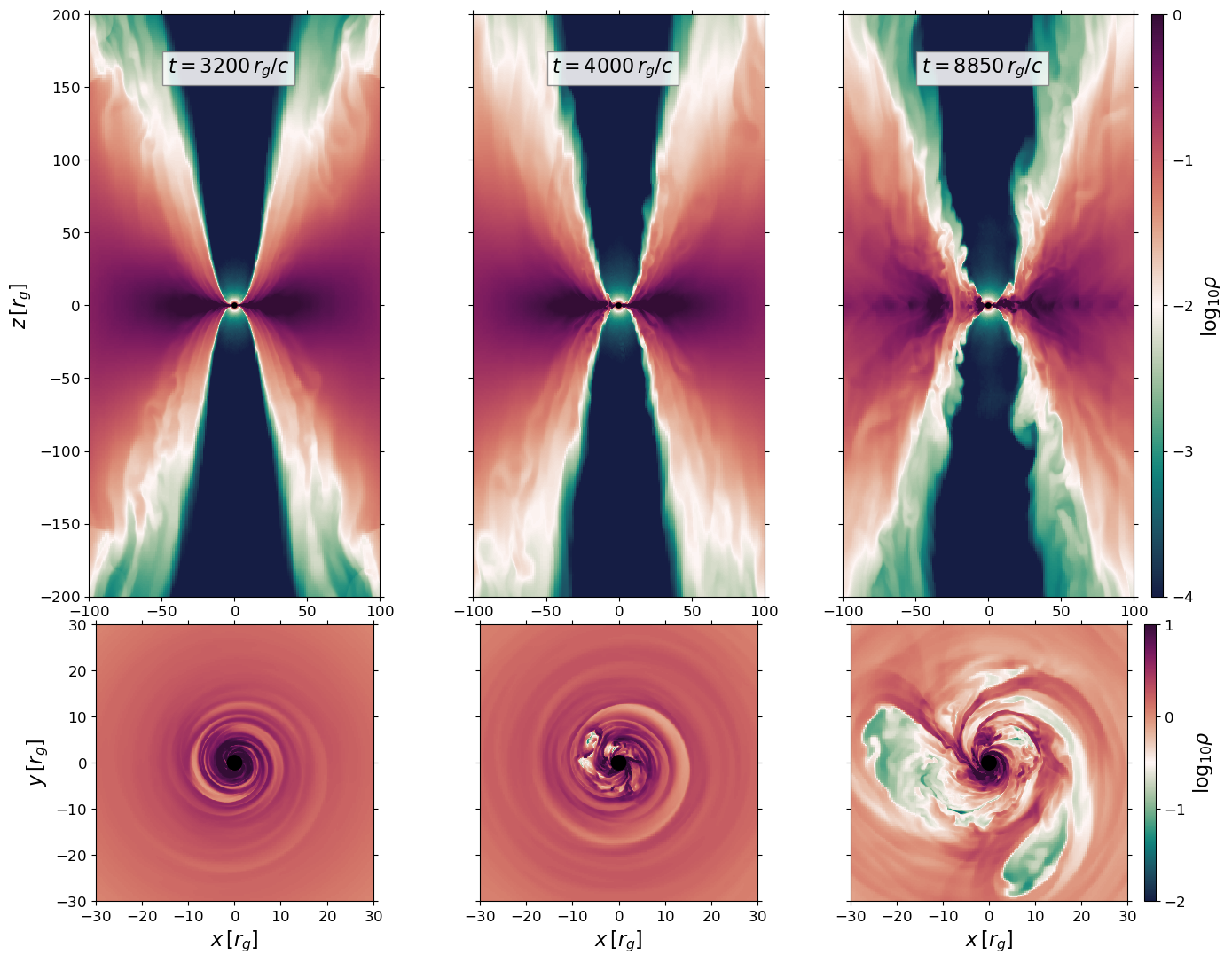}
\caption{Density profiles in the x-z plane (top row) and x-y plane (bottom row, notice different axis scaling compared to the top row). The left column shows the simulation at $t=3200 r_{\rm g}/c$, well before the magnetic flux is saturated and a MAD state is reached. At this point, the jet-wind shear layer is more laminar. Middle column: at $t=4000 r_{\rm g}/c$, the accretion flow reaches the MAD state for the first time. In the bottom panel, small under-density regions are visible, indicative of potential flux eruptions, which can be seen as drops in the integrated $\Phi_{\rm B}$ time series shown in Figure \ref{fig-grmhd} A first sign of waves can be seen in the jet-wind shear layer. Right column: simulation snapshot at $t=8850 r_{\rm g}/c$, large under-densities in the bottom panel is seen near the horizon, making the accretion flow non-axisymmetric. The under-densities correlate with large dissipation events of $\Phi_{\rm B}$; see again middle panel Fig. 1. Additionally, large-scale waves are present in the jet-wind shear layer.}
   \label{fig-panel}
\end{figure*}

In Figure \ref{fig-panel}, we show two-dimensional maps of the logarithm of density sliced along the spin axis, top row, or along the equatorial plane, bottom row. Initially, the jet is more laminar in the left column as flux is still building up on the horizon, and no eruptions have occurred. In the middle column, after the system has reached the MAD limit, flux tubes can be seen in the x-y plane, indicated by lower densities in the disk. The exhaust from magnetic flux eruptions generates these flux tubes. During the eruptions, magnetic energy is dissipated via large equatorial current sheets generated when the disk becomes magnetically arrested, and the northern and southern jets get in direct contact \citep{ripperda2022}. The exhaust of these sheets forms flux tubes containing vertical magnetic fields that spiral outwards in the disk before dissipating due to Rayleigh-Taylor mixing \citep{zhdankin2023}.

\begin{figure}
\centering
    \includegraphics[width=0.5\textwidth]{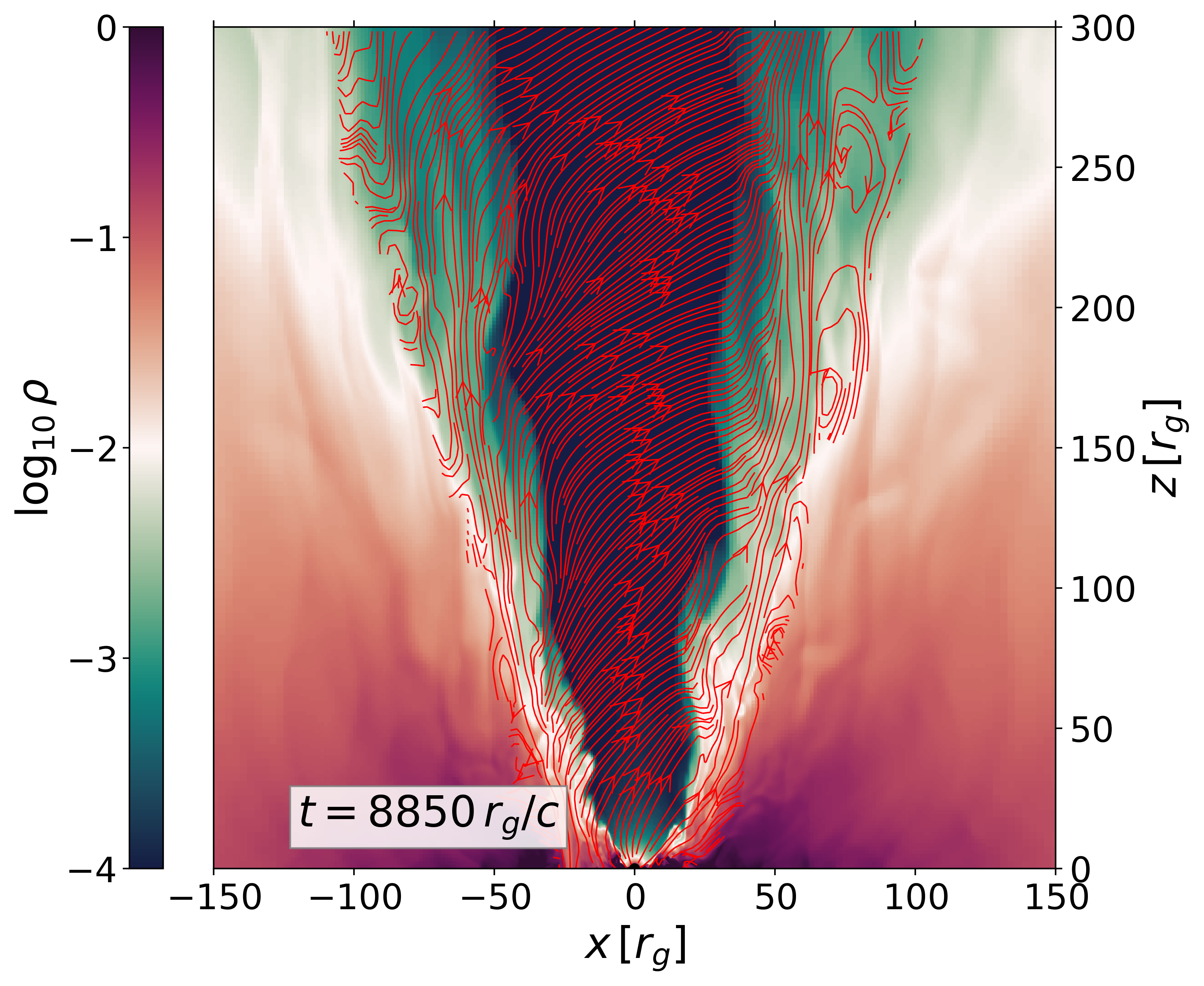}
    \caption{The density profile in the x-z plane overplotted with the magnetic field lines in the same plane. The magnetic field lines are sheared in the jet-wind shear layer, and roll-ups and magnetic islands are visible.}
   \label{fig-field-lines}
\end{figure}

In the middle panel, small amplitude waves propagate along the shear layer, interfacing the higher-density disk wind and low-density jet in the top panel. Large amplitude waves propagate outwards in the right column because the system produces strong magnetic flux eruptions; see Figure \ref{fig-grmhd} at $t=8800 r_{\rm g}/c$. The panels correspond to the vertical lines in the middle panel of Fig. \ref{fig-grmhd}. The variability introduced by the flux eruptions at the base of the jet acts like a forced oscillator, introducing waves that propagate and grow from near the event horizon to the shear layer between the jet and the disk. The waves propagate to large scales, growing in size while shearing magnetic field lines and generating field reversals, shown in Fig. \ref{fig-field-lines}.

\begin{figure}
\centering
    \includegraphics[width=0.5\textwidth]{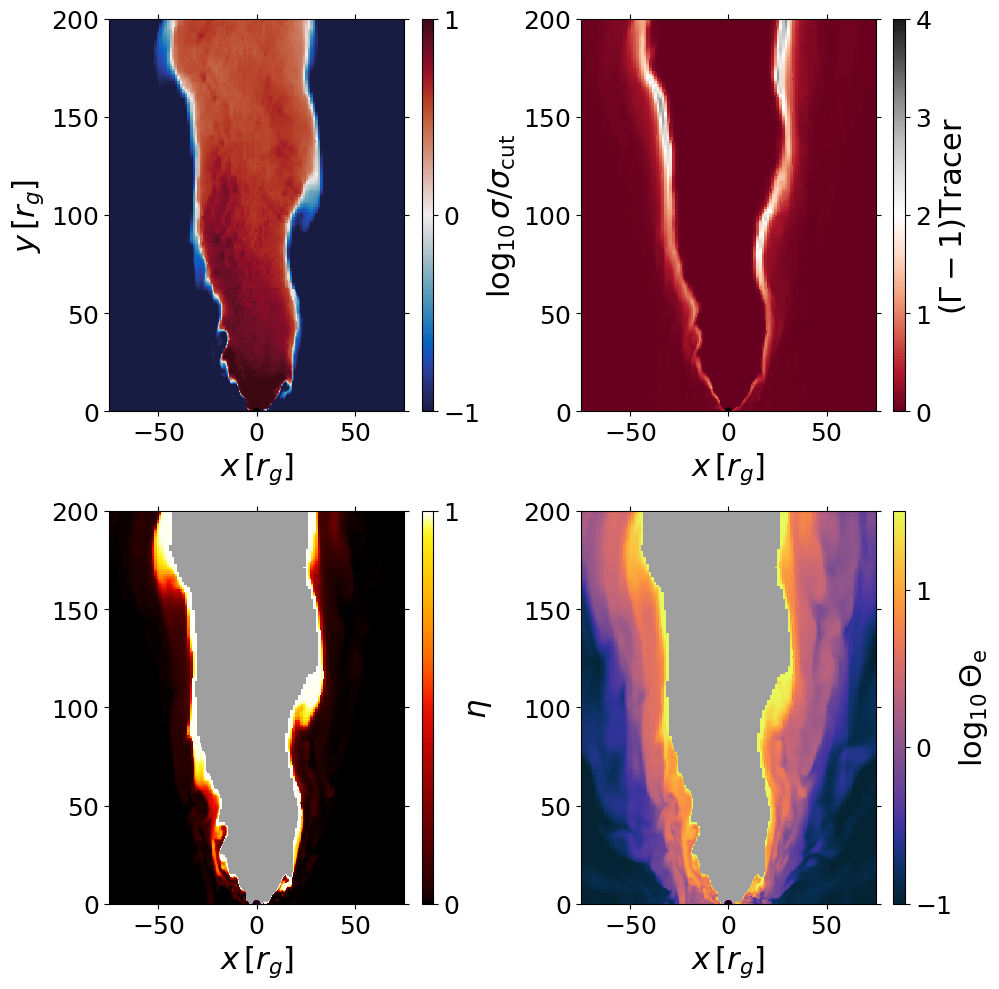}
    \caption{Simulation snapshot at $t=8850~\rg/c$ slice along the x-z plane of various quantities used for the radiative transfer calculation. Top left: cold magnetization $\sigma=b^2/\rho$. The inner core of the jet, which is at $\sigma\geq \sigma_{\rm cut}=5$, is dominated by the simulation floors, also visible in Figure \ref{fig-panel} by the low-density regions in the jet. Top right: the specific kinetic energy $\Gamma-1$ multiplied by our passive scalar. The scalar is set to zero for the floors and unity for disk matter. At the jet-wind shear layer, we see an increase in $\Gamma-1$ around the locations of the waves, indicating that matter originating from the disk is mixed in the shear layer and accelerated to high bulk Lorentz factor. Bottom left: acceleration efficiency $\eta$, where $\eta=1$ means $\kappa$-DF only, whereas $\eta=0$ is thermal-DF only. The jet-wind shear layer shows $\eta=1$, coinciding with the high temperatures in the right bottom panel. Grey region indicates $\sigma\geq5$, which is excluded from our GRRT computations. Bottom right: the electron temperature $\Theta_{\rm e}$ as prescribed by Eqn. 1b. The largest temperatures are found in the jet-wind shear layer. Grey region indicates $\sigma>5$.}
   \label{fig-radtrans-quan}

\end{figure}

To assess the potential effect of the waves on the polarized emission, we show slices along the x-z plane in Figure \ref{fig-radtrans-quan} at $t=8850~r_{\rm g}/c$ (right panel in Figure \ref{fig-panel}) of several quantities relevant to the radiation transport. We find that distinct patches of magnetization close to $\sigma\sim1$ can be seen along the jet-wind shear layer coinciding with the location of the waves in the top right panel in Fig. \ref{fig-grmhd}. The higher magnetization is important since particle acceleration is more efficient at higher magnetization values \citep{Sironi2014}. We use a passively advected tracer to study the acceleration of wind-based material to relativistic speeds at the jet-wind surface. The tracer is set to zero when the density or pressure is set to floors and is set to one in the initial torus. We then evolve this quantity as a passive scalar, tracing the advection of disk-based material into regions that were at least once set to floors. We find that matter from the disk (un-floored material) around the location of the jet-wind shear layer waves is accelerated to high bulk Lorentz factor, and emission generated in the waves is emitted from on un-floored matter originating from the accretion disk, shown by non-zero tracer and $\Gamma\geq 2$ in the right-top panel in Fig. 4. Looking at the bottom left panel, the waves also correlate with regions with a large fraction of non-thermal electrons since $\eta\sim1$, which is an evident result of the high magnetization (and low plasma-$\beta$), and our choice of electron distribution model (Eqn. 6). Finally, the electrons are relativistically hot (i.e., $\Theta_{\rm e} > 1$), as shown in the bottom right panel.  We associate the relativistic temperatures and the large fraction of non-thermal electrons with the heating of the plasma by the waves due to the dissipation of magnetic energy, as predicted in \cite{sironi2021}. In summary, the shear layer that is at the surface of the jet interfacing with the wind has a magnetization of order unity, has a high relativistic electron temperature, is moving at relativistic bulk Lorentz factors ($\Gamma\sim3$), and is likely dominated by non-thermal electrons.  

% \begin{figure}[h!]
%   \centering
%     \includegraphics[width=0.45\textwidth]{waves_327.png}
%     \caption{}
%    \label{fig-waves}
% \end{figure}

\subsection{Spectral distribution functions}\label{spectra}

\begin{figure}
  \centering
    \includegraphics[width=0.49\textwidth]{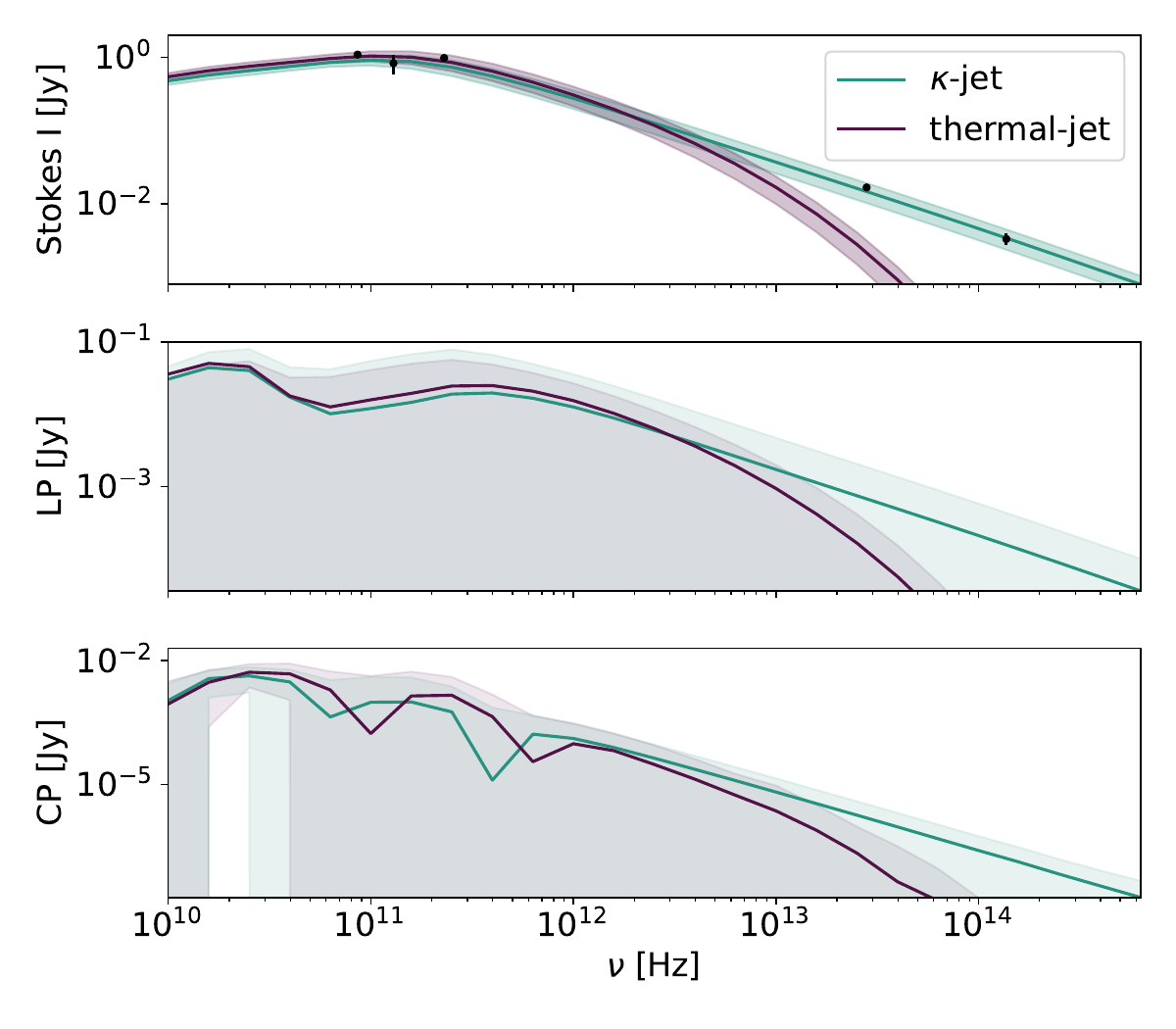}
    \caption{Spectra of time-averaged Stokes $\mathcal{I}$ (top panel), linearly polarized (LP) flux (middle panel), and circular polarized (CP) flux (bottom panel) for both the $\kappa$-jet (cyan) as well as thermal-jet (purple) models. The shaded area shows a standard deviation on the time-averaged fluxes. The thermal-jet under-produces the NIR emission compared to observations for Stokes $\mathcal{I}$, while the $\kappa$-jet recovers the observed spectral slope. The $\kappa$-jet model obtains similar LP and CP as the thermal-jet at lower frequencies ($\nu \lesssim 10^{13}$ Hz), while at higher frequencies it produces higher fluxes.}
   \label{fig-sed}
\end{figure}

In Figure \ref{fig-sed}, we show the spectral distribution functions of our thermal-DF model (thermal-jet) and $\kappa$-DF model ($\kappa$-jet) models. The top panel shows the total intensity (Stokes $\mathcal{I}$) as a function of frequency. The thermal-jet model recovers the low-frequency part of the spectrum accurately, e.g., $\nu < 10^{12}$ GHz, however at higher frequencies, it drops off too fast, which is consistent with \cite{davelaar2019}. In the case of the $\kappa$-jet model, the high-frequency emission is enhanced and obtains a spectral slope of $\alpha\approx-1$, consistent with the observations. In contrast, the thermal-jet model underestimates the near-infrared flux. The $\kappa$-jet model predicts that non-thermal electrons emit energetic photons pre-dominantly in the jet boundary and are, therefore, a probe of dissipation of magnetic energy due to wave dynamics. To match the observed flux at 86 GHz $F_{86 {\rm GHz}} = 1 ~{\rm Jy}$ we set the mass scaling to $\mathcal{M}=1.5\times10^{25}$ g for both the thermal and $\kappa$-jet models. These units of mass correspond to a mass accretion rate of $\approx 10^{-4} M_\odot/{\rm year}$ and a jet power of $\approx 10^{42-43} {\rm ergs/s}$, both similar to values obtained in previous works \citep{chael2019,eventhorizontelescopecollaboration2019d,cruz-osorio2022}, and consistent with jet powers inferred from observations of M87 \citep{Prieto2016}.

The two bottom panels show linear polarization (LP) and circular polarization (CP). For LP, the thermal-jet achieves similar fluxes as the $\kappa$-jet at a lower frequency ($\nu \lesssim 10^{13}$ Hz), while at a higher frequency, a clear power-law is visible in the $\kappa$-jet case. A similar power-law is visible for CP, but at a lower frequency, the $\kappa$-jet model is comparable to the thermal case. Subsequent analysis is done at $86$ GHz. This frequency was chosen since it probes emission structures at the base of the jet \citep{hada2016a,walker2018a}. 

\subsection{Total intensity}\label{total-intensity}

\begin{figure}
  \centering
    \includegraphics[width=0.5\textwidth]{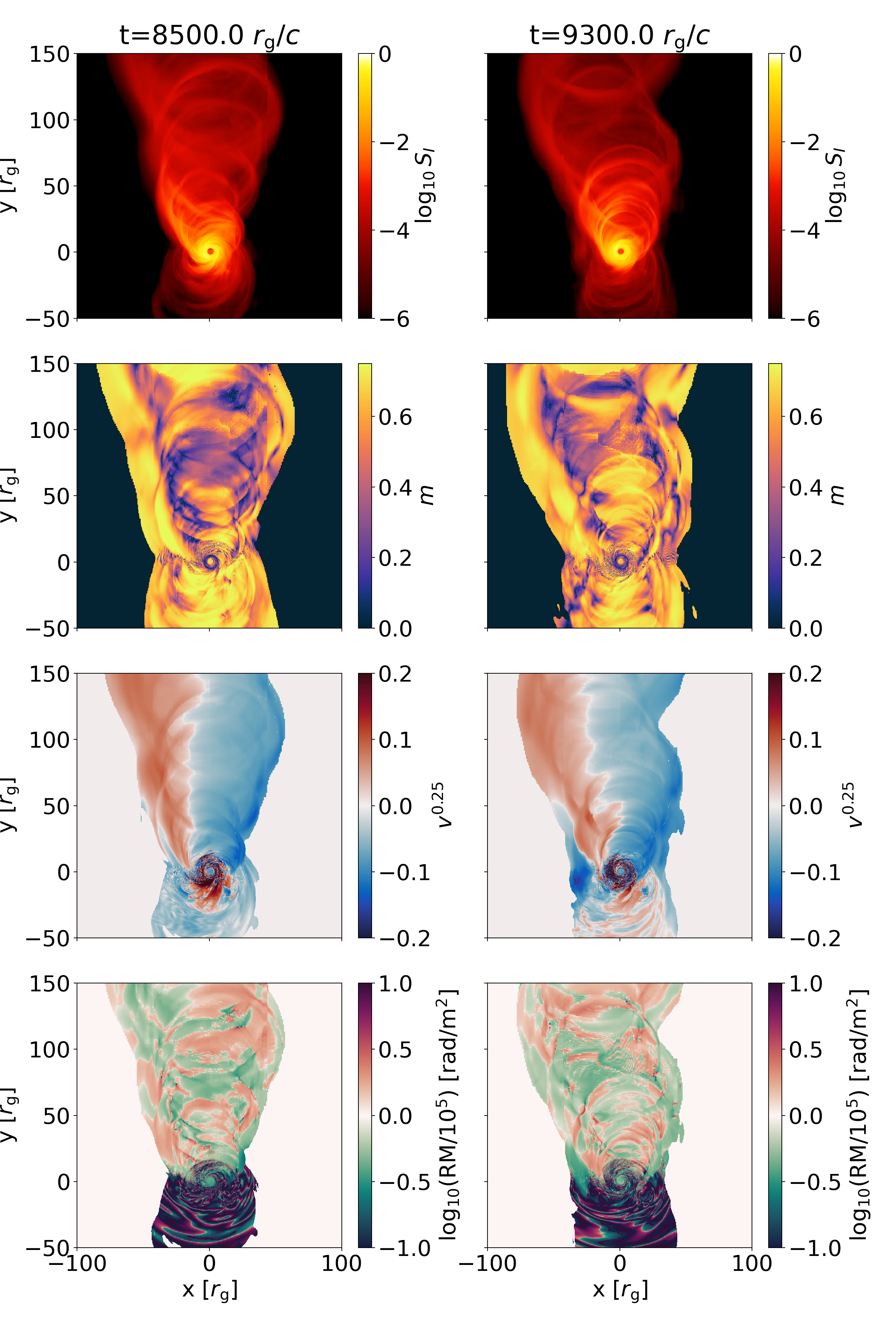}
    \caption{Synthetic synchrotron images at 86 GHz of two GRMHD simulation snapshots at $t=8500 r_{\rm g}/c$ and $t=9000 r_{\rm g}/c$. Top row: total intensity. Second row: LP fraction $m$, Eqn \ref{eqn-lp}. Third row: CP fraction $v$,  Eqn. \ref{eqn-cp}. Bottom row: Rotation Measure, Eqn \ref{eqn-rm}, normalized by $10^5$ rad/m$^2$. Large values of linear polarization at the edges of the jet are artificial since they are caused by regions of very low total intensity.}
    \label{fig-grrt-panel}
\end{figure}

In the top panels of Figure \ref{fig-grrt-panel}, we show total intensity maps for our $\kappa$-jet model at $86$ GHz. The images correspond to $t=8500~r_{\rm g}/c$ and $t=9300~r_{\rm g}/c$. At the core of the image, a darkening is visible, corresponding to the "black hole shadow" \citep{Luminet1979,Falcke2000}, recently observed at 86 GHz by \cite{Lu2023}. At the later stage (right panel), when the surface waves are prominently visible in the GRMHD simulations, helical wave-like substructures can be distinguished within the jet at larger scales.

\subsection{Linear polarization}\label{linpol}

This subsection summarizes the LP results computed at $86$ GHz. In unresolved LP fraction, we find that enhanced emission regions generate loops in a Stokes $\mathcal{Q}-\mathcal{U}$ diagram during the magnetic flux eruptions. The resolved LP fraction is then inversely proportional to the magnetic flux on the horizon, resulting in an enhancement of the fraction as the flux goes down. Furthermore, we find that the waves seen in the GRMHD simulation imprint themselves in LP maps as they lower the LP fraction. The effect of non-thermal $\kappa$-DF on the LP is minor, although we see a slight decrease compared to the thermal model for the LP fraction in the jet.

\subsubsection{Core}

\begin{figure}
  \centering
    \includegraphics[width=0.49\textwidth]{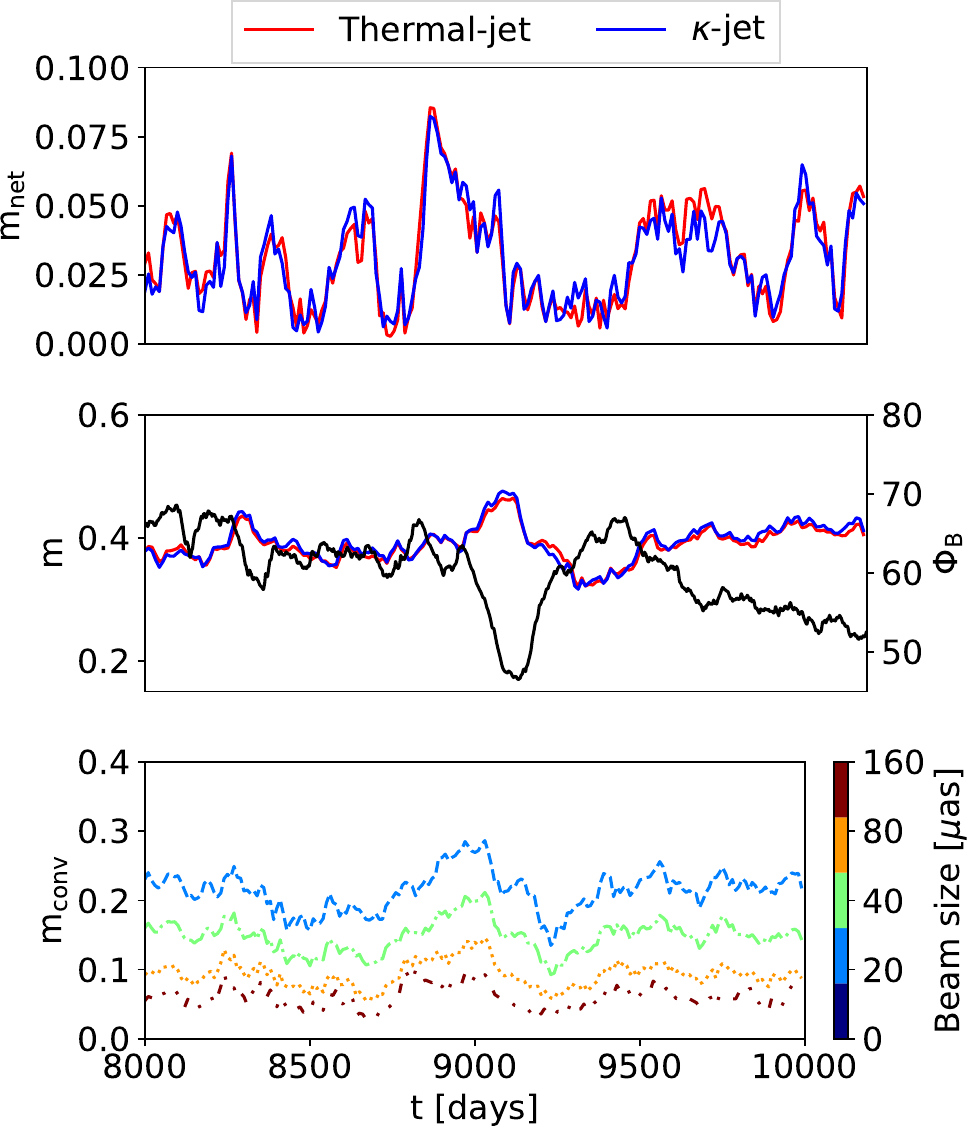}
    \caption{LP fraction as a function of time. Top panel, net LP fraction $m_{\rm net}$, for the thermal and $\kappa$-jet models, showing an average $m_{\rm net}\sim0.026$. Middle panel: resolved LP fraction $m$, overplotted with the magnetic flux on the horizon $\Phi_{\rm B}$ (green curve). Both models show a substantially higher resolved LP fraction. Additionally, at the moment of a magnetic flux eruption, $t=9000~r_{\rm g}/c$, an increase in $m$ is visible. Bottom panel: LP fraction as function as telescope beam size. The LP fraction decreases with increasing LP fraction due to incoherent addition.}
   \label{fig-lp-timeseries}
\end{figure}

In Figure \ref{fig-lp-timeseries} top panel, we show a time series of the unresolved LP fraction $m_{\rm net}$ for both the thermal-jet (solid lines) as well as the $\kappa$-jet models (dashed lines), defined as

\begin{equation}
m_{\rm net} = \frac{\sqrt{ (\sum_{\rm pixels} S_Q)^2 + (\sum_{\rm pixels}S_U)^2}}{\sum_{\rm pixels} S_I}.
\end{equation}
The image integrated net LP fraction does not depend on telescope beam size since it is incoherent addition of the Stokes parameters. We obtain an average value of $m_{\rm net}=0.026$, consistent with the low values found by observations \citep{hada2016a,walker2018a}. The thermal and the $\kappa$-jet models show almost identical values. 

In Figure \ref{fig-lp-timeseries} bottom panel we show the resolved LP fraction, $m$, defined as

\begin{equation}\label{eqn-lp}
m = \frac{ \sum_{\rm pixels}  \sqrt{S_Q^2 + S_U^2}}{  \sum_{\rm pixels} S_I}.
\end{equation}

Here we follow the definition of resolved LP fraction from Eqn. 8 of \citep{eventhorizontelescopecollaboration2021}, which is an image-averaged linear polarization fraction, taking into account some telescope beam size. As a first step we assumed that the telescope fully resolves the image, taking the beam size to be much smaller than the intrinsic features we are interested in. Due to the coherent addition the resulting resolved LP fraction $m$ is substantially higher than the unresolved LP fraction $m_{\rm net}$. The reason for this is that we preserve the sign of $\mathcal{Q}$ and $\mathcal{U}$ in the summation, in the case of $m_{\rm net}$, but the sign is dropped for $m$. In reality, $\mathcal{Q}$ and $\mathcal{U}$ will be convolved with the telescope's beam, resulting in incoherent addition. This effect can be seen in the bottom panel of Fig. \ref{fig-lp-timeseries}, where we compute the convolved LP fraction $m_{\rm conv}$ for varying telescope beam sizes $20, 40, 80, 160 ~\mu$as. This computation is done by blurring the original images with a Gaussian filter where the full-width half maximum represents the beam size. As the telescope beam size increases, the underlying substructure in both $\mathcal{Q}$ and $\mathcal{U}$ is being averaged out, resulting in an overall drop in LP fraction asymptotically approaching $m_{\rm net}$ as the beam becomes comparable to the core size.

In the fully resolved LP fraction $m$ as well as the $20-80$ ~$\mu$as $m_{\rm conv}$ cases in Fig. \ref{fig-lp-timeseries}, at $t=9000 ~r_{\rm g}/c$, an increase in LP fraction is visible. This increase coincides with a flux eruption at the horizon, visible from the over-plotted $\Phi_{\rm B}$ curve (solid green line, identical to middle panel of Figure \ref{fig-grmhd}). The $m$ shows a fractional increase by  $20$\%, while $\Phi_{\rm B}$ shows a fractional decrease of 20\%, indicating an inverse relationship between the two quantities. A similar correlation can also be seen at the smaller eruption at $t=8400 ~r_{\rm g}/c$. The $\kappa$-jet model reaches identical LP fractions as the thermal-jet.  

Close to the flux eruption starting around $t\sim8700 ~r_{\rm g}/c$, we show the unresolved Stokes parameters in a $\mathcal{Q}/\mathcal{I}-\mathcal{U}/\mathcal{I}$ diagram. After the onset of flux eruption, we see a clear clockwise loop with an LP excess of $m_{\rm net}\sim0.06$. The loop we find is similar to the loops found by \cite{Marrone2008,wielgus22} at $230$ GHz during X-ray flares for Sagittarius A*. \cite{najafi2023} finds evidence that these loops could be generated by enhanced polarized emission in the accretion disk by orbiting flux bundles ejected into the disk after a flux eruption. The enhanced emission leads to a local polarization excess: as the emission increases, the polarized emission increases. This enhanced emission region, often called a hot spot, orbits through the local magnetic field. Since the spot only lights up a small region of the accretion disk, this hot spot acts as a probe for the underlying magnetic field geometry. The magnetic field geometry close to the jet base is mostly poloidal, therefore, Stokes $\mathcal{Q}$ and $\mathcal{U}$ generate four quadrants, like a spoke wheel pattern, in the image plane, which alternate in sign, see, e.g., \cite{Narayan2021,Gravity2023}. Since the magnetic field orientation periodically varies, Stokes $\mathcal{Q}$ and $\mathcal{U}$ will show sinusoidal behavior in the total integrated $\mathcal{Q}$ and $\mathcal{U}$ since either an excess of positive or negative $\mathcal{Q}$ and $\mathcal{U}$ is seen, see for more details \cite{Vos2022,najafi2023}.

\begin{figure}
  \centering
    \includegraphics[width=0.49\textwidth]{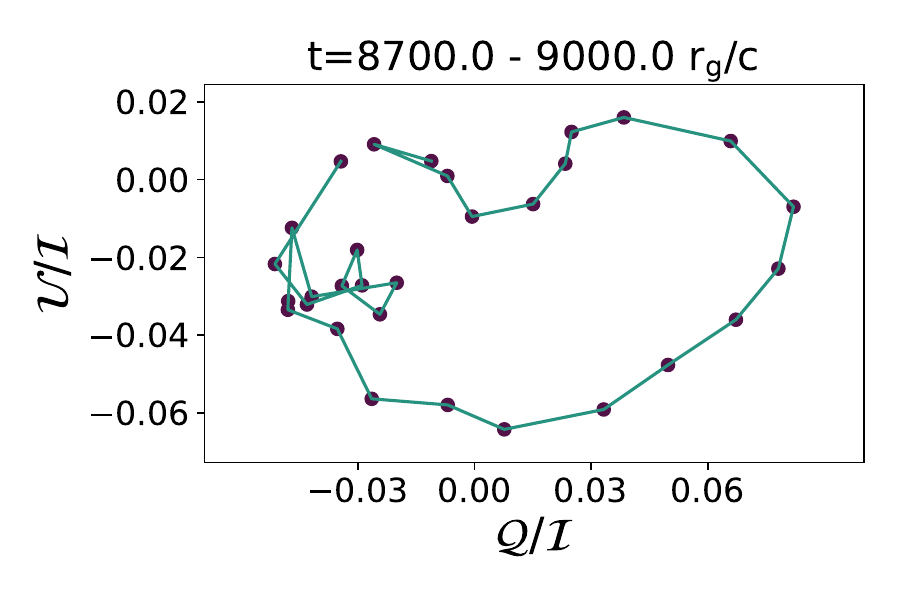}
    \caption{$\mathcal{Q}-\mathcal{U}$ diagram at 86 GHz. At the strongest magnetic flux eruption during the duration of our simulation at $t\sim 9000 r_{\rm g}/c$, we observe a pattern that manifests as a loop moving in a clockwise motion in a $\mathcal{Q}-\mathcal{U}$ diagram with a linear polarization excess of $m_{\rm net}\sim0.06$.}
   \label{fig-qu-loop}
\end{figure}

\subsubsection{Jet}

\begin{figure}
  \centering
    \includegraphics[width=0.49\textwidth]{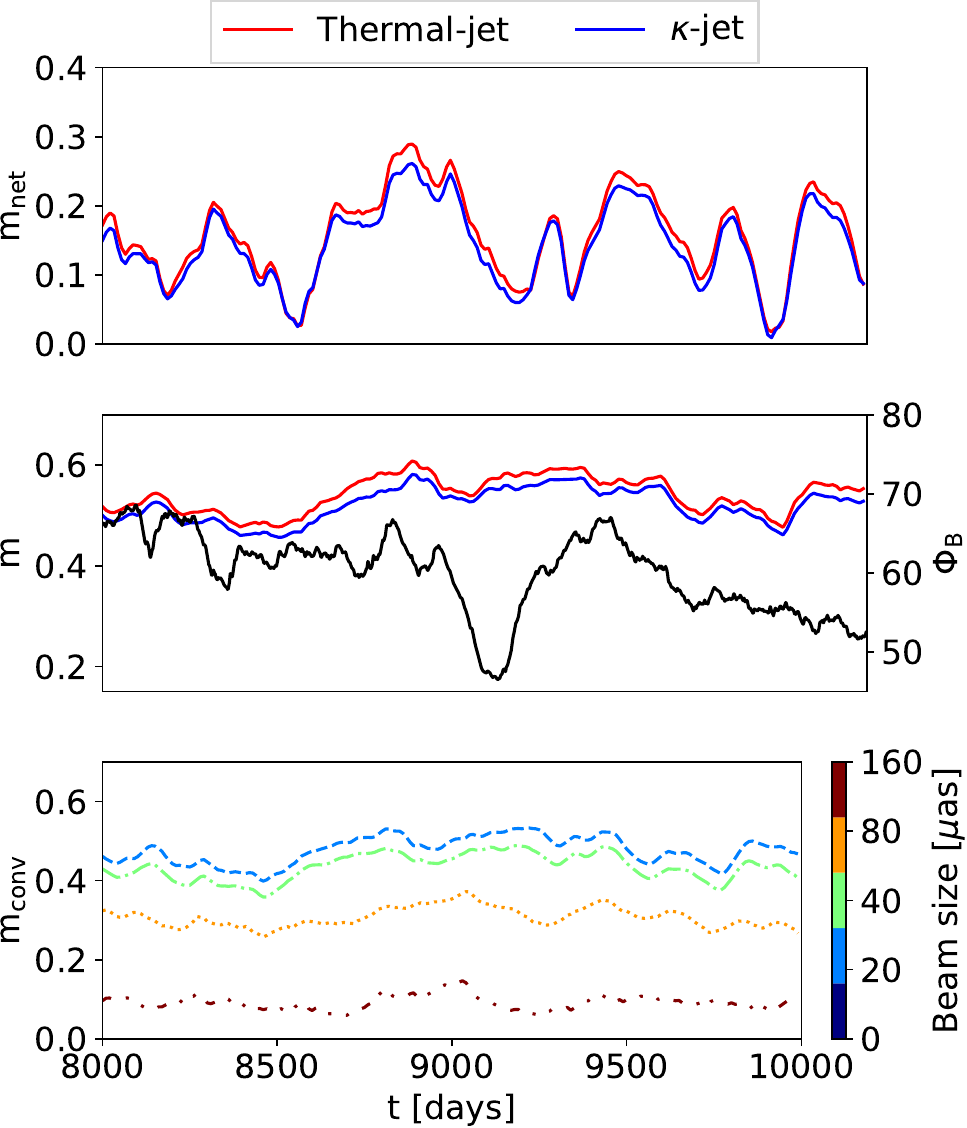}
    \caption{Identical to Figure \ref{fig-lp-timeseries} but now with the inner $30~\rg$ excised from the image plane of $m_{\rm net}$, $m$ and $m_{\rm conv}$} to compute the jet contribution only. Also for the excised LP fraction, both models obtain similar fractions.
   \label{fig-lp-jet-timeseries}
\end{figure}

To study the large-scale jet only, we exclude the image plane's inner $30 ~\rg$; in other words, we exclude the near-horizon emission. The resulting $m_{\rm net}$ and $m$ are shown in Figure \ref{fig-lp-jet-timeseries}. The unresolved LP fraction $m_{\rm net}$ ranges from 5-20\%. The resolved fraction, $m$, reaches values of around 50\%. The $\kappa$-jet model shows slightly lower values than the thermal-jet model in the case of $m$. In the second row of Figure \ref{fig-grrt-panel}, we show a map of $m$. In these maps, alternating regions of high and low linear polarization are visible, coinciding with enhanced emission in the total intensity panels in the top row. This indicates a potential correlation between LP fraction and the presence of the waves seen in the GRMHD simulation. This potential correlation will be investigated further in Section \ref{sec-waves}. The LP substructure seen in our simulation in Cartesian coordinates is absent in a low-resolution MAD simulation in spherical coordinates at an effective resolution of $192\times96\times96$ cells in $r,\theta,\phi$ respectively, see Appendix \ref{app-B}.

The jet stands out as a high-intensity emission region, while the accretion disk does not contribute to the emission at larger scales ($r>30~\rg$). Comparing the total intensity map with the LP fraction map, the foreground disk surrounding the jet shows high values of $m$ (close to unity). To understand this behavior we evaluate the asymptotic limit of our fitting formula for the emission coefficients $J_\mathcal{S}$ (with $\mathcal{S}$ indicating one of the Stokes parameters), see Eqn. 31 in \cite{pandya2016}. Given that the disk has weak magnetic fields and low temperatures; we have $\nu/\nu_{\rm c}\ll 1$, with $\nu_{\rm c} = eB/(2 \pi m_e c^2)$, and $\Theta_{\rm e}\ll1$, we find that $J_\mathcal{I}/|J_\mathcal{Q}| \rightarrow 1.0$. This makes physical sense since due to the low temperature the thermal distribution function is narrow, which means that there is a quasi-mono-energetic population of electrons that is causing the emission so the polarization fraction should go to unity. This, however, does not alter the computation of our image integrated $m$, $m_{\rm conv}$ and $m_{\rm net}$, since this outer region has low intensity both polarized as well as Stokes $\mathcal{I}$, so they don't contribute to the numerator and denominator of Eqn 7 and 8. Additionally, we also exclude regions of very low intensity from our map, where we set the Stokes parameters of a pixel to zero if $S_\mathcal{I}/{\rm max}(S_\mathcal{I}) < 10^{-6}$.

\begin{figure}
  \centering
    \includegraphics[width=0.49\textwidth]{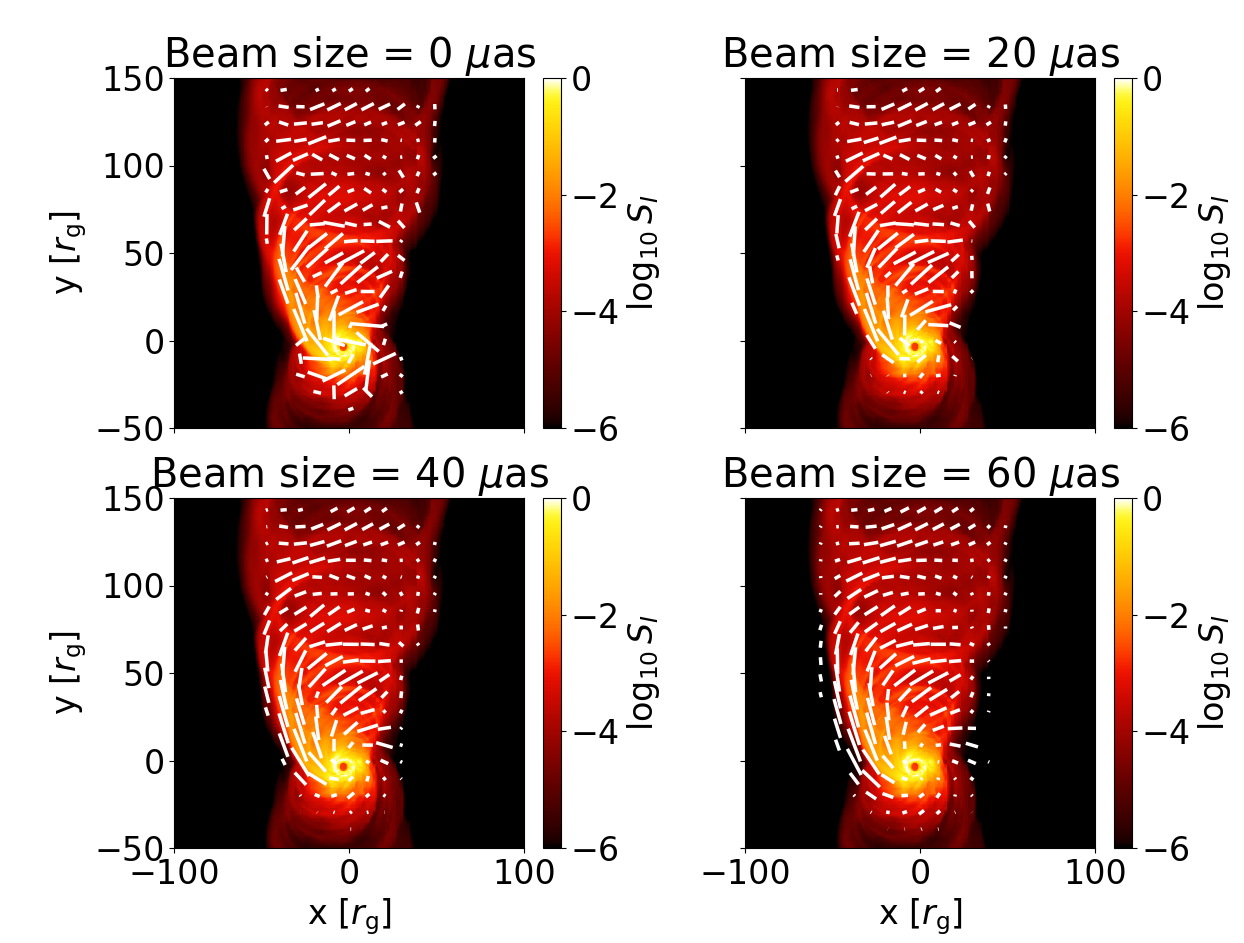}
    \caption{Polarization angle maps as a function of beam size, overplotted on the total intensity map of the top left panel in Fig. \ref{fig-grrt-panel}. Top left, zero beam size, the polarization vector shows a clear correlation with the ridges seen in total intensity. For increasing beam size, this correlation becomes weaker, however, some reorientation of the vector is still visible, see e.g. $x\sim 50 ~\rg$, $y\sim 50 ~\rg$.}
   \label{fig-multi-angle}
\end{figure}

Lastly, we compute polarization angle maps as a function of beam width. These can be seen in Fig. \ref{fig-multi-angle}, where we overplotted ticks of the polarization vector on top of the total intensity map from the top left in Fig \ref{fig-grrt-panel}. The length of the ticks is set by the linear polarization fraction, while the angle is set by the EVPA, as defined by $\chi = \frac{1}{2} \tan^{-1}(S_\mathcal{Q}/S_\mathcal{U})$, here we also exclude linear polarization fractions when $S_\mathcal{I}/{\rm max}(S_\mathcal{I}) < 10^{-6}$, so tick lengths are set to zero. For the case of zero beam size the polarization vector clearly follows the ridges of enhanced intensity. As the beam size increases, the correlation becomes weaker, however re-orientation of the polarization vector is also here visible, e.g. at $x\sim 0 ~\rg$ $y\sim 50 ~\rg$  the polarization pattern switches from diagonal, to horizontal, and back to diagonal. We only show beam sizes up to $60 ~\mu$as which is the expected resolution of the ng-EHT at 86 GHz \citep{Issaoun2023}, assuming identical baselines to the current EHT array (using, $\theta \sim 20 \mu {\rm as} (\lambda / 1 {\rm mm})$ with $\lambda = 3$ mm).

\subsection{Circular polarization}\label{circpol}

In this subsection, we study the CP fractions computed at $86$ GHz. We find that the resolved CP fraction decreases during a flux eruption as the inner accretion disk is ejected, resulting in a more dilute plasma to perform Faraday conversion. Overall, we find low unresolved and resolved CP fractions for our thermal and $\kappa$-jet models. The CP maps show sign reversals, indicative of alternating magnetic field orientation that coincides with the features seen in the linear polarization maps.  

\subsubsection{Core}

\begin{figure}
  \centering
    \includegraphics[width=0.49\textwidth]{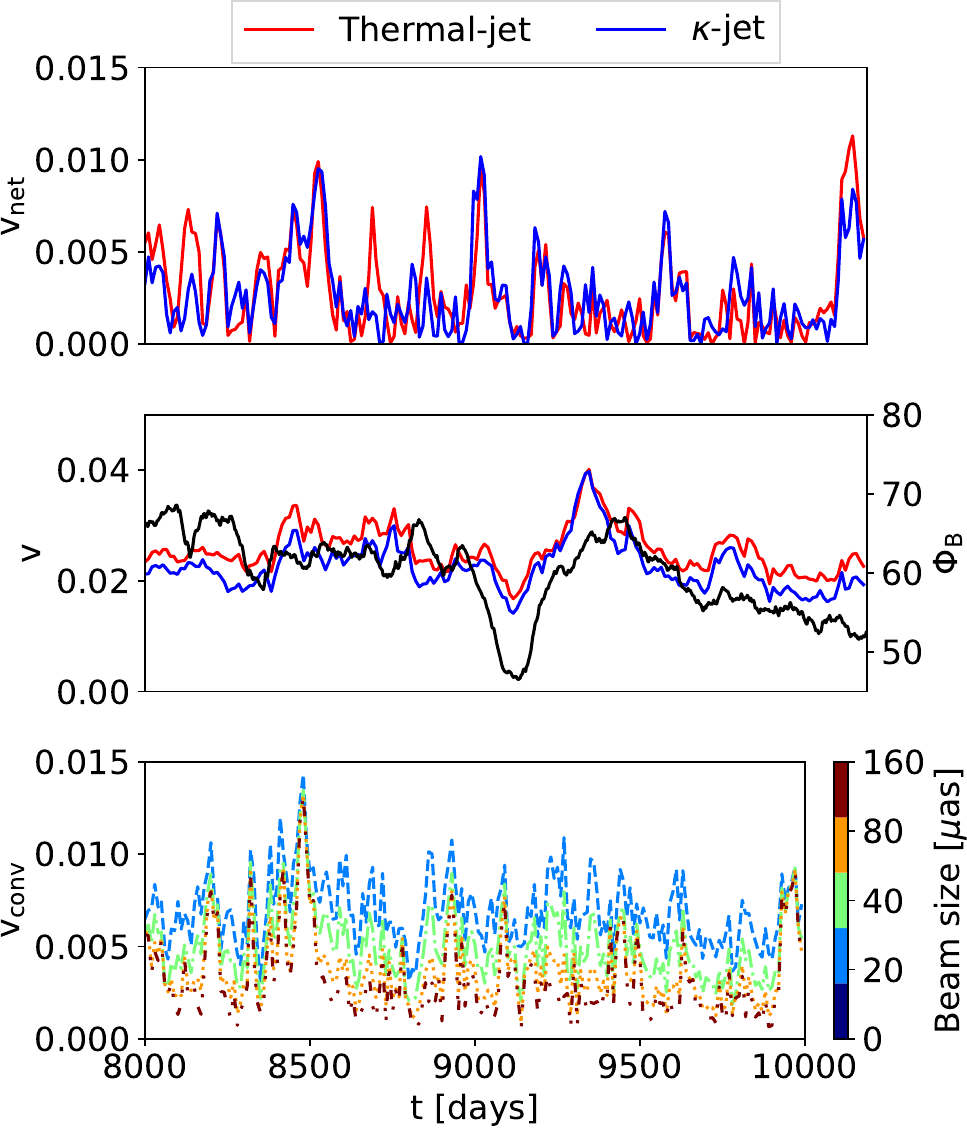}
    \caption{CP fraction as a function of time. Top panel, net CP fraction $v_{\rm net}$, for the thermal and $\kappa$-jet models, showing an average {$v_{\rm net}\sim0.026$}. Middle panel: resolved CP fraction $v$, overplotted with the magnetic flux on the horizon $\Phi_{\rm B}$ (green curve).The thermal-jet model and $\kappa$-jet model obtain similar CP fractions. Additionally, at the moment of a magnetic flux eruption, $t=9000~r_{\rm g}/c$, a decrease in $v$ is visible. Bottom panel: cP fraction as function as telescope beam size. The CP fraction decreases with increasing beam size due to incoherent addition.}
   \label{fig-cp-timeseries}
\end{figure}

In Figure \ref{fig-cp-timeseries} we show the unresolved and resolved CP fractions, $v_{\rm net}$ and $v$, defined as 

\begin{eqnarray}\label{eqn-cp}
v_{\rm net} = \frac{\sqrt{ (\sum_{\rm pixels} S_V)^2}}{\sum_{\rm pixels} S_I},\\
v = \frac{  \sum_{\rm pixels} \sqrt{S_V^2}}{  \sum_{\rm pixels} S_I}.
\end{eqnarray}

Both $v_{\rm net}$ and $v$ are small in value, as expected from synchrotron radiation \citep{rybicki1979}. In Figure \ref{fig-cp-timeseries} bottom panel, during the strongest flux eruption at $t=9000 ~r_{\rm g}/c$, both models show a slight decrease in CP fraction. The accretion disk enhances the amount of circularly polarized emission as Faraday conversion converts linear to circular polarization. Due to the ejection of the inner part of the accretion disk during a flux eruption, there is more dilute plasma in this region, resulting in a drop in the CP fraction.

\subsubsection{Jet}

\begin{figure}
  \centering
    \includegraphics[width=0.45\textwidth]{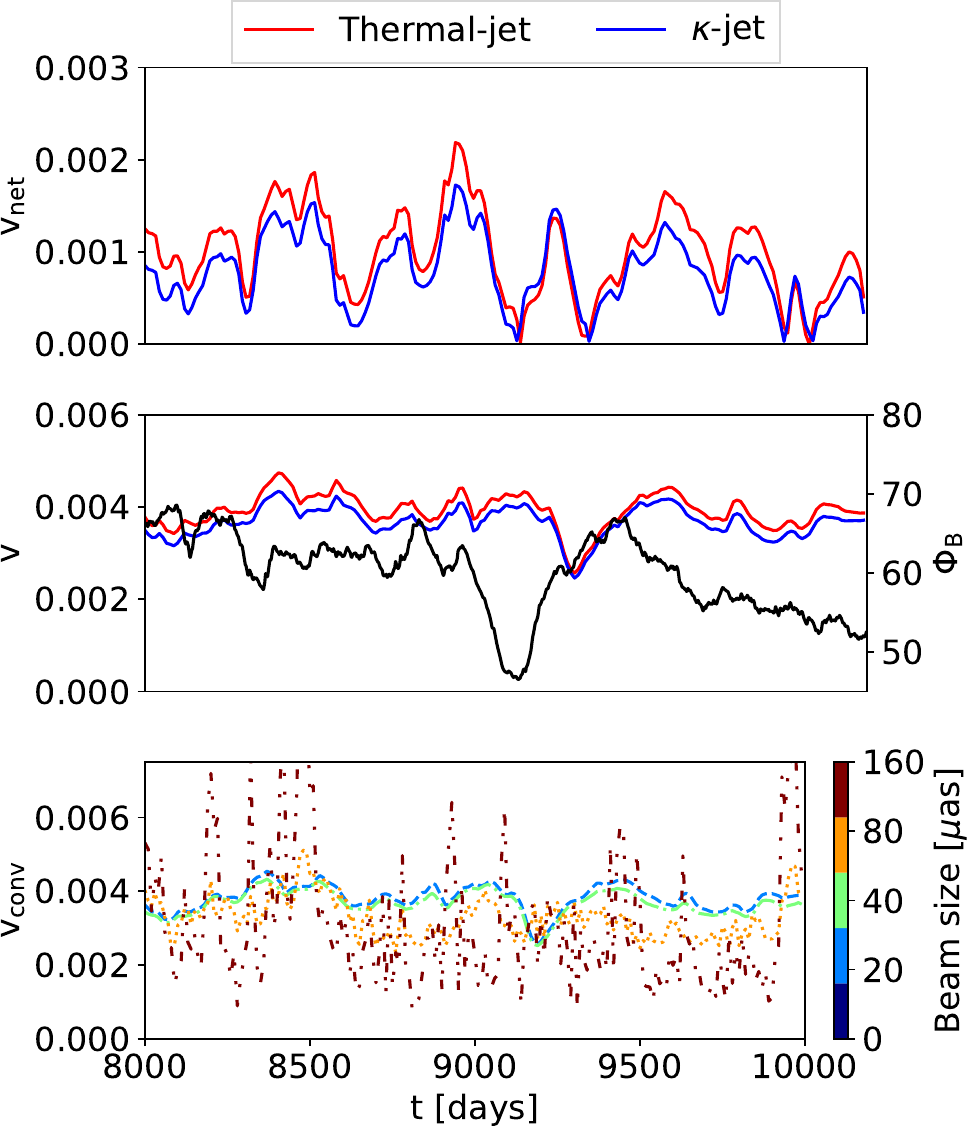}
    \caption{Identical to Figure \ref{fig-cp-timeseries} but now with the inner $30~\rg$ excised from the computation of $v_{\rm net}$, $v$ and $v_{\rm conv}$}. While for $v_{\rm net}$, both models are similar, for $v$, the $\kappa$-jet model similar CP fractions as the thermal-jet model.
   \label{fig-cp-jet-timeseries}
\end{figure}

In Figure \ref{fig-cp-jet-timeseries}, we compute $v_{\rm net}$ and $v$, but now also exclude the inner $30 \rg$ to exclude the near horizon emission and focus on the larger scale jet only. This exclusion results in smaller fractions than the entire image-integrated values since Faraday conversion happens in high-density regions. The third row of Fig. \ref{fig-grrt-panel} shows maps of $v$ where the CP fraction is smaller at larger radii. In the map, we preserved the sign of Stokes $\mathcal{V}$, which is set by the direction of the magnetic field along the line of sight. The image shows reversals in the sign of $v$ and additional ridges of low CP fraction. Since Stokes $\mathcal{V}$ carries information on the direction of the magnetic field along the line of sight, this indicates a potential orientation switch in the underlying magnetic field geometry. This reversal could be caused by the waves, as shown in Figure \ref{fig-field-lines}; we will further investigate the sign reversal in Section \ref{sec-waves}. 

\text{ }

\subsection{Faraday rotation}\label{RM}

Figure \ref{fig-grrt-panel} bottom row shows maps of rotation measures computed between 80 and 100 GHz. The rotation measure is defined as 

\begin{equation}\label{eqn-rm}
    RM = \frac{\chi_{1}-\chi_{2}}{\lambda_1^{2}-\lambda_2^{2}}
\end{equation}

where $\chi_\nu$ is the electric vector position angle (EVPA) computed at a specific frequency $\nu$ defined as $\chi=\frac{1}{2}\tan^{-1}({S_\mathcal{Q}/S_\mathcal{U}})$, and $\lambda$ the wavelength. Large RMs are visible along the jet, and in the core, we find values as large as $10^4-10^5$ rad/m$^2$. However, the core dominates the total RM, which is expected since the Faraday depth is larger due to higher density and lower temperatures. The relatively large value for the RM in the jet is somewhat surprising, given that the jet does not exhibit large Faraday depths. Given the small Faraday depth, the change in EVPA is not caused by Faraday rotation but is caused by the transverse gradients in $n_{\rm e}$, $\Theta_{\rm e}$ and $B$ in the emitting shear layer. These gradients result in emission at different frequencies to peak at different depths in the shear layer, which have different plasma properties, e.g., different magnetic field orientations will result in different orientations of the EVPA, giving rise to non-zero $RM$ values. 

\subsection{Properties along a ray}\label{rayprop}

\begin{figure}
  \centering
    \includegraphics[width=0.49\textwidth]{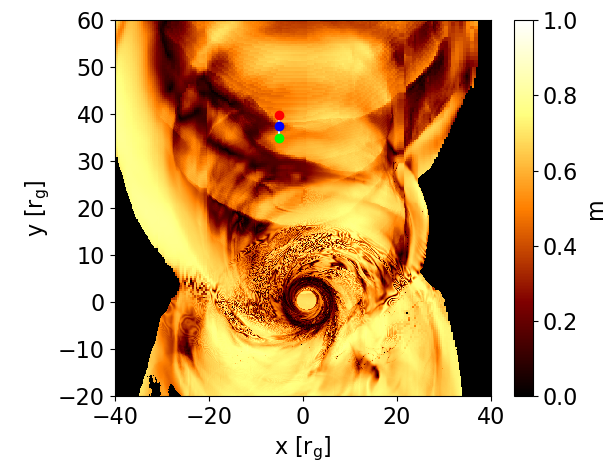}
    \includegraphics[width=0.49\textwidth]{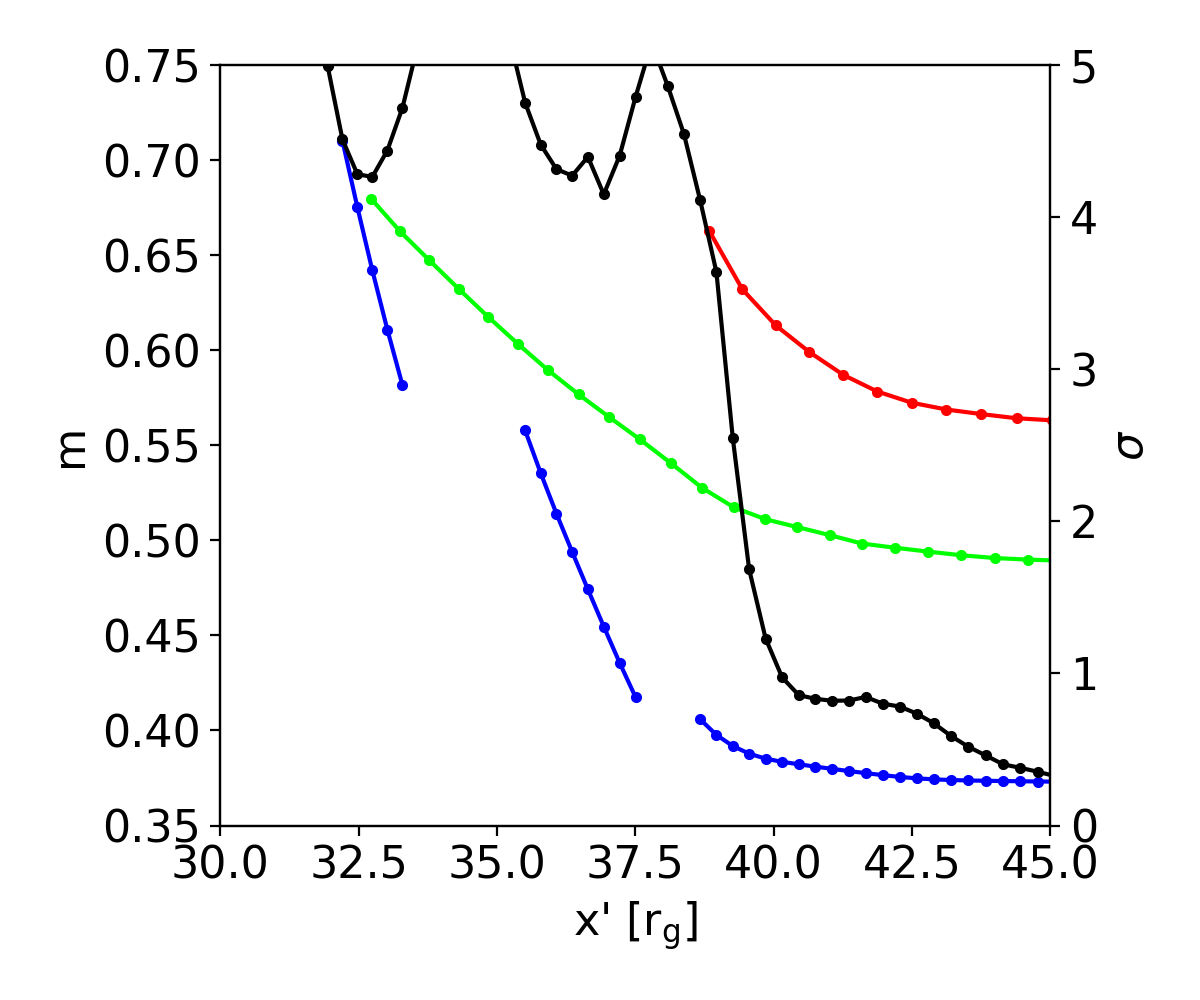}
    \caption{In the top panel we show a linear polarization map, $m$. Red, blue, and green dots indicate the pixels along which we compute the linear polarization as a function of the geodesics $x'$. The dependence on $x'$ is shown in corresponding colors in the bottom panel. The black line indicate magnetization along the pixel corresponding to the green dot (axis on the right). }
   \label{fig-rays}

\end{figure}

To test if the waves cause linear depolarization, we identified representative light rays showing high or low polarization fractions. The selected geodesics are indicated with the red, blue, and green dots in the top panel of Figure \ref{fig-rays}. We then compute the linear polarization fraction as a function of the Cartesian Kerr-Schild $x'$ coordinate, meaning smaller values of $x'$ are closer to the spin axis. The result of this is shown in the bottom of Figure \ref{fig-rays}. We show the geodesics only as they approach the jet-wind surface and only show segments when the local magnetization $\sigma<5$, meaning no radiation transfer is applied when the geodesic is inside the jet. The red-colored ray is terminated early, meaning that for larger $x'$ its net polarization fraction is higher, indicating that the shear layer at that point is thinner. The blue and green-colored rays have a larger travel path, meaning the polarization starts to drop. In the case of the green-colored ray, the situation is even more interesting. The line of this ray is interrupted twice, indicating it crossed into two regions of high magnetization but then left. We interpret this as the ray crossing through a rolling wave, similar to the wave seden at $y=-100~r_{\rm g}$ in Figure \ref{fig-panel}. We test this by also computing the magnetization along the ray, this is shown also in the bottom panel of Fig. \ref{fig-rays} by the black line. This line crosses our magnetization threshold ($\sigma > 5$) twice. In general, the waves alter the thickness of the jet-wind shear layer, and their presence results in varying path lengths inside the shear layer for different rays, which affects the linear polarization fraction. 

\subsection{Correlation with the jet-wind shear waves}\label{sec-waves}

\begin{figure}
\centering
    \includegraphics[width=0.49\textwidth]{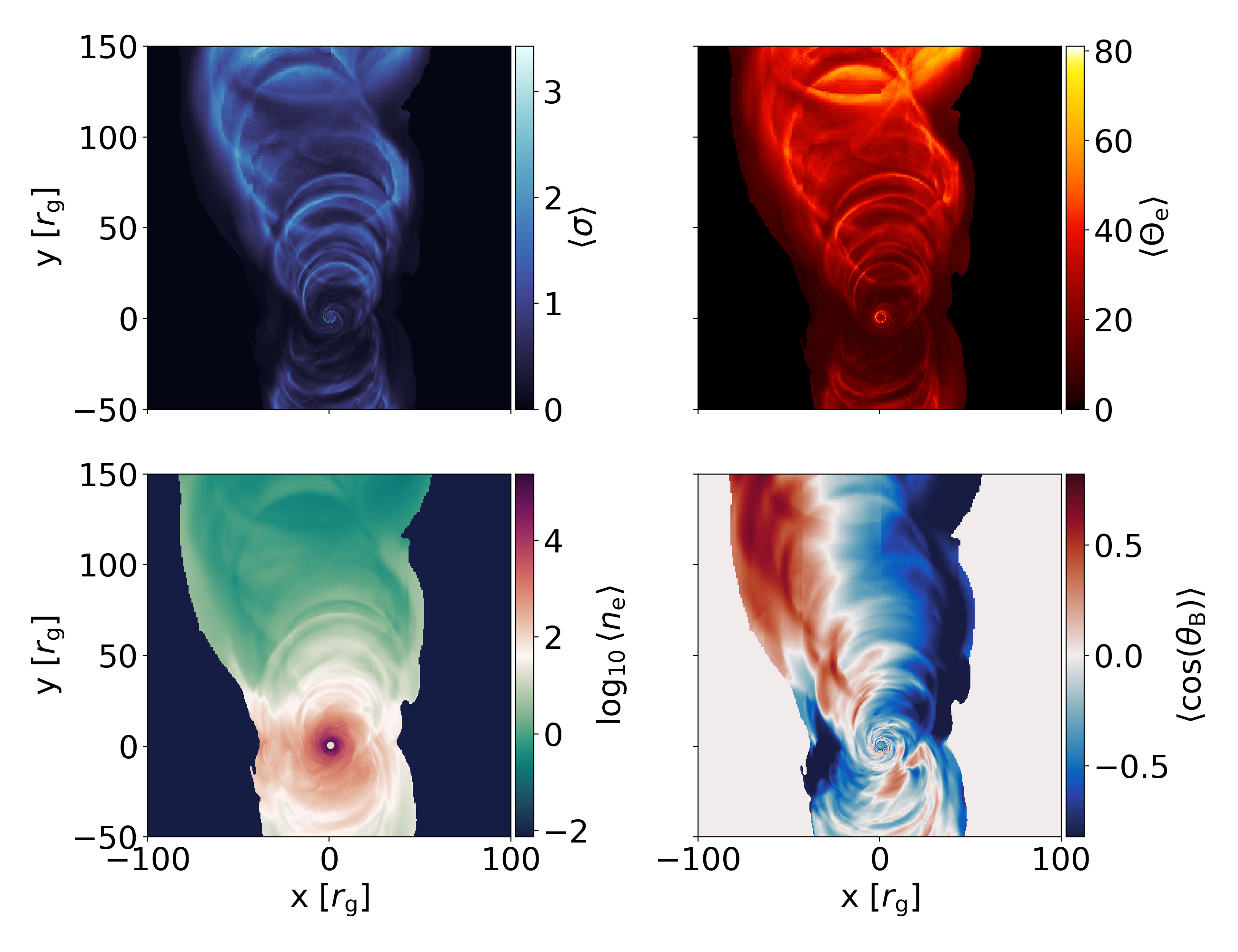}
    \caption{Emissivity weighted averages of the magnetization $\sigma$ (top left), electron temperature $\Theta_{\rm e}$(top right), electron number density $n_{\rm e}$ (bottom left), and  the angle between the magnetic field and wave vector $\cos(\theta_{\rm B})$ (bottom right).}
   \label{fig-vol-panel}

\end{figure}

To connect the waves in GRMHD with the structures seen in the synthetic 86 GHz images, we compute emissivity weighted averages of the magnetization $\sigma$, the pitch angle between the wave vector and the magnetic field orientation $\cos(\theta_{\rm B})$, the electron number density $n_{\rm e}$, and the electron temperature $\Theta_{\rm e}$, via

\begin{equation}
    \langle q \rangle = \frac{\int j_{\nu} q d\lambda_{\rm aff}}{\int j_{\nu} d\lambda_{\rm aff}},
\end{equation}

where $q$ represents the weighted quantity, the result of this computation is shown in Figure \ref{fig-vol-panel}. The top left panel, $\sigma$, shows that the images' higher intensity features also have a larger magnetization. This agrees with the waves having a larger magnetization in Figure \ref{fig-radtrans-quan}. Additionally, the same patterns are visible in the higher electron temperatures (top right panel in Figure \ref{fig-vol-panel}), and the over-densities (bottom left panel in Figure \ref{fig-vol-panel}), also in agreement with the properties of the waves, as shown in Section 3.1.

For Stokes $\mathcal{V}$, we finally compare the emissivity weighted average of $\cos(\theta_{\rm B})$ (bottom right panel in Figure \ref{fig-vol-panel}). The overall map of $\cos(\theta_{\rm B})$, where $\theta_{\rm B}$ is the angle between the wave vector and the magnetic field vector, shows the same sign as Stokes $\mathcal{V}$, implying that reversals in Stokes $\mathcal{V}$, are caused by the shearing of the magnetic fields leading to reversed field orientations. 

\section{Discussion and conclusion}

In this Letter, we present a global GRMHD simulation in Cartesian Kerr-Schild coordinates in the MAD regime that shows the formation of waves along the jet-wind shear layer. We post-process this simulation with our polarized radiative transfer code and compute polarized spectral energy distributions, times series of polarization quantities, and synthetic images at 86 GHz. We find observational signatures of the surface waves seen in the GRMHD simulation in the polarization information. As the waves propagate outwards, they show up as bright features along the jet that alter the polarization signature of the jet at larger scales. As the waves alter the orientation of the magnetic field lines, the linear polarization fraction drops due to the cancellation of subsequently rotated Stokes vectors.

During magnetic flux eruptions, we find an inversed relation between $\Phi_{\rm B}$ and the LP fraction, meaning that as $\Phi_{\rm B}$ drops, the LP fraction increases. We see the opposite for the CP, where the fraction decreases as $\Phi_{\rm B}$ decreases. Both effects can be explained by the flux eruption ejecting the disk near the horizon; as the strong poloidal field arrests parts of the disk, the density drops, and the disk becomes more optically and Faraday thin, leading to lower Faraday rotation and conversion. This results in a decrease in CP and an increased LP fraction.

At the largest magnetic flux eruption in our simulation, we find that the unresolved Stokes $\mathcal{Q}$ and $\mathcal{U}$ at 3 mm show a clockwise loop in a $\mathcal{Q}-\mathcal{U}$ diagram with a polarization excess of $m_{\rm net}\sim0.06$. Loops like this were previously identified in the case of our galactic supermassive black hole SgrA*, either observationally \citep{Marrone2008,wielgus22,Gravity2023}, or in theoretical works, e.g. \citep{Vos2022,najafi2023}. A key difference is that the time scales in the case of M87 are longer, which puts the period of our loop at $\sim 3$ months, compared to $\sim 1$ hour in the case of SgrA*.

Our model recovers resolved polarization fractions that are too high compared to the ones measured by \cite{hada2016a}. However, when convolved with a more realistic telescope beam, we show that the LP fraction substantially drops since $\mathcal{Q}$ and $\mathcal{U}$ are averaged out due to patches within the beam having opposing signs. We find consistent rotation measures without invoking an external Faraday screen, and we recover the observed spectral shape from radio to optical frequencies \citep{ehtmwlscienceworkinggroup2021}. Although we limit ourselves to M87, our results generally apply to other LLAGNs reaching the MAD state since the waves result from the underlying flow geometry and the flux eruptions typical for such systems. We expect these polarization signatures to be independent of black hole masses and accretion rates. 

In the literature, studies of wave instability at jet-wind shear layers are typically limited to analytical studies, e.g., linear analysis \citep{ferrari1978,sobacchi2018a,chow2022a} or with numerical MHD/Particle-in-Cell studies of local idealized setups \citep{hardee2007,Perucho2010,sironi2021}. The overall conclusions of these works are that jets, if in the right conditions, can be prone to the excitation of waves due to linear instability, e.g., KH waves. These waves are asymmetric, meaning they have different plasma properties on either side of the shear layer. Previous work by \cite{sironi2021} showed that particles can be accelerated to high energies in mildly relativistic, magnetized asymmetric shear flows. However, evidence of wave instabilities in global simulations is sparse and often underresolved in 3D simulations due to the restrictions on the large-scale resolution in spherical coordinates; see \cite{chatterjee2019,wong2021}. Observationally, some evidence for wave-like perturbations at large distances from the central engine is found by, e.g., \cite{Perucho2007,Pasetto2021,Issaoun2022}.

In this work, we did not perform a rigorous linear analysis to determine if the waves could be grown from linear scales and what instability is driving them. Visually, the jet is initially stable and shows no waves, while when the system reaches the MAD state and the first flux eruptions occur, waves travel outwards along the jet-wind boundary. Therefore, the waves we see are more likely to grow by forced oscillations of the jet base due to accretion variability and ejecta from magnetic flux eruptions. The waves are already non-linear within a few gravitational radii, which would require short linear growth times. A more likely scenario is that the jet base's variability efficiently drives the waves' growth and becomes non-linear at larger scales. A study of the conditions under which these waves are growing by either applying linear analysis \citep{chow2022a} to local conditions extracted from our simulation or by performing local idealized simulations will be done in future works. 

Compared to previous global simulations of LLAGN jets, our simulations stand out due to the Cartesian nature of the grid, allowing us to resolve the jet to larger distances compared to more standard simulations with spherical grids as used in, e.g., \cite{eventhorizontelescopecollaboration2019e}. This higher resolution enables us to follow the perturbations at the jet base to larger scales. However, the waves we see in our simulations are not a result of our choice of coordinates and can be found in spherical simulations if run at sufficiently high resolution \citep{ripperda2022}, as shown in Appendix \ref{app-A}. 

The evidence we find for the shear layer waves in our simulation may have implications for particle energization. The waves could introduce a source of turbulence or reconnection in the shear layer \citep{sironi2021}. These processes could lead to electron acceleration, resulting in non-thermal emission that could explain the edge brightening of AGN jets. Additionally, reconnection induced by the waves could drive injection of high energy ions originating from the disk into shear-driven acceleration, potentially producing ultra high energy Cosmic Rays, see, e.g., \cite{caprioli2015,rieger2019,Mbarek2021}.

In this work, we discovered a correlation between jet-wind surface waves and polarized emission properties. We find evidence that the substructure in the jet, in the form of waves, imprints itself on the Stokes $\mathcal{I}$ and LP maps. We identify ridges and alternating low and high linear polarization fractions as tell-tale signatures of these waves. Although currently below the achievable resolution of VLBI arrays, this effect might be resolvable by future next-generate arrays such as the next-generation EHT (ng-EHT) \citep{Ricarte2023,Issaoun2023}. If the ng-EHT operates at 86 GHz, it will achieve a resolution of $\sim 60 ~\mu$as, or $20 ~r_{\rm g}$ scaled to M87, which would be sufficient for resolving the features we find in this study.   
 
\acknowledgments

J.D. is supported by a Joint Columbia University and Flatiron Institute Postdoctoral Fellowship. Research at the Flatiron Institute is supported by the Simons Foundation. B.R. is supported by the Natural Sciences \& Engineering Research Council of Canada (NSERC). L.S. acknowledges support from the Cottrell Scholars Award and DoE Early Career Award DE-SC0023015. L.S. and J.D. acknowledge support from NSF AST-2108201. A.P.~ acknowledges support by NASA ATP grant 80NSSC22K1054. H.O. was supported by a Virtual Institute of Accretion (VIA) postdoctoral fellowship from the Netherlands Research School for Astronomy (NOVA). K.C. acknowledges support from grants from the Gordon and Betty Moore Foundation and the John Templeton Foundation to the Black Hole Initiative at Harvard University, and from NSF award OISE-1743747. This work was supported by a grant from the Simons Foundation (MP-SCMPS-00001470) to LS and AP, and facilitated by Multimessenger Plasma Physics Center (MPPC), NSF grants PHY-2206609 and PHY-2206610.

The computational resources and services used in this work were partially provided by facilities supported by the Scientific Computing Core at the Flatiron Institute, a division of the Simons Foundation; and by the VSC (Flemish Supercomputer Center), funded by the Research Foundation Flanders (FWO) and the Flemish Government – department EWI. This research was enabled by support provided by grant no. NSF PHY-1125915 along with a INCITE program award PHY129, using resources from the Oak Ridge Leadership Computing Facility, Summit, which is a US Department of Energy office of Science User Facility supported under contract DE-AC05- 00OR22725, as well as Calcul Quebec (http://www.calculquebec.ca) and Compute Canada (http://www.computecanada.ca). The computational resources and services used in this work were partially provided by facilities supported by the Scientific Computing Core at the Flatiron Institute, a division of the Simons Foundation. This research is part of the Frontera (\citealt{Frontera}) computing project at the Texas Advanced Computing Center (LRAC-AST20008). Frontera is made possible by National Science Foundation award OAC-1818253.

\software{{\tt python} \citep{oliphant2007,millman2011}, {\tt scipy} \citep{jones2001}, {\tt numpy} \citep{vanderwalt2011}, {\tt matplotlib} \citep{hunter2007}, {\tt RAPTOR} \citep{Bronzwaer2018,Bronzwaer2020}, {\tt BHAC} \citep{porth2017,olivares2019}}

\newpage
\appendix
  % \vspace{-0.9cm}

%\end{figure}

\section{GRMHD coordinate comparison} \label{app-A}

To assess if the waves shown in our simulation are a robust feature, we cross-compared our Cartesian Kerr-Schild simulation with a set of Modified Kerr-Schild (MKS) simulations in a Spherical coordinate basis at varying resolution ran with the {\tt H-AMR} code \citep{Liska2022} and presented in \cite{ripperda2022}. All simulations are initialized with identical initial conditions to our CKS simulation. In Table \ref{tab-resolutions}, we show the cells in the $\theta$ and $\phi$ direction on the horizon and the number of cells per jet radius. We use a jet radius of approximately $50 ~r_{\rm g}$, where we find typical wave structures in our simulations.

\begin{table}[ht]
    \centering
    \begin{tabular}{c|c|c|c}
        Model & $N_{\rm r}$ x $N_{\theta}$ x $N_{\phi}$ & Cells on horizon & Cells per jet radius at $50 ~r_{\rm g}$\\
        \hline
        {\tt BHAC} CKS & - & $54\times107$ & 150\\
        {\tt HAMR} MKS low & $288\times128\times128$ & $128\times128$ & 40\\
        {\tt HAMR} MKS standard & $580\times288\times256$ & $288\times256$  & 91 \\
        {\tt HAMR} MKS high & $2240\times1056\times1024$ & $1056\times1024$  & 332\\
        {\tt HAMR} MKS extreme & $5376\times2304\times2304$ & $2304\times2304$ & 733
    \end{tabular}
    \caption{Summary of the number of cells, horizon resolution, and jet resolution at $r=50 ~r_{\rm g}$ for our {\tt BHAC} and {\tt HAMR} models.}
    \label{tab-resolutions}
\end{table}

\begin{wrapfigure}{r}{0.45\linewidth}
%\begin{figure}[ht]
  \centering
   \vspace{-0.6cm}
    \includegraphics[width=0.4\textwidth]{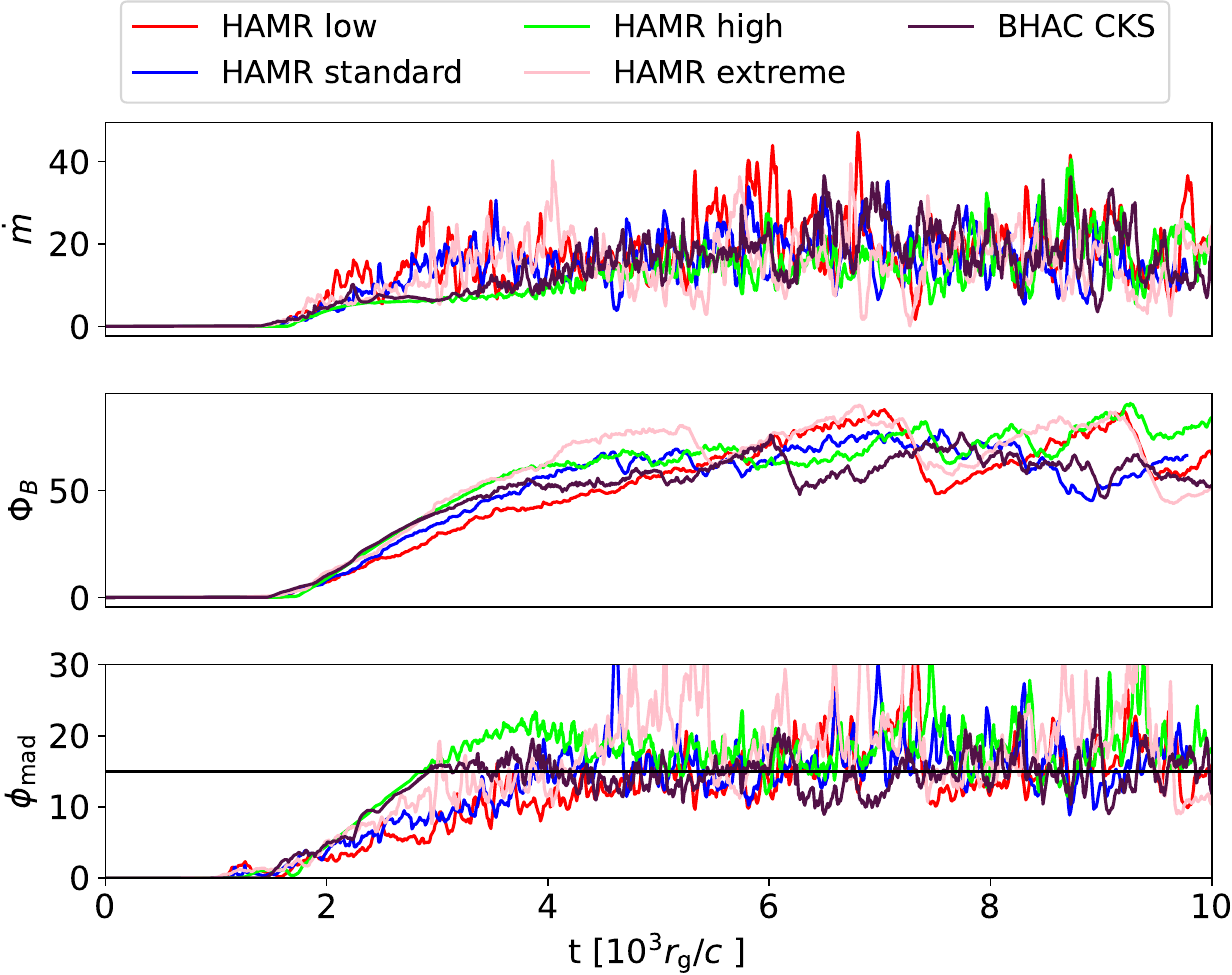}
    \caption{Comparison between {\tt BHAC} CKS and {\tt HAMR} MKS for varying resolutions. Shown are, the mass accretion rate $\dot{m}$ (top), magnetic flux $\Phi_{\rm B}$ (middle), and the MAD parameter $\phi_{\rm mad}$ (bottom).}
   \label{fig-hamr-comp}
%   \vspace{-1.25cm}
\end{wrapfigure}

\begin{figure}[ht]
         \centering
        \begin{subfigure}[b]{0.45\textwidth}
            \centering
            \includegraphics[width=0.9\textwidth,trim={2cm 0 0 0}]{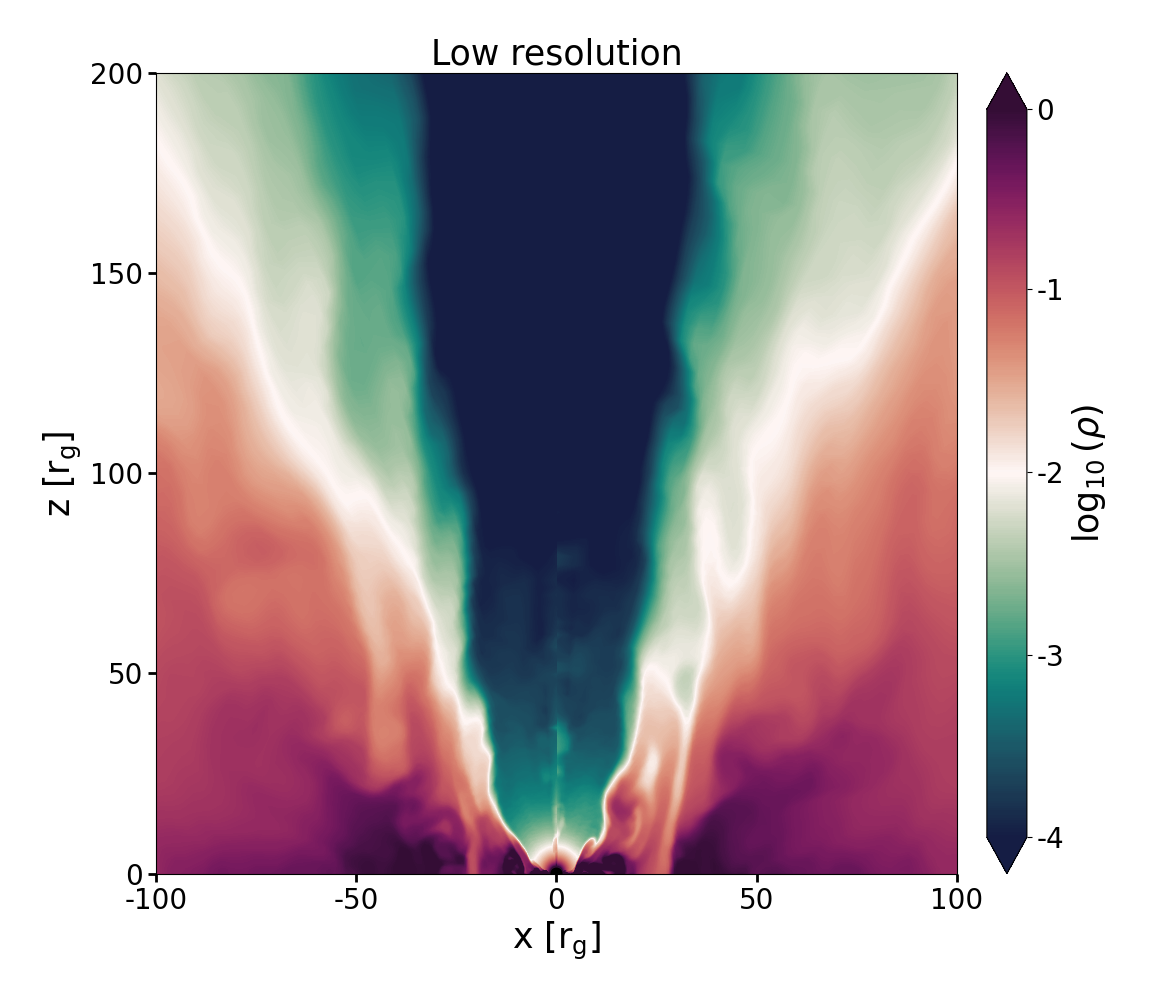}
        \end{subfigure}
       % \hfill
        \begin{subfigure}[b]{0.45\textwidth}  
            \centering 
            \includegraphics[width=0.9\textwidth,trim={2cm 0 0 0}]{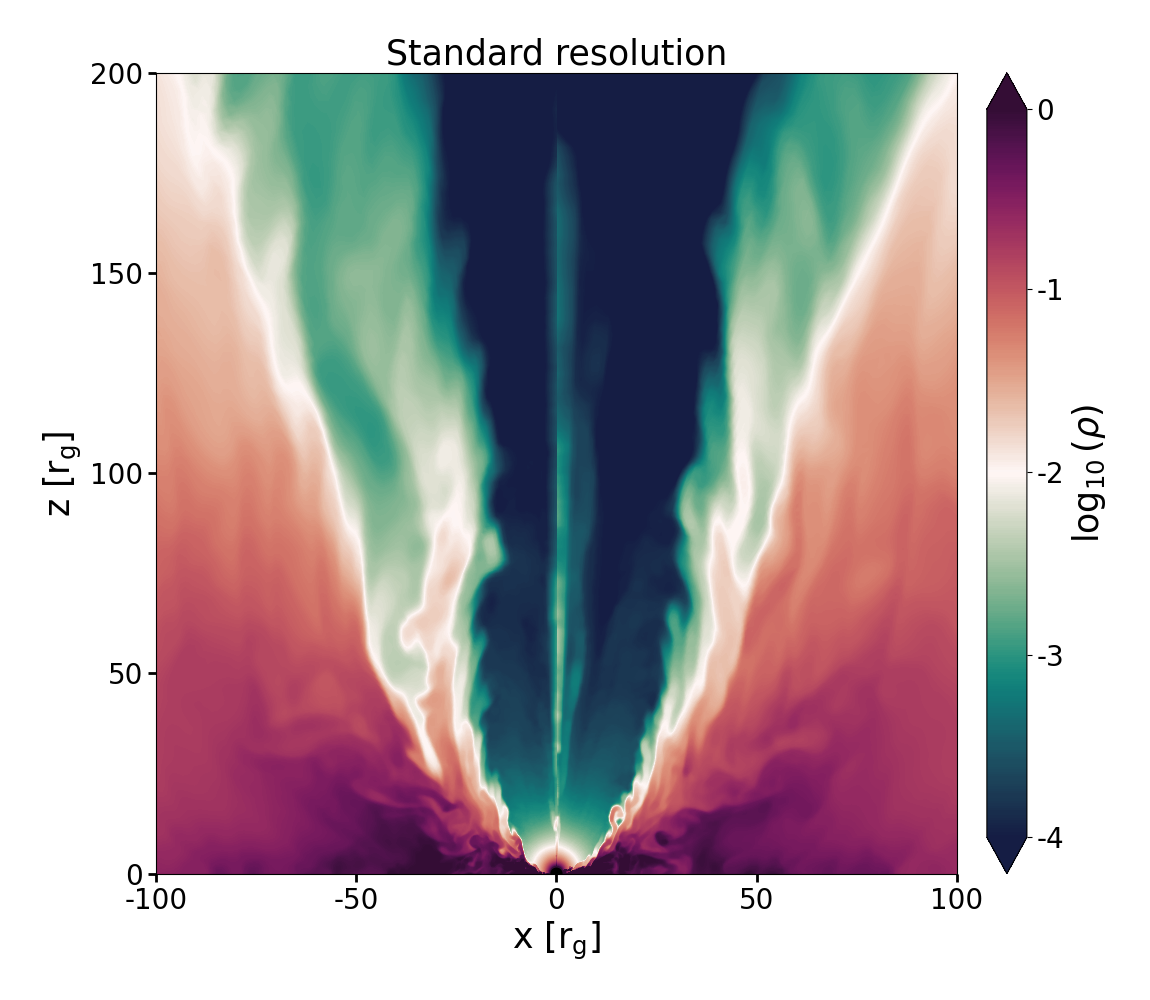}
        \end{subfigure}
        % \hfill
        % \begin{subfigure}[b]{0.305\textwidth}  
        %     \centering 
        %     \includegraphics[width=\textwidth,trim={10cm 0 0 0}]{rho_512.png}
        % \end{subfigure}
         \vspace{-0.7cm}
        \vskip\baselineskip
        \begin{subfigure}[b]{0.45\textwidth}   
            \centering 
            \includegraphics[width=0.9\textwidth,trim={2cm 0 0 0}]{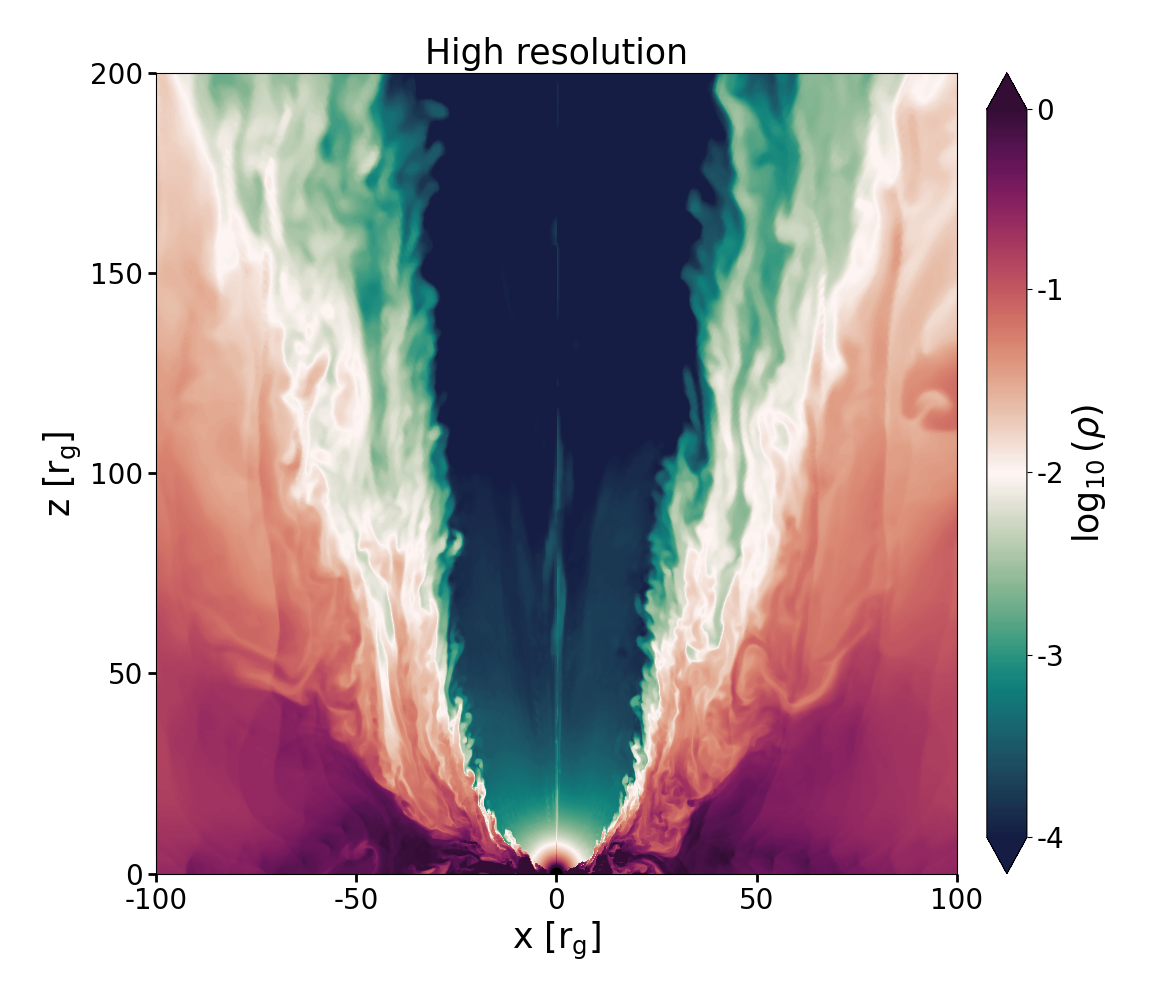}
        \end{subfigure}
      %  \hfill
        \begin{subfigure}[b]{0.45\textwidth}   
            \centering 
            \includegraphics[width=0.9\textwidth,trim={2cm 0 0 0}]{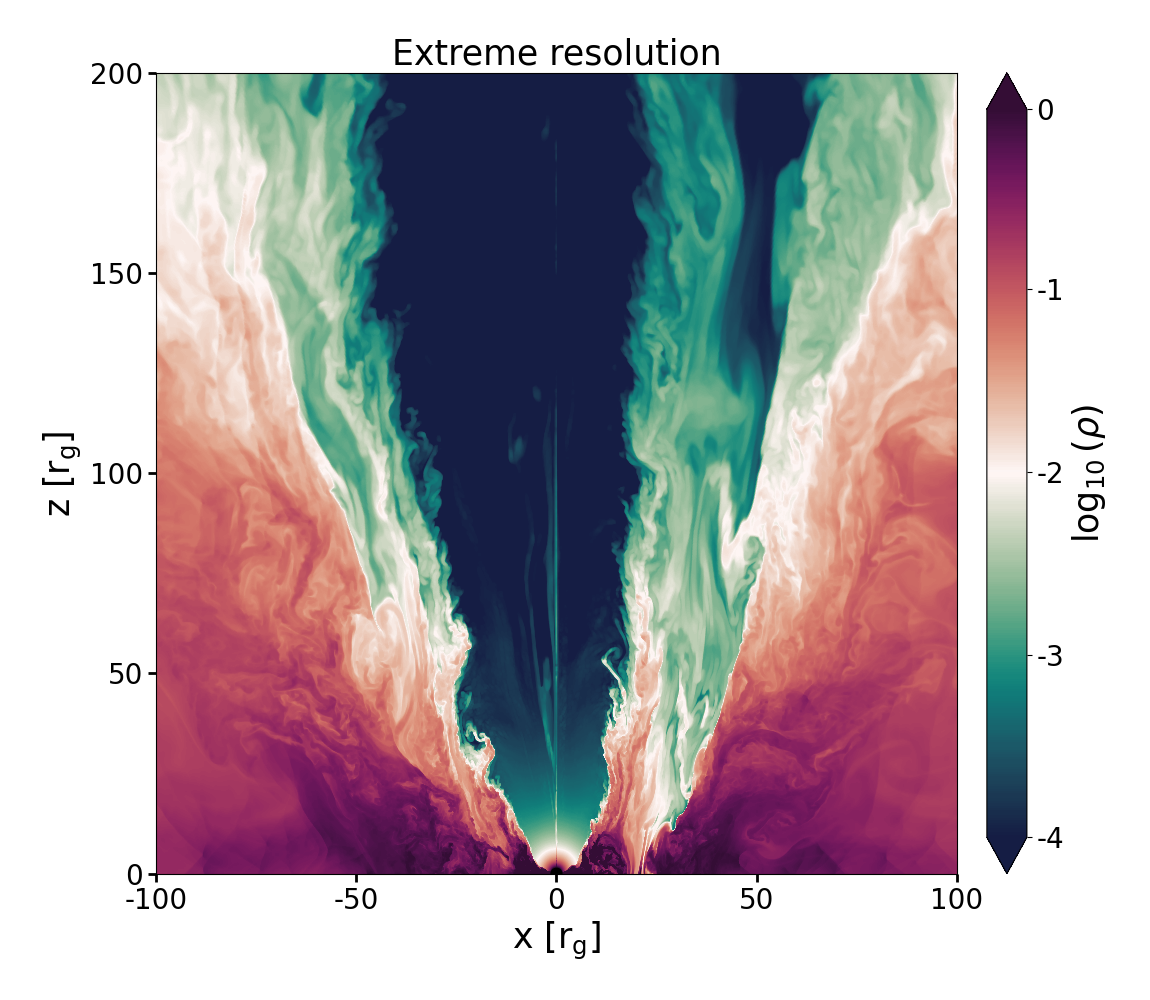}
        \end{subfigure}
        \caption{Simulation snapshots at $t=10,000~r_{\rm g}/c$ for the four {\tt HAMR} MKS models at varying resolution. Shown are slices along the $x-z$ plane of the logarithm of density. As the resolution increases from left to right and top to bottom, the simulations show a more extended substructure in the form of waves along the jet-wind interface, where the lowest two resolutions are more diffuse than the highest resolutions.}
        \label{hamr-slices-comp}
\end{figure}

In Fig. \ref{fig-hamr-comp}, we compare the mass accretion rate computed at $r=5~r_{\rm g}$(top panel) as well as magnetic flux threading the horizon (bottom panel), and the MAD parameter $\phi_{\rm mad}=\Phi_{\rm B}/\sqrt{\dot{m}}$, for all four resolutions as well as the {\tt BHAC} CKS run. All simulations reach, on average, similar values of all three quantities and obtain a MAD state at $t\sim3000-4000 r_{\rm g}/c$. During the remainder of the simulations, multiple flux eruptions are visible in $\Phi_{\rm B}$, which look similar among all simulations, e.g., similar slopes when $\Phi_{\rm B}$ drops and similar fractional decreases. The horizon integrated quantities, therefore, indicate a convergence of the global dynamics typical for a MAD accretion flow. 
Finally, we show slices along the $x-z$ axis of the {\tt HAMR} simulations in Figure \ref{hamr-slices-comp}. The jet of four simulations in Spherical coordinates shows similar opening angles as our Cartesian simulation in Fig. \ref{fig-grmhd}. As the resolution increases, we see more and more substructure in the form of waves along the jet-wind interface, whereas the lowest two resolutions are more diffuse. Compared to Figure \ref{fig-panel}, our simulation falls somewhere between the Standard and High-resolution runs, which is unsurprising given the jet resolutions of our CKS run are also between these two spherical runs. However, due to the larger horizon cells, the CKS simulation has a substantially lower computational cost, $\approx 500,000$ CPU hours, similar to the low-resolution case run times. Additionally, we resort to post-processing the {\tt BHAC} CKS simulation over the higher resolution {\tt HAMR} simulation due to a more practical reason: as of to date, no radiation transfer code is fully coupled to the AMR-based grid structure of {\tt HAMR} and are unable handle the extreme data volume that these simulations have, we, however, aim to develop an extended more efficient version of {\tt RAPTOR} that is fully coupled to the {\tt HAMR} data format that will be capable of ray tracing the full extreme resolution simulation in the future.

\vspace{0.3cm}

\section{Comparison to low-resolution Spherical Kerr-Schild} \label{app-B}

\vspace{0.3cm}

\begin{figure}[ht]
  
  %      \vskip\baselineskip
\begin{subfigure}{.33\textwidth}
  \centering
  \includegraphics[width=\textwidth]{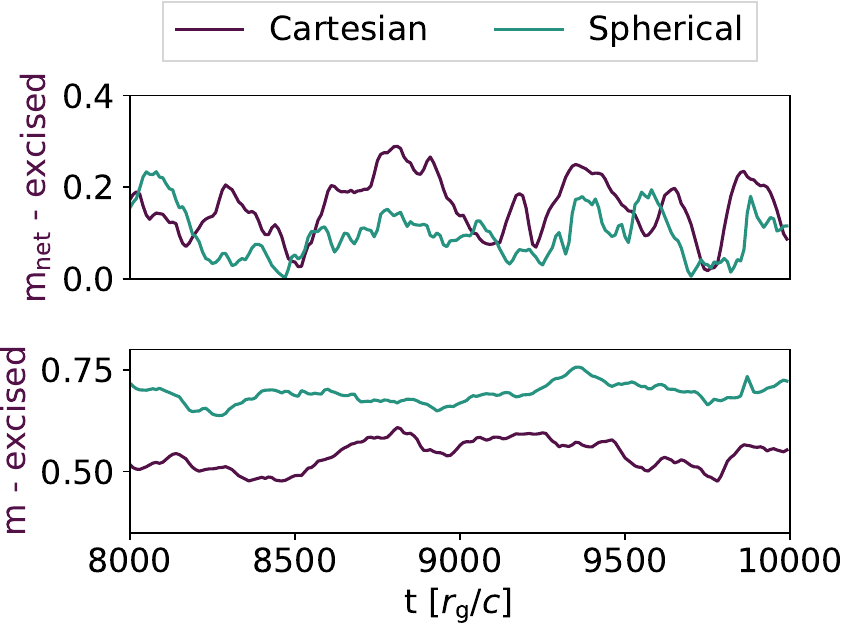}
\end{subfigure}%
\begin{subfigure}{.66\textwidth}
  \centering
  \includegraphics[width=0.9\textwidth]{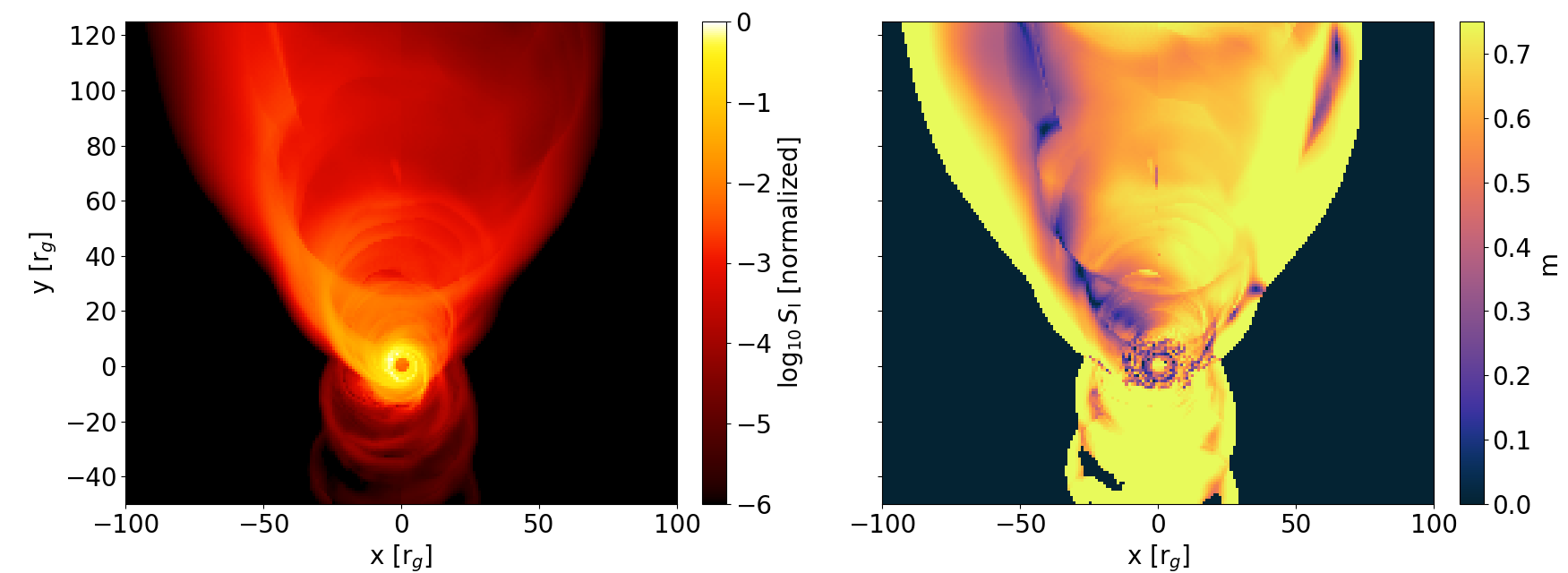}
\end{subfigure}
    \caption{Left: LP fraction as a function of time for both the Cartesian and the Spherical simulations. Both panels exclude the core emission within $30 ~r_{\rm g}$ of the image origin. Top: net LP fraction $m_{\rm net}$, showing similar values. Bottom: resolved LP fraction $m$, showing a clear factor 1.5 difference. Middle: Stokes $\mathcal{I}$ map of the Spherical simulation, which shows limited substructure. Right: LP fraction map of Spherical simulation. A limited amount of structure is visible compared to the Cartesian case and shows a substantially higher LP fraction, $m\sim0.75$ versus $m\sim0.5$. }
   \label{fig-mks-cks}
\end{figure}

We cross-compared our CKS MAD simulation with a low-resolution spherical MKS simulation to further strengthen our conclusions. The MKS simulation has a base resolution of [96,48,48] in $r, \theta, \phi$, and one additional level of AMR. The simulation was run up to $t=10,000 ~r_{\rm g}/c$ with {\tt BHAC}. Due to the low resolution in this simulation's jet region, no waves are present along the jet-wind surface. Note that due to the low resolution, the physical solution of this simulation is far from resolved and, therefore, in an unrealistically low regime of Reynolds number. We only use it here to compare a laminar flow to a flow where the jet-wind surface shows wave instabilities. We ray trace the spherical simulation over the final $2000~r_{\rm g}/c$, with the same model and camera parameters as the $\kappa$-jet model presented in the main manuscript. Comparing the resolved linear polarization fraction with the higher resolution Cartesian case shows a substantially higher fraction, namely at $m\sim0.7$, compared to $m\sim0.5$, see Figure \ref{fig-mks-cks} left panel. Looking at synthetic images of linear polarization, Figure \ref{fig-mks-cks} right panel, also no wave-like substructure, as seen in the Cartesian case, is visible along the jet-wind shear layer. This comparison, therefore, further confirms our hypothesis that the jet-wind shear layer waves, which are only captured with sufficiently high resolution, lead to the drop in LP fraction.
 
\vspace{-0.3cm}

\newpage

\begin{thebibliography}{}
\expandafter\ifx\csname natexlab\endcsname\relax\def\natexlab#1{#1}\fi
\providecommand{\url}[1]{\href{#1}{#1}}
\providecommand{\dodoi}[1]{doi:~\href{http://doi.org/#1}{\nolinkurl{#1}}}
\providecommand{\doeprint}[1]{\href{http://ascl.net/#1}{\nolinkurl{http://ascl.net/#1}}}
\providecommand{\doarXiv}[1]{\href{https://arxiv.org/abs/#1}{\nolinkurl{https://arxiv.org/abs/#1}}}

\bibitem[{{Bronzwaer} {et~al.}(2018){Bronzwaer}, {Davelaar}, {Younsi},
  {Mo{\'s}cibrodzka}, {Falcke}, {Kramer}, \& {Rezzolla}}]{Bronzwaer2018}
{Bronzwaer}, T., {Davelaar}, J., {Younsi}, Z., {et~al.} 2018, \aap, 613, A2,
  \dodoi{10.1051/0004-6361/201732149}

\bibitem[{{Bronzwaer} {et~al.}(2020){Bronzwaer}, {Younsi}, {Davelaar}, \&
  {Falcke}}]{Bronzwaer2020}
{Bronzwaer}, T., {Younsi}, Z., {Davelaar}, J., \& {Falcke}, H. 2020, \aap, 641,
  A126, \dodoi{10.1051/0004-6361/202038573}

\bibitem[{Caprioli(2015)}]{caprioli2015}
Caprioli, D. 2015, The Astrophysical Journal, 811, L38,
  \dodoi{10.1088/2041-8205/811/2/L38}

\bibitem[{{Chael} {et~al.}(2019){Chael}, {Narayan}, \& {Johnson}}]{chael2019}
{Chael}, A., {Narayan}, R., \& {Johnson}, M.~D. 2019, \mnras, 486, 2873,
  \dodoi{10.1093/mnras/stz988}

\bibitem[{Chatterjee {et~al.}(2019)Chatterjee, Liska, Tchekhovskoy, \&
  Markoff}]{chatterjee2019}
Chatterjee, K., Liska, M., Tchekhovskoy, A., \& Markoff, S.~B. 2019, Monthly
  Notices of the Royal Astronomical Society, 490, 2200,
  \dodoi{10.1093/mnras/stz2626}

\bibitem[{Chow {et~al.}(2022)Chow, Davelaar, \& Sironi}]{chow2022a}
Chow, A., Davelaar, J., \& Sironi, L. 2022, The {{Kelvin-Helmholtz}}
  Instability at the Boundary of Relativistic Magnetized Jets,  {arXiv}.
\newblock \doarXiv{2209.13699}

\bibitem[{{Cruz-Osorio} {et~al.}(2022){Cruz-Osorio}, Fromm, Mizuno, Nathanail,
  Younsi, Porth, Davelaar, Falcke, Kramer, \& Rezzolla}]{cruz-osorio2022}
{Cruz-Osorio}, A., Fromm, C.~M., Mizuno, Y., {et~al.} 2022, Nature Astronomy,
  6, 103, \dodoi{10.1038/s41550-021-01506-w}

\bibitem[{Davelaar \& Haiman(2022)}]{davelaar2022}
Davelaar, J., \& Haiman, Z. 2022, Physical Review D, 105, 103010,
  \dodoi{10.1103/PhysRevD.105.103010}

\bibitem[{Davelaar {et~al.}(2018)Davelaar, Mo{\'s}cibrodzka, Bronzwaer, \&
  Falcke}]{davelaar2018a}
Davelaar, J., Mo{\'s}cibrodzka, M., Bronzwaer, T., \& Falcke, H. 2018,
  Astronomy \&amp; Astrophysics, Volume 612, id.A34, 16 pp., 612, A34,
  \dodoi{10.1051/0004-6361/201732025}

\bibitem[{Davelaar {et~al.}(2019)Davelaar, Olivares, Porth, Bronzwaer, Janssen,
  Roelofs, Mizuno, Fromm, Falcke, \& Rezzolla}]{davelaar2019}
Davelaar, J., Olivares, H., Porth, O., {et~al.} 2019, Astronomy \&amp;
  Astrophysics, Volume 632, id.A2, 16 pp., 632, A2,
  \dodoi{10.1051/0004-6361/201936150}

\bibitem[{{Dexter}(2016)}]{Dexter2016}
{Dexter}, J. 2016, \mnras, 462, 115, \dodoi{10.1093/mnras/stw1526}

\bibitem[{{Dexter} {et~al.}(2012){Dexter}, {McKinney}, \& {Agol}}]{dexter2012}
{Dexter}, J., {McKinney}, J.~C., \& {Agol}, E. 2012, \mnras, 421, 1517,
  \dodoi{10.1111/j.1365-2966.2012.20409.x}

\bibitem[{Dexter {et~al.}(2020)Dexter, Tchekhovskoy, {Jim{\'e}nez-Rosales},
  Ressler, Baub{\"o}ck, Dallilar, {de Zeeuw}, Eisenhauer, {von Fellenberg},
  Gao, Genzel, Gillessen, Habibi, Ott, Stadler, Straub, \&
  Widmann}]{dexter2020}
Dexter, J., Tchekhovskoy, A., {Jim{\'e}nez-Rosales}, A., {et~al.} 2020, Monthly
  Notices of the Royal Astronomical Society, 497, 4999,
  \dodoi{10.1093/mnras/staa2288}

\bibitem[{{EHT MWL Science Working Group} {et~al.}(2021){EHT MWL Science
  Working Group}, Algaba, Anczarski, Asada, Balokovi{\'c}, Chandra, Cui,
  Falcone, Giroletti, Goddi, Hada, Haggard, Jorstad, Kaur, Kawashima, Keating,
  Kim, Kino, Komossa, Kravchenko, Krichbaum, Lee, Lu, Lucchini, Markoff,
  Neilsen, Nowak, Park, Principe, Ramakrishnan, Reynolds, Sasada, Savchenko,
  Williamson, {Event Horizon Telescope Collaboration}, Akiyama, Alberdi, Alef,
  Anantua, Azulay, Baczko, Ball, Barrett, Bintley, Benson, Blackburn, Blundell,
  Boland, Bouman, Bower, Boyce, Bremer, Brinkerink, Brissenden, Britzen,
  Broderick, Broguiere, Bronzwaer, Byun, Carlstrom, Chael, Chan, Chatterjee,
  Chatterjee, Chen, Chen, Chesler, Cho, Christian, Conway, Cordes, Crawford,
  Crew, {Cruz-Osorio}, Davelaar, {de Laurentis}, Deane, Dempsey, Desvignes,
  Dexter, Doeleman, Eatough, Falcke, Farah, Fish, Fomalont, Ford,
  {Fraga-Encinas}, Friberg, Fromm, Fuentes, Galison, Gammie, Garc{\'i}a,
  Gentaz, Georgiev, Gold, G{\'o}mez, {G{\'o}mez-Ruiz}, Gu, Gurwell, Hecht,
  Hesper, Ho, Ho, Honma, Huang, Huang, Hughes, Ikeda, Inoue, Issaoun, James,
  Jannuzi, Janssen, Jeter, Jiang, {Jim{\'e}nez-Rosales}, Johnson, Jung, Karami,
  Karuppusamy, Kettenis, Kim, Kim, Kim, Koay, Kofuji, Koch, Koyama, Kramer,
  Kramer, Kuo, Lauer, Levis, Li, Li, Lindqvist, Lico, Lindahl, Liu, Liu,
  Liuzzo, Lo, Lobanov, Loinard, Lonsdale, MacDonald, Mao, Marchili, Marrone,
  Marscher, {Mart{\'i}-Vidal}, Matsushita, Matthews, Medeiros, Menten, Mizuno,
  Mizuno, Moran, Moriyama, Moscibrodzka, M{\"u}ller, Musoke, Mej{\'i}as, Nagai,
  Nagar, Nakamura, Narayan, Narayanan, Natarajan, Nathanail, Neri, Ni, Noutsos,
  Okino, Olivares, {Ortiz-Le{\'o}n}, Oyama, {\"O}zel, Palumbo, Patel, Pen,
  Pesce, Pi{\'e}tu, Plambeck, Popstefanija, Porth, P{\"o}tzl, Prather,
  {Preciado-L{\'o}pez}, Psaltis, Pu, Rao, Rawlings, Raymond, Rezzolla, Ricarte,
  Ripperda, Roelofs, Rogers, Ros, Rose, Roshanineshat, Rottmann, Roy, Ruszczyk,
  Rygl, S{\'a}nchez, {S{\'a}nchez-Arguelles}, Savolainen, Schloerb, Schuster,
  Shao, Shen, Small, Sohn, Soohoo, Sun, Tazaki, Tetarenko, Tiede, Tilanus,
  Titus, Toma, Torne, Trent, Traianou, Trippe, {van Bemmel}, {van Langevelde},
  {van Rossum}, Wagner, {Ward-Thompson}, Wardle, Weintroub, Wex, Wharton,
  Wielgus, Wong, Wu, Yoon, Young, Young, Younsi, Yuan, Yuan, Zensus, Zhao,
  Zhao, {Fermi Large Area Telescope Collaboration}, Principe, Giroletti,
  D'Ammando, Orienti, {H. E. S. S. Collaboration}, Abdalla, Adam, Aharonian,
  Benkhali, Ang{\"u}ner, Arcaro, Armand, Armstrong, Ashkar, Backes, Baghmanyan,
  Barbosa~Martins, Barnacka, Barnard, Becherini, Berge, Bernl{\"o}hr, Bi,
  B{\"o}ttcher, Boisson, Bolmont, {de Lavergne}, Breuhaus, Brun, Brun, Bryan,
  B{\"u}chele, Bulik, Bylund, Caroff, Carosi, Casanova, Chand, Chen, Cotter,
  Cury{\l}o, Damascene~Mbarubucyeye, Davids, Davies, Deil, Devin, Dewilt,
  Dirson, {Djannati-Ata{\"i}}, Dmytriiev, Donath, Doroshenko, Duffy, Dyks,
  Egberts, Eichhorn, Einecke, Emery, Ernenwein, Feijen, Fegan, Fiasson, {de
  Clairfontaine}, Fontaine, Funk, F{\"u}{\ss}ling, Gabici, Gallant, Giavitto,
  Giunti, Glawion, Glicenstein, Gottschall, Grondin, Hahn, Haupt, Hermann,
  Hinton, Hofmann, Hoischen, Holch, Holler, H{\"o}rbe, Horns, Huber, Jamrozy,
  Jankowsky, Jankowsky, {Jardin-Blicq}, Joshi, {Jung-Richardt}, Kasai,
  Kastendieck, Katarzy{\'n}ski, Katz, Khangulyan, Kh{\'e}lifi, Klepser,
  Klu{\'z}niak, Komin, Konno, Kosack, Kostunin, Kreter, Lamanna, Lemi{\`e}re,
  {Lemoine-Goumard}, Lenain, Levy, Lohse, Lypova, Mackey, Majumdar, Malyshev,
  Malyshev, Marandon, Marchegiani, Marcowith, Mares, {Mart{\'i}-Devesa}, Marx,
  Maurin, Meintjes, Meyer, Moderski, Mohamed, Mohrmann, Montanari, Moore,
  Morris, Moulin, Muller, Murach, Nakashima, Nayerhoda, {de Naurois},
  Ndiyavala, Niederwanger, Niemiec, Oakes, O'Brien, Odaka, Ohm,
  {Olivera-Nieto}, {de Ona Wilhelmi}, Ostrowski, Panter, Panny, Parsons, Peron,
  Peyaud, Piel, Pita, Poireau, Noel, Prokhorov, Prokoph, P{\"u}hlhofer, Punch,
  Quirrenbach, Rauth, Reichherzer, Reimer, Reimer, Remy, Renaud, Rieger,
  Rinchiuso, Romoli, Rowell, Rudak, {Ruiz-Velasco}, Sahakian, Sailer, Sanchez,
  Santangelo, Sasaki, Scalici, Schutte, Schwanke, Schwemmer, {Seglar-Arroyo},
  Senniappan, Seyffert, Shafi, Shiningayamwe, Simoni, Sinha, Sol, Specovius,
  Spencer, {Spir-Jacob}, Stawarz, Sun, Steenkamp, Stegmann, Steinmassl, Steppa,
  Takahashi, Tavernier, Taylor, Terrier, Tiziani, Tluczykont, Tomankova,
  Trichard, Tsirou, Tuffs, Uchiyama, {van der Walt}, {van Eldik}, {van
  Rensburg}, {van Soelen}, Vasileiadis, Veh, Venter, Vincent, Vink, V{\"o}lk,
  Vuillaume, Wadiasingh, Wagner, Watson, Werner, White, Wierzcholska, Wong,
  Yusafzai, Zacharias, Zanin, Zargaryan, Zdziarski, Zech, Zhu, Zorn, Zouari,
  {\.Z}ywucka, {MAGIC Collaboration}, Acciari, Ansoldi, Antonelli, Engels,
  Artero, Asano, Baack, Babi{\'c}, Baquero, {de Almeida}, Barrio,
  Becerra~Gonz{\'a}lez, Bednarek, Bellizzi, Bernardini, Bernardos, Berti,
  Besenrieder, Bhattacharyya, Bigongiari, Biland, Blanch, Bonnoli, Bo{\v
  s}njak, Busetto, Carosi, Ceribella, Cerruti, Chai, Chilingarian, Cikota,
  Colak, Colombo, Contreras, Cortina, Covino, D'Amico, D'Elia, {da Vela},
  Dazzi, {de Angelis}, {de Lotto}, Delfino, Delgado, Delgado~Mendez, Depaoli,
  {di Pierro}, {di Venere}, Do~Souto~Espi{\~n}eira, Dominis~Prester, Donini,
  Dorner, Doro, Elsaesser, Ramazani, Fattorini, Ferrara, Fonseca, Font, Fruck,
  Fukami, Garc{\'i}a~L{\'o}pez, Garczarczyk, Gasparyan, Gaug, Giglietto,
  Giordano, Gliwny, Godinovi{\'c}, Green, Green, Hadasch, Hahn, Heckmann,
  Herrera, Hoang, Hrupec, H{\"u}tten, Inada, Inoue, Ishio, Iwamura,
  Jim{\'e}nez, Jormanainen, Jouvin, Kajiwara, Karjalainen, Kerszberg,
  Kobayashi, Kubo, Kushida, Lamastra, Lelas, Leone, Lindfors, Lombardi, Longo,
  {L{\'o}pez-Coto}, {L{\'o}pez-Moya}, {L{\'o}pez-Oramas}, Loporchio, {Machado
  de Oliveira Fraga}, Maggio, Majumdar, Makariev, Mallamaci, Maneva, Manganaro,
  Mannheim, Maraschi, Mariotti, Mart{\'i}nez, Mazin, Menchiari, Mender,
  Mi{\'c}anovi{\'c}, Miceli, Miener, Minev, Miranda, Mirzoyan, Molina,
  Moralejo, Morcuende, Moreno, Moretti, Neustroev, Nigro, Nilsson, Nishijima,
  Noda, Nozaki, Ohtani, Oka, {Otero-Santos}, Paiano, Palatiello, Paneque,
  Paoletti, Paredes, Pavleti{\'c}, Pe{\~n}il, Perennes, Persic, Moroni,
  Prandini, Priyadarshi, Puljak, Rhode, Rib{\'o}, Rico, Righi, Rugliancich,
  Saha, Sahakyan, Saito, Sakurai, Satalecka, Saturni, Schleicher, Schmidt,
  Schweizer, Sitarek, {\v S}nidari{\'c}, Sobczynska, Spolon, Stamerra, Strom,
  Strzys, Suda, Suri{\'c}, Takahashi, Tavecchio, Temnikov, Terzi{\'c}, Teshima,
  Tosti, Truzzi, Tutone, Ubach, {van Scherpenberg}, Vanzo, Vazquez~Acosta,
  Ventura, Verguilov, Vigorito, Vitale, Vovk, Will, Wunderlich, Zari{\'c},
  {VERITAS Collaboration}, Adams, Benbow, Brill, Capasso, Christiansen,
  Chromey, Daniel, Errando, Farrell, Feng, Finley, Fortson, Furniss, Gent,
  Giuri, Hassan, Hervet, Holder, Hughes, Humensky, Jin, Kaaret, Kertzman,
  Kieda, Kumar, Lang, Lundy, Maier, Moriarty, Mukherjee, Nieto,
  {Nievas-Rosillo}, O'Brien, Ong, Otte, Patel, Pfrang, Pohl, Prado, Pueschel,
  Quinn, Ragan, Reynolds, Ribeiro, Richards, Roache, Rulten, Ryan, Santander,
  Sembroski, Shang, Weinstein, Williams, Williamson, {Eavn Collaboration},
  Hirota, Cui, Niinuma, Ro, Sakai, {Sawada-Satoh}, Wajima, Wang, Liu, \&
  Yonekura}]{ehtmwlscienceworkinggroup2021}
{EHT MWL Science Working Group}, Algaba, J.~C., Anczarski, J., {et~al.} 2021,
  The Astrophysical Journal, 911, L11, \dodoi{10.3847/2041-8213/abef71}

\bibitem[{{Event Horizon Telescope Collaboration}
  {et~al.}(2019{\natexlab{a}}){Event Horizon Telescope Collaboration}, Akiyama,
  Alberdi, Alef, Asada, Azulay, Baczko, Ball, Balokovi{\'c}, Barrett, Bintley,
  Blackburn, Boland, Bouman, Bower, Bremer, Brinkerink, Brissenden, Britzen,
  Broderick, Broguiere, Bronzwaer, Byun, Carlstrom, Chael, Chan, Chatterjee,
  Chatterjee, Chen, Chen, Cho, Christian, Conway, Cordes, Crew, Cui, Davelaar,
  De~Laurentis, Deane, Dempsey, Desvignes, Dexter, Doeleman, Eatough, Falcke,
  Fish, Fomalont, {Fraga-Encinas}, Freeman, Friberg, Fromm, G{\'o}mez, Galison,
  Gammie, Garc{\'i}a, Gentaz, Georgiev, Goddi, Gold, Gu, Gurwell, Hada, Hecht,
  Hesper, Ho, Ho, Honma, Huang, Huang, Hughes, Ikeda, Inoue, Issaoun, James,
  Jannuzi, Janssen, Jeter, Jiang, Johnson, Jorstad, Jung, Karami, Karuppusamy,
  Kawashima, Keating, Kettenis, Kim, Kim, Kim, Kino, Koay, Koch, Koyama,
  Kramer, Kramer, Krichbaum, Kuo, Lauer, Lee, Li, Li, Lindqvist, Liu, Liuzzo,
  Lo, Lobanov, Loinard, Lonsdale, Lu, MacDonald, Mao, Markoff, Marrone,
  Marscher, {Mart{\'i}-Vidal}, Matsushita, Matthews, Medeiros, Menten, Mizuno,
  Mizuno, Moran, Moriyama, Moscibrodzka, M{\"u}ller, Nagai, Nagar, Nakamura,
  Narayan, Narayanan, Natarajan, Neri, Ni, Noutsos, Okino, Olivares,
  {Ortiz-Le{\'o}n}, Oyama, {\"O}zel, Palumbo, Patel, Pen, Pesce, Pi{\'e}tu,
  Plambeck, PopStefanija, Porth, Prather, {Preciado-L{\'o}pez}, Psaltis, Pu,
  Ramakrishnan, Rao, Rawlings, Raymond, Rezzolla, Ripperda, Roelofs, Rogers,
  Ros, Rose, Roshanineshat, Rottmann, Roy, Ruszczyk, Ryan, Rygl, S{\'a}nchez,
  {S{\'a}nchez-Arguelles}, Sasada, Savolainen, Schloerb, Schuster, Shao, Shen,
  Small, Sohn, SooHoo, Tazaki, Tiede, Tilanus, Titus, Toma, Torne, Trent,
  Trippe, Tsuda, {van Bemmel}, {van Langevelde}, {van Rossum}, Wagner, Wardle,
  Weintroub, Wex, Wharton, Wielgus, Wong, Wu, Young, Young, Younsi, Yuan, Yuan,
  Zensus, Zhao, Zhao, Zhu, Algaba, Allardi, Amestica, Anczarski, Bach,
  Baganoff, Beaudoin, Benson, Berthold, Blanchard, Blundell, Bustamente,
  Cappallo, {Castillo-Dom{\'i}nguez}, Chang, Chang, Chang, Chen, Chilson,
  Chuter, C{\'o}rdova~Rosado, Coulson, Crawford, Crowley, David, Derome,
  Dexter, Dornbusch, Dudevoir, Dzib, Eckart, Eckert, Erickson, Everett, Faber,
  Farah, Fath, Folkers, Forbes, Freund, {G{\'o}mez-Ruiz}, Gale, Gao, Geertsema,
  Graham, Greer, Grosslein, Gueth, Haggard, Halverson, Han, Han, Hao, Hasegawa,
  Henning, {Hern{\'a}ndez-G{\'o}mez}, {Herrero-Illana}, Heyminck, Hirota, Hoge,
  Huang, Impellizzeri, Jiang, Kamble, Keisler, Kimura, Kono, Kubo, Kuroda,
  Lacasse, Laing, Leitch, Li, Lin, Liu, Liu, Lu, Marson, {Martin-Cocher},
  Massingill, Matulonis, McColl, McWhirter, Messias, {Meyer-Zhao}, Michalik,
  Monta{\~n}a, Montgomerie, {Mora-Klein}, Muders, Nadolski, Navarro, Neilsen,
  Nguyen, Nishioka, Norton, Nowak, Nystrom, Ogawa, Oshiro, Oyama, Parsons,
  Paine, Pe{\~n}alver, Phillips, Poirier, Pradel, Primiani, Raffin, Rahlin,
  Reiland, Risacher, Ruiz, {S{\'a}ez-Mada{\'i}n}, Sassella, Schellart, Shaw,
  Silva, Shiokawa, Smith, Snow, Souccar, Sousa, Sridharan, Srinivasan, Stahm,
  Stark, Story, Timmer, Vertatschitsch, Walther, Wei, Whitehorn, Whitney,
  Woody, Wouterloot, Wright, Yamaguchi, Yu, Zeballos, Zhang, \&
  Ziurys}]{eventhorizontelescopecollaboration2019}
{Event Horizon Telescope Collaboration}, Akiyama, K., Alberdi, A., {et~al.}
  2019{\natexlab{a}}, The Astrophysical Journal, 875, L1,
  \dodoi{10.3847/2041-8213/ab0ec7}

\bibitem[{{Event Horizon Telescope Collaboration}
  {et~al.}(2019{\natexlab{b}}){Event Horizon Telescope Collaboration}, Akiyama,
  Alberdi, Alef, Asada, Azulay, Baczko, Ball, Balokovi{\'c}, Barrett, Bintley,
  Blackburn, Boland, Bouman, Bower, Bremer, Brinkerink, Brissenden, Britzen,
  Broderick, Broguiere, Bronzwaer, Byun, Carlstrom, Chael, Chan, Chatterjee,
  Chatterjee, Chen, Chen, Cho, Christian, Conway, Cordes, Crew, Cui, Davelaar,
  De~Laurentis, Deane, Dempsey, Desvignes, Dexter, Doeleman, Eatough, Falcke,
  Fish, Fomalont, {Fraga-Encinas}, Friberg, Fromm, G{\'o}mez, Galison, Gammie,
  Garc{\'i}a, Gentaz, Georgiev, Goddi, Gold, Gu, Gurwell, Hada, Hecht, Hesper,
  Ho, Ho, Honma, Huang, Huang, Hughes, Ikeda, Inoue, Issaoun, James, Jannuzi,
  Janssen, Jeter, Jiang, Johnson, Jorstad, Jung, Karami, Karuppusamy,
  Kawashima, Keating, Kettenis, Kim, Kim, Kim, Kino, Koay, Koch, Koyama,
  Kramer, Kramer, Krichbaum, Kuo, Lauer, Lee, Li, Li, Lindqvist, Liu, Liuzzo,
  Lo, Lobanov, Loinard, Lonsdale, Lu, MacDonald, Mao, Markoff, Marrone,
  Marscher, {Mart{\'i}-Vidal}, Matsushita, Matthews, Medeiros, Menten, Mizuno,
  Mizuno, Moran, Moriyama, Moscibrodzka, Mu{\"l}ler, Nagai, Nagar, Nakamura,
  Narayan, Narayanan, Natarajan, Neri, Ni, Noutsos, Okino, Olivares, Oyama,
  {\"O}zel, Palumbo, Patel, Pen, Pesce, Pi{\'e}tu, Plambeck, PopStefanija,
  Porth, Prather, {Preciado-L{\'o}pez}, Psaltis, Pu, Ramakrishnan, Rao,
  Rawlings, Raymond, Rezzolla, Ripperda, Roelofs, Rogers, Ros, Rose,
  Roshanineshat, Rottmann, Roy, Ruszczyk, Ryan, Rygl, S{\'a}nchez,
  {S{\'a}nchez-Arguelles}, Sasada, Savolainen, Schloerb, Schuster, Shao, Shen,
  Small, Sohn, SooHoo, Tazaki, Tiede, Tilanus, Titus, Toma, Torne, Trent,
  Trippe, Tsuda, {van Bemmel}, {van Langevelde}, {van Rossum}, Wagner, Wardle,
  Weintroub, Wex, Wharton, Wielgus, Wong, Wu, Young, Young, Younsi, Yuan, Yuan,
  Zensus, Zhao, Zhao, Zhu, Anczarski, Baganoff, Eckart, Farah, Haggard,
  {Meyer-Zhao}, Michalik, Nadolski, Neilsen, Nishioka, Nowak, Pradel, Primiani,
  Souccar, Vertatschitsch, Yamaguchi, \&
  Zhang}]{eventhorizontelescopecollaboration2019d}
---. 2019{\natexlab{b}}, The Astrophysical Journal, 875, L5,
  \dodoi{10.3847/2041-8213/ab0f43}

\bibitem[{{Event Horizon Telescope Collaboration}
  {et~al.}(2019{\natexlab{c}}){Event Horizon Telescope Collaboration},
  {Akiyama}, {Alberdi}, {Alef}, {Asada}, {Azulay}, {Baczko}, {Ball},
  {Balokovi{\'c}}, {Barrett}, {Bintley}, {Blackburn}, {Boland}, {Bouman},
  {Bower}, {Bremer}, {Brinkerink}, {Brissenden}, {Britzen}, {Broderick},
  {Broguiere}, {Bronzwaer}, {Byun}, {Carlstrom}, {Chael}, {Chan}, {Chatterjee},
  {Chatterjee}, {Chen}, {Chen}, {Cho}, {Christian}, {Conway}, {Cordes}, {Crew},
  {Cui}, {Davelaar}, {De Laurentis}, {Deane}, {Dempsey}, {Desvignes}, {Dexter},
  {Doeleman}, {Eatough}, {Falcke}, {Fish}, {Fomalont}, {Fraga-Encinas},
  {Friberg}, {Fromm}, {G{\'o}mez}, {Galison}, {Gammie}, {Garc{\'\i}a},
  {Gentaz}, {Georgiev}, {Goddi}, {Gold}, {Gu}, {Gurwell}, {Hada}, {Hecht},
  {Hesper}, {Ho}, {Ho}, {Honma}, {Huang}, {Huang}, {Hughes}, {Ikeda}, {Inoue},
  {Issaoun}, {James}, {Jannuzi}, {Janssen}, {Jeter}, {Jiang}, {Johnson},
  {Jorstad}, {Jung}, {Karami}, {Karuppusamy}, {Kawashima}, {Keating},
  {Kettenis}, {Kim}, {Kim}, {Kim}, {Kino}, {Koay}, {Koch}, {Koyama}, {Kramer},
  {Kramer}, {Krichbaum}, {Kuo}, {Lauer}, {Lee}, {Li}, {Li}, {Lindqvist}, {Liu},
  {Liuzzo}, {Lo}, {Lobanov}, {Loinard}, {Lonsdale}, {Lu}, {MacDonald}, {Mao},
  {Markoff}, {Marrone}, {Marscher}, {Mart{\'\i}-Vidal}, {Matsushita},
  {Matthews}, {Medeiros}, {Menten}, {Mizuno}, {Mizuno}, {Moran}, {Moriyama},
  {Moscibrodzka}, {Mul{\ensuremath{\ddot{}}}ler}, {Nagai}, {Nagar}, {Nakamura},
  {Narayan}, {Narayanan}, {Natarajan}, {Neri}, {Ni}, {Noutsos}, {Okino},
  {Olivares}, {Oyama}, {{\"O}zel}, {Palumbo}, {Patel}, {Pen}, {Pesce},
  {Pi{\'e}tu}, {Plambeck}, {PopStefanija}, {Porth}, {Prather},
  {Preciado-L{\'o}pez}, {Psaltis}, {Pu}, {Ramakrishnan}, {Rao}, {Rawlings},
  {Raymond}, {Rezzolla}, {Ripperda}, {Roelofs}, {Rogers}, {Ros}, {Rose},
  {Roshanineshat}, {Rottmann}, {Roy}, {Ruszczyk}, {Ryan}, {Rygl},
  {S{\'a}nchez}, {S{\'a}nchez-Arguelles}, {Sasada}, {Savolainen}, {Schloerb},
  {Schuster}, {Shao}, {Shen}, {Small}, {Sohn}, {SooHoo}, {Tazaki}, {Tiede},
  {Tilanus}, {Titus}, {Toma}, {Torne}, {Trent}, {Trippe}, {Tsuda}, {van
  Bemmel}, {van Langevelde}, {van Rossum}, {Wagner}, {Wardle}, {Weintroub},
  {Wex}, {Wharton}, {Wielgus}, {Wong}, {Wu}, {Young}, {Young}, {Younsi},
  {Yuan}, {Yuan}, {Zensus}, {Zhao}, {Zhao}, {Zhu}, {Anczarski}, {Baganoff},
  {Eckart}, {Farah}, {Haggard}, {Meyer-Zhao}, {Michalik}, {Nadolski},
  {Neilsen}, {Nishioka}, {Nowak}, {Pradel}, {Primiani}, {Souccar},
  {Vertatschitsch}, {Yamaguchi}, \&
  {Zhang}}]{eventhorizontelescopecollaboration2019e}
{Event Horizon Telescope Collaboration}, {Akiyama}, K., {Alberdi}, A., {et~al.}
  2019{\natexlab{c}}, \apjl, 875, L5, \dodoi{10.3847/2041-8213/ab0f43}

\bibitem[{{Event Horizon Telescope Collaboration} {et~al.}(2021){Event Horizon
  Telescope Collaboration}, Akiyama, Algaba, Alberdi, Alef, Anantua, Asada,
  Azulay, Baczko, Ball, Balokovi{\'c}, Barrett, Benson, Bintley, Blackburn,
  Blundell, Boland, Bouman, Bower, Boyce, Bremer, Brinkerink, Brissenden,
  Britzen, Broderick, Broguiere, Bronzwaer, Byun, Carlstrom, Chael, Chan,
  Chatterjee, Chatterjee, Chen, Chen, Chesler, Cho, Christian, Conway, Cordes,
  Crawford, Crew, {Cruz-Osorio}, Cui, Davelaar, De~Laurentis, Deane, Dempsey,
  Desvignes, Dexter, Doeleman, Eatough, Falcke, Farah, Fish, Fomalont, Ford,
  {Fraga-Encinas}, Freeman, Friberg, Fromm, Fuentes, Galison, Gammie,
  Garc{\'i}a, Gentaz, Georgiev, Goddi, Gold, G{\'o}mez, {G{\'o}mez-Ruiz}, Gu,
  Gurwell, Hada, Haggard, Hecht, Hesper, Ho, Ho, Honma, Huang, Huang, Hughes,
  Ikeda, Inoue, Issaoun, James, Jannuzi, Janssen, Jeter, Jiang,
  {Jimenez-Rosales}, Johnson, Jorstad, Jung, Karami, Karuppusamy, Kawashima,
  Keating, Kettenis, Kim, Kim, Kim, Kim, Kino, Koay, Kofuji, Koch, Koyama,
  Kramer, Kramer, Krichbaum, Kuo, Lauer, Lee, Levis, Li, Li, Lindqvist, Lico,
  Lindahl, Liu, Liu, Liuzzo, Lo, Lobanov, Loinard, Lonsdale, Lu, MacDonald,
  Mao, Marchili, Markoff, Marrone, Marscher, {Mart{\'i}-Vidal}, Matsushita,
  Matthews, Medeiros, Menten, Mizuno, Mizuno, Moran, Moriyama, Moscibrodzka,
  M{\"u}ller, Musoke, Mej{\'i}as, Michalik, Nadolski, Nagai, Nagar, Nakamura,
  Narayan, Narayanan, Natarajan, Nathanail, Neilsen, Neri, Ni, Noutsos, Nowak,
  Okino, Olivares, {Ortiz-Le{\'o}n}, Oyama, {\"O}zel, Palumbo, Park, Patel,
  Pen, Pesce, Pi{\'e}tu, Plambeck, PopStefanija, Porth, P{\"o}tzl, Prather,
  {Preciado-L{\'o}pez}, Psaltis, Pu, Ramakrishnan, Rao, Rawlings, Raymond,
  Rezzolla, Ricarte, Ripperda, Roelofs, Rogers, Ros, Rose, Roshanineshat,
  Rottmann, Roy, Ruszczyk, Rygl, S{\'a}nchez, {S{\'a}nchez-Arguelles}, Sasada,
  Savolainen, Schloerb, Schuster, Shao, Shen, Small, Sohn, SooHoo, Sun, Tazaki,
  Tetarenko, Tiede, Tilanus, Titus, Toma, Torne, Trent, Traianou, Trippe, {van
  Bemmel}, {van Langevelde}, {van Rossum}, Wagner, {Ward-Thompson}, Wardle,
  Weintroub, Wex, Wharton, Wielgus, Wong, Wu, Yoon, Young, Young, Younsi, Yuan,
  Yuan, Zensus, Zhao, \& Zhao}]{eventhorizontelescopecollaboration2021}
{Event Horizon Telescope Collaboration}, Akiyama, K., Algaba, J.~C., {et~al.}
  2021, The Astrophysical Journal, 910, L12, \dodoi{10.3847/2041-8213/abe71d}

\bibitem[{{Event Horizon Telescope Collaboration} {et~al.}(2022){Event Horizon
  Telescope Collaboration}, Akiyama, Alberdi, Alef, Algaba, Anantua, Asada,
  Azulay, Bach, Baczko, Ball, Balokovi{\'c}, Barrett, Baub{\"o}ck, Benson,
  Bintley, Blackburn, Blundell, Bouman, Bower, Boyce, Bremer, Brinkerink,
  Brissenden, Britzen, Broderick, Broguiere, Bronzwaer, Bustamante, Byun,
  Carlstrom, Ceccobello, Chael, Chan, Chatterjee, Chatterjee, Chen, Chen,
  Cheng, Cho, Christian, Conroy, Conway, Cordes, Crawford, Crew, {Cruz-Osorio},
  Cui, Davelaar, De~Laurentis, Deane, Dempsey, Desvignes, Dexter, Dhruv,
  Doeleman, Dougal, Dzib, Eatough, Emami, Falcke, Farah, Fish, Fomalont, Ford,
  {Fraga-Encinas}, Freeman, Friberg, Fromm, Fuentes, Galison, Gammie,
  Garc{\'i}a, Gentaz, Georgiev, Goddi, Gold, {G{\'o}mez-Ruiz}, G{\'o}mez, Gu,
  Gurwell, Hada, Haggard, Haworth, Hecht, Hesper, Heumann, Ho, Ho, Honma,
  Huang, Huang, Hughes, Ikeda, Violette~Impellizzeri, Inoue, Issaoun, James,
  Jannuzi, Janssen, Jeter, Jiang, {Jim{\'e}nez-Rosales}, Johnson, Jorstad,
  Joshi, Jung, Karami, Karuppusamy, Kawashima, Keating, Kettenis, Kim, Kim,
  Kim, Kim, Kino, Koay, Kocherlakota, Kofuji, Koch, Koyama, Kramer, Kramer,
  Krichbaum, Kuo, La~Bella, Lauer, Lee, Lee, Leung, Levis, Li, Lico, Lindahl,
  Lindqvist, Lisakov, Liu, Liu, Liuzzo, Lo, Lobanov, Loinard, Lonsdale, Lu,
  Mao, Marchili, Markoff, Marrone, Marscher, {Mart{\'i}-Vidal}, Matsushita,
  Matthews, Medeiros, Menten, Michalik, Mizuno, Mizuno, Moran, Moriyama,
  Moscibrodzka, M{\"u}ller, Mus, Musoke, Myserlis, Nadolski, Nagai, Nagar,
  Nakamura, Narayan, Narayanan, Natarajan, Nathanail, Navarro~Fuentes, Neilsen,
  Neri, Ni, Noutsos, Nowak, Oh, Okino, Olivares, {Ortiz-Le{\'o}n}, Oyama,
  {\"O}zel, Palumbo, Filippos~Paraschos, Park, Parsons, Patel, Pen, Pesce,
  Pi{\'e}tu, Plambeck, PopStefanija, Porth, P{\"o}tzl, Prather,
  {Preciado-L{\'o}pez}, Psaltis, Pu, Ramakrishnan, Rao, Rawlings, Raymond,
  Rezzolla, Ricarte, Ripperda, Roelofs, Rogers, Ros, {Romero-Ca{\~n}izales},
  Roshanineshat, Rottmann, Roy, Ruiz, Ruszczyk, Rygl, S{\'a}nchez,
  {S{\'a}nchez-Arg{\"u}elles}, {S{\'a}nchez-Portal}, Sasada, Satapathy,
  Savolainen, Schloerb, Schonfeld, Schuster, Shao, Shen, Small, Sohn, SooHoo,
  Souccar, Sun, Tazaki, Tetarenko, Tiede, Tilanus, Titus, Torne, Traianou,
  Trent, Trippe, Turk, {van Bemmel}, {van Langevelde}, {van Rossum}, Vos,
  Wagner, {Ward-Thompson}, Wardle, Weintroub, Wex, Wharton, Wielgus, Wiik,
  Witzel, Wondrak, Wong, Wu, Yamaguchi, Yoon, Young, Young, Younsi, Yuan, Yuan,
  Zensus, Zhang, Zhao, Zhao, Chan, Qiu, Ressler, \&
  White}]{eventhorizontelescopecollaboration2022b}
{Event Horizon Telescope Collaboration}, Akiyama, K., Alberdi, A., {et~al.}
  2022, The Astrophysical Journal, 930, L16, \dodoi{10.3847/2041-8213/ac6672}

\bibitem[{{Falcke} {et~al.}(2000){Falcke}, {Melia}, \& {Agol}}]{Falcke2000}
{Falcke}, H., {Melia}, F., \& {Agol}, E. 2000, \apjl, 528, L13,
  \dodoi{10.1086/312423}

\bibitem[{{Ferrari} {et~al.}(1978){Ferrari}, {Trussoni}, \&
  {Zaninetti}}]{ferrari1978}
{Ferrari}, A., {Trussoni}, E., \& {Zaninetti}, L. 1978, \aap, 64, 43

\bibitem[{Fishbone \& Moncrief(1976)}]{fishbone1976}
Fishbone, L.~G., \& Moncrief, V. 1976, The Astrophysical Journal, 207, 962,
  \dodoi{10.1086/154565}

\bibitem[{Fromm {et~al.}(2022)Fromm, {Cruz-Osorio}, Mizuno, Nathanail, Younsi,
  Porth, Olivares, Davelaar, Falcke, Kramer, \& Rezzolla}]{fromm2022}
Fromm, C.~M., {Cruz-Osorio}, A., Mizuno, Y., {et~al.} 2022, Astronomy \&
  Astrophysics, 660, A107, \dodoi{10.1051/0004-6361/202142295}

\bibitem[{Giovannini {et~al.}(2018)Giovannini, Savolainen, Orienti, Nakamura,
  Nagai, Kino, Giroletti, Hada, Bruni, Kovalev, Anderson, D'Ammando, Hodgson,
  Honma, Krichbaum, Lee, Lico, Lisakov, Lobanov, Petrov, Sohn, Sokolovsky,
  Voitsik, Zensus, \& Tingay}]{giovannini2018}
Giovannini, G., Savolainen, T., Orienti, M., {et~al.} 2018, Nature Astronomy,
  2, 472, \dodoi{10.1038/s41550-018-0431-2}

\bibitem[{Hada {et~al.}(2016)Hada, Kino, Doi, Nagai, Honma, Akiyama, Tazaki,
  Lico, Giroletti, Giovannini, Orienti, \& Hagiwara}]{hada2016a}
Hada, K., Kino, M., Doi, A., {et~al.} 2016, The Astrophysical Journal, 817,
  131, \dodoi{10.3847/0004-637X/817/2/131}

\bibitem[{Hakobyan {et~al.}(2023)Hakobyan, Ripperda, \&
  Philippov}]{hakobyan2023}
Hakobyan, H., Ripperda, B., \& Philippov, A.~A. 2023, The Astrophysical
  Journal, 943, L29, \dodoi{10.3847/2041-8213/acb264}

\bibitem[{Hardee {et~al.}(2007)Hardee, Mizuno, \& Nishikawa}]{hardee2007}
Hardee, P., Mizuno, Y., \& Nishikawa, K.-I. 2007, Astrophysics and Space
  Science, Volume 311, Issue 1-3, pp. 281-286, 311, 281,
  \dodoi{10.1007/s10509-007-9529-1}

\bibitem[{Hunter(2007)}]{hunter2007}
Hunter, J.~D. 2007, Computing in Science and Engineering, 9, 90,
  \dodoi{10.1109/MCSE.2007.55}

\bibitem[{Igumenshchev {et~al.}(2003)Igumenshchev, Narayan, \&
  Abramowicz}]{igumenshchev2003}
Igumenshchev, I.~V., Narayan, R., \& Abramowicz, M.~A. 2003, The Astrophysical
  Journal, 592, 1042, \dodoi{10.1086/375769}

\bibitem[{{Issaoun} {et~al.}(2022){Issaoun}, {Wielgus}, {Jorstad}, {Krichbaum},
  {Blackburn}, {Janssen}, {Chan}, {Pesce}, {G{\'o}mez}, {Akiyama},
  {Mo{\'s}cibrodzka}, {Mart{\'\i}-Vidal}, {Chael}, {Lico}, {Liu},
  {Ramakrishnan}, {Lisakov}, {Fuentes}, {Zhao}, {Moriyama}, {Broderick},
  {Tiede}, {MacDonald}, {Mizuno}, {Traianou}, {Loinard}, {Davelaar}, {Gurwell},
  {Lu}, {Alberdi}, {Alef}, {Algaba}, {Anantua}, {Asada}, {Azulay}, {Bach},
  {Baczko}, {Ball}, {Balokovi{\'c}}, {Barrett}, {Baub{\"o}ck}, {Benson},
  {Bintley}, {Blundell}, {Boland}, {Bouman}, {Bower}, {Boyce}, {Bremer},
  {Brinkerink}, {Brissenden}, {Britzen}, {Broguiere}, {Bronzwaer},
  {Bustamante}, {Byun}, {Carlstrom}, {Ceccobello}, {Chatterjee}, {Chatterjee},
  {Chen}, {Chen}, {Cho}, {Christian}, {Conroy}, {Conway}, {Cordes}, {Crawford},
  {Crew}, {Cruz-Osorio}, {Cui}, {De Laurentis}, {Deane}, {Dempsey},
  {Desvignes}, {Dexter}, {Doeleman}, {Dhruv}, {Dzib Quijano}, {Eatough},
  {Emami}, {Falcke}, {Farah}, {Fish}, {Fomalont}, {Ford}, {Fraga-Encinas},
  {Freeman}, {Friberg}, {Fromm}, {Galison}, {Gammie}, {Garc{\'\i}a}, {Gentaz},
  {Georgiev}, {Goddi}, {Gold}, {G{\'o}mez-Ruiz}, {Gu}, {Hada}, {Haggard},
  {Hecht}, {Hesper}, {Ho}, {Ho}, {Honma}, {Huang}, {Huang}, {Hughes}, {Ikeda},
  {Impellizzeri}, {Inoue}, {James}, {Jannuzi}, {Jeter}, {Jiang},
  {Jimenez-Rosales}, {Johnson}, {Joshi}, {Jung}, {Karami}, {Karuppusamy},
  {Kawashima}, {Keating}, {Kettenis}, {Kim}, {Kim}, {Kim}, {Kim}, {Kino},
  {Koay}, {Kocherlakota}, {Kofuji}, {Koch}, {Koyama}, {Kramer}, {Kramer},
  {Kuo}, {La Bella}, {Lauer}, {Lee}, {Lee}, {Leung}, {Levis}, {Li}, {Lico},
  {Lindahl}, {Lindqvist}, {Liu}, {Liuzzo}, {Lo}, {Lobanov}, {Lonsdale}, {Mao},
  {Marchili}, {Markoff}, {Marrone}, {Marscher}, {Matsushita}, {Matthews},
  {Medeiros}, {Menten}, {Michalik}, {Mizuno}, {Mizuno}, {Moran}, {M{\"u}ller},
  {Mus}, {Musoke}, {Myserlis}, {Nadolski}, {Nagai}, {Nagar}, {Nakamura},
  {Narayan}, {Narayanan}, {Natarajan}, {Nathanail}, {Neilsen}, {Neri}, {Ni},
  {Noutsos}, {Nowak}, {Oh}, {Okino}, {Olivares}, {Ortiz-Le{\'o}n}, {Oyama},
  {{\"O}zel}, {Palumbo}, {Paraschos}, {Park}, {Parsons}, {Patel}, {Pen},
  {Pi{\'e}tu}, {Plambeck}, {PopStefanija}, {Porth}, {P{\"o}tzl}, {Prather},
  {Preciado-L{\'o}pez}, {Psaltis}, {Pu}, {Rao}, {Rawlings}, {Raymond},
  {Rezzolla}, {Ricarte}, {Ripperda}, {Roelofs}, {Rogers}, {Ros},
  {Romero-Canizales}, {Roshanineshat}, {Rottmann}, {Roy}, {Ruiz}, {Ruszczyk},
  {Rygl}, {S{\'a}nchez}, {S{\'a}nchez-Arguelles}, {Sanchez-Portal}, {Sasada},
  {Satapathy}, {Savolainen}, {Schloerb}, {Schuster}, {Shao}, {Shen}, {Small},
  {Sohn}, {SooHoo}, {Souccar}, {Sun}, {Tazaki}, {Tetarenko}, {Tiede},
  {Tilanus}, {Titus}, {Torne}, {Trent}, {Trippe}, {van Bemmel}, {van
  Langevelde}, {van Rossum}, {Vos}, {Wagner}, {Ward-Thompson}, {Wardle},
  {Weintroub}, {Wex}, {Wharton}, {Wiik}, {Witzel}, {Wondrak}, {Wong}, {Wu},
  {Yamaguchi}, {Yoon}, {Young}, {Young}, {Younsi}, {Yuan}, {Yuan}, {Zensus},
  {Zhang}, \& {Zhao}}]{Issaoun2022}
{Issaoun}, S., {Wielgus}, M., {Jorstad}, S., {et~al.} 2022, \apj, 934, 145,
  \dodoi{10.3847/1538-4357/ac7a40}

\bibitem[{{Issaoun} {et~al.}(2023){Issaoun}, {Pesce}, {Roelofs}, {Chael},
  {Dodson}, {Rioja}, {Akiyama}, {Aran}, {Blackburn}, {Doeleman}, {Fish},
  {Fitzpatrick}, {Johnson}, {Narayanan}, {Raymond}, \& {Tilanus}}]{Issaoun2023}
{Issaoun}, S., {Pesce}, D.~W., {Roelofs}, F., {et~al.} 2023, Galaxies, 11, 28,
  \dodoi{10.3390/galaxies11010028}

\bibitem[{Janssen {et~al.}(2021)Janssen, Falcke, Kadler, Ros, Wielgus, Akiyama,
  Balokovi{\'c}, Blackburn, Bouman, Chael, Chan, Chatterjee, Davelaar, Edwards,
  Fromm, G{\'o}mez, Goddi, Issaoun, Johnson, Kim, Koay, Krichbaum, Liu, Liuzzo,
  Markoff, Markowitz, Marrone, Mizuno, M{\"u}ller, Ni, Pesce, Ramakrishnan,
  Roelofs, Rygl, {van Bemmel}, {Event Horizon Telescope Collaboration},
  Alberdi, Alef, Algaba, Anantua, Asada, Azulay, Baczko, Ball, Ball, Barrett,
  Benson, Bintley, Bintley, Blundell, Boland, Boland, Bower, Boyce, Bremer,
  Brinkerink, Brissenden, Britzen, Broderick, Broguiere, Bronzwaer, Byun,
  Carlstrom, Chatterjee, Chen, Chen, Chesler, Cho, Christian, Conway, Cordes,
  Crawford, Crew, {Cruz-Osorio}, Cui, Cui, De~Laurentis, Deane, Dempsey,
  Desvignes, Dexter, Doeleman, Eatough, Farah, Farah, Fish, Fomalont, Ford,
  {Fraga-Encinas}, Friberg, Friberg, Fuentes, Galison, Gammie, Garc{\'i}a,
  Gelles, Gentaz, Georgiev, Georgiev, Gold, Gold, {G{\'o}mez-Ruiz}, Gu,
  Gurwell, Hada, Haggard, Hecht, Hesper, Himwich, Ho, Ho, Honma, Huang, Huang,
  Hughes, Ikeda, Inoue, Inoue, James, Jannuzi, Jannuzi, Jeter, Jiang,
  {Jimenez-Rosales}, {Jimenez-Rosales}, Jorstad, Jung, Karami, Karuppusamy,
  Kawashima, Keating, Kettenis, Kim, Kim, Kim, Kim, Kino, Kino, Kofuji, Koyama,
  Kramer, Kramer, Kramer, Kuo, Lauer, Lee, Levis, Li, Li, Lindqvist, Lico,
  Lindahl, Lindahl, Liu, Liu, Lo, Lobanov, Loinard, Lonsdale, Lu, MacDonald,
  Mao, Marchili, Marchili, Marchili, Marscher, {Mart{\'i}-Vidal}, Matsushita,
  Matthews, Medeiros, Menten, Mizuno, Mizuno, Moran, Moriyama, Moscibrodzka,
  Moscibrodzka, Musoke, Mej{\'i}as, Nagai, Nagar, Nakamura, Narayan, Narayanan,
  Natarajan, Nathanail, Neilsen, Neri, Neri, Noutsos, Nowak, Okino, Olivares,
  {Ortiz-Le{\'o}n}, Oyama, {\"O}zel, Palumbo, Park, Patel, Pen, Pen, Pi{\'e}tu,
  Plambeck, PopStefanija, Porth, P{\"o}tzl, Prather, {Preciado-L{\'o}pez},
  Psaltis, Pu, Pu, Rao, Rawlings, Raymond, Rezzolla, Ricarte, Ripperda,
  Ripperda, Rogers, Rogers, Rose, Roshanineshat, Rottmann, Roy, Ruszczyk,
  Ruszczyk, S{\'a}nchez, {S{\'a}nchez-Arguelles}, Sasada, Savolainen, Schloerb,
  Schuster, Shao, Shen, Small, Sohn, SooHoo, Sun, Tazaki, Tetarenko, Tiede,
  Tilanus, Titus, Torne, Trent, Traianou, Trippe, {van Bemmel}, {van
  Langevelde}, {van Rossum}, Wagner, {Ward-Thompson}, Wardle, Weintroub, Wex,
  Wharton, Wharton, Wong, Wu, Yoon, Young, Young, Younsi, Yuan, Yuan, Zensus,
  Zhao, \& Zhao}]{janssen2021}
Janssen, M., Falcke, H., Kadler, M., {et~al.} 2021, Nature Astronomy, 5, 1017,
  \dodoi{10.1038/s41550-021-01417-w}

\bibitem[{Jones {et~al.}(2001)Jones, Oliphant, Peterson, {et~al.}}]{jones2001}
Jones, E., Oliphant, T., Peterson, P., {et~al.} 2001, {{SciPy}}: {{Open}}
  Source Scientific Tools for {{Python}}

\bibitem[{Kim {et~al.}(2018)Kim, Krichbaum, Lu, Ros, Bach, Bremer, {de
  Vicente}, Lindqvist, \& Zensus}]{kim2018a}
Kim, J.-Y., Krichbaum, T.~P., Lu, R.-S., {et~al.} 2018, Astronomy \&amp;
  Astrophysics, Volume 616, id.A188, 13 pp., 616, A188,
  \dodoi{10.1051/0004-6361/201832921}

\bibitem[{{Liska} {et~al.}(2022){Liska}, {Chatterjee}, {Issa}, {Yoon}, {Kaaz},
  {Tchekhovskoy}, {van Eijnatten}, {Musoke}, {Hesp}, {Rohoza}, {Markoff},
  {Ingram}, \& {van der Klis}}]{Liska2022}
{Liska}, M.~T.~P., {Chatterjee}, K., {Issa}, D., {et~al.} 2022, \apjs, 263, 26,
  \dodoi{10.3847/1538-4365/ac9966}

\bibitem[{{Lu} {et~al.}(2023){Lu}, {Asada}, {Krichbaum}, {Park}, {Tazaki},
  {Pu}, {Nakamura}, {Lobanov}, {Hada}, {Akiyama}, {Kim}, {Marti-Vidal},
  {G{\'o}mez}, {Kawashima}, {Yuan}, {Ros}, {Alef}, {Britzen}, {Bremer},
  {Broderick}, {Doi}, {Giovannini}, {Giroletti}, {Ho}, {Honma}, {Hughes},
  {Inoue}, {Jiang}, {Kino}, {Koyama}, {Lindqvist}, {Liu}, {Marscher},
  {Matsushita}, {Nagai}, {Rottmann}, {Savolainen}, {Schuster}, {Shen}, {de
  Vicente}, {Walker}, {Yang}, {Zensus}, {Algaba}, {Allardi}, {Bach},
  {Berthold}, {Bintley}, {Byun}, {Casadio}, {Chang}, {Chang}, {Chang}, {Chen},
  {Chen}, {Chilson}, {Chuter}, {Conway}, {Crew}, {Dempsey}, {Dornbusch},
  {Faber}, {Friberg}, {Garc{\'\i}a}, {Garrido}, {Han}, {Han}, {Hasegawa},
  {Herrero-Illana}, {Huang}, {Huang}, {Impellizzeri}, {Jiang}, {Jinchi},
  {Jung}, {Kallunki}, {Kirves}, {Kimura}, {Koay}, {Koch}, {Kramer}, {Kraus},
  {Kubo}, {Kuo}, {Li}, {Lin}, {Liu}, {Liu}, {Lo}, {Lu}, {MacDonald},
  {Martin-Cocher}, {Messias}, {Meyer-Zhao}, {Minter}, {Nair}, {Nishioka},
  {Norton}, {Nystrom}, {Ogawa}, {Oshiro}, {Patel}, {Pen}, {Pidopryhora},
  {Pradel}, {Raffin}, {Rao}, {Ruiz}, {Sanchez}, {Shaw}, {Snow}, {Sridharan},
  {Srinivasan}, {Tercero}, {Torne}, {Traianou}, {Wagner}, {Walther}, {Wei},
  {Yang}, \& {Yu}}]{Lu2023}
{Lu}, R.-S., {Asada}, K., {Krichbaum}, T.~P., {et~al.} 2023, \nat, 616, 686,
  \dodoi{10.1038/s41586-023-05843-w}

\bibitem[{{Luminet}(1979)}]{Luminet1979}
{Luminet}, J.~P. 1979, \aap, 75, 228

\bibitem[{Marrone {et~al.}(2008)Marrone, Baganoff, Morris, Moran, Ghez,
  Hornstein, Dowell, Mu{\~n}oz, Bautz, Ricker, Brandt, Garmire, Lu, Matthews,
  Zhao, Rao, \& Bower}]{Marrone2008}
Marrone, D.~P., Baganoff, F.~K., Morris, M.~R., {et~al.} 2008, The
  Astrophysical Journal, 682, 373, \dodoi{10.1086/588806}

\bibitem[{{Marszewski} {et~al.}(2021){Marszewski}, {Prather}, {Joshi},
  {Pandya}, \& {Gammie}}]{Marszewski2021}
{Marszewski}, A., {Prather}, B.~S., {Joshi}, A.~V., {Pandya}, A., \& {Gammie},
  C.~F. 2021, \apj, 921, 17, \dodoi{10.3847/1538-4357/ac1b28}

\bibitem[{{Mbarek} \& {Caprioli}(2021)}]{Mbarek2021}
{Mbarek}, R., \& {Caprioli}, D. 2021, \apj, 921, 85,
  \dodoi{10.3847/1538-4357/ac1da8}

\bibitem[{Millman \& Aivazis(2011)}]{millman2011}
Millman, K.~J., \& Aivazis, M. 2011, Computing in Science \& Engineering, 13,
  9, \dodoi{10.1109/MCSE.2011.36}

\bibitem[{Mo{\'s}cibrodzka {et~al.}(2017)Mo{\'s}cibrodzka, Dexter, Davelaar, \&
  Falcke}]{moscibrodzka2017}
Mo{\'s}cibrodzka, M., Dexter, J., Davelaar, J., \& Falcke, H. 2017, Monthly
  Notices of the Royal Astronomical Society, 468, 2214,
  \dodoi{10.1093/mnras/stx587}

\bibitem[{Mo{\'s}cibrodzka {et~al.}(2016)Mo{\'s}cibrodzka, Falcke, \&
  Shiokawa}]{moscibrodzka2016}
Mo{\'s}cibrodzka, M., Falcke, H., \& Shiokawa, H. 2016, Astronomy \&amp;
  Astrophysics, 586, A38, \dodoi{10.1051/0004-6361/201526630}

\bibitem[{{Najafi-Ziyazi} {et~al.}(2023){Najafi-Ziyazi}, {Davelaar}, {Mizuno},
  \& {Porth}}]{najafi2023}
{Najafi-Ziyazi}, M., {Davelaar}, J., {Mizuno}, Y., \& {Porth}, O. 2023, arXiv
  e-prints, arXiv:2308.16740, \dodoi{10.48550/arXiv.2308.16740}

\bibitem[{Narayan {et~al.}(2003)Narayan, Igumenshchev, \&
  Abramowicz}]{narayan2003}
Narayan, R., Igumenshchev, I.~V., \& Abramowicz, M.~A. 2003, Publications of
  the Astronomical Society of Japan, 55, L69, \dodoi{10.1093/pasj/55.6.L69}

\bibitem[{{Narayan} {et~al.}(2021){Narayan}, {Palumbo}, {Johnson}, {Gelles},
  {Himwich}, {Chang}, {Ricarte}, {Dexter}, {Gammie}, {Chael}, {Event Horizon
  Telescope Collaboration}, {Akiyama}, {Alberdi}, {Alef}, {Algaba}, {Anantua},
  {Asada}, {Azulay}, {Baczko}, {Ball}, {Balokovi{\'c}}, {Barrett}, {Benson},
  {Bintley}, {Blackburn}, {Blundell}, {Boland}, {Bouman}, {Bower}, {Boyce},
  {Bremer}, {Brinkerink}, {Brissenden}, {Britzen}, {Broderick}, {Broguiere},
  {Bronzwaer}, {Byun}, {Carlstrom}, {Chan}, {Chatterjee}, {Chatterjee}, {Chen},
  {Chen}, {Chesler}, {Cho}, {Christian}, {Conway}, {Cordes}, {Crawford},
  {Crew}, {Cruz-Osorio}, {Cui}, {Davelaar}, {De Laurentis}, {Deane}, {Dempsey},
  {Desvignes}, {Doeleman}, {Eatough}, {Falcke}, {Farah}, {Fish}, {Fomalont},
  {Ford}, {Fraga-Encinas}, {Friberg}, {Fromm}, {Fuentes}, {Galison},
  {Garc{\'\i}a}, {Gentaz}, {Georgiev}, {Goddi}, {Gold}, {G{\'o}mez},
  {G{\'o}mez-Ruiz}, {Gu}, {Gurwell}, {Hada}, {Haggard}, {Hecht}, {Hesper},
  {Ho}, {Ho}, {Honma}, {Huang}, {Huang}, {Hughes}, {Ikeda}, {Inoue}, {Issaoun},
  {James}, {Jannuzi}, {Janssen}, {Jeter}, {Jiang}, {Jimenez-Rosales},
  {Jorstad}, {Jung}, {Karami}, {Karuppusamy}, {Kawashima}, {Keating},
  {Kettenis}, {Kim}, {Kim}, {Kim}, {Kim}, {Kino}, {Koay}, {Kofuji}, {Koch},
  {Koyama}, {Kramer}, {Kramer}, {Krichbaum}, {Kuo}, {Lauer}, {Lee}, {Levis},
  {Li}, {Li}, {Lindqvist}, {Lico}, {Lindahl}, {Liu}, {Liu}, {Liuzzo}, {Lo},
  {Lobanov}, {Loinard}, {Lonsdale}, {Lu}, {MacDonald}, {Mao}, {Marchili},
  {Markoff}, {Marrone}, {Marscher}, {Mart{\'\i}-Vidal}, {Matsushita},
  {Matthews}, {Medeiros}, {Menten}, {Mizuno}, {Mizuno}, {Moran}, {Moriyama},
  {Moscibrodzka}, {M{\"u}ller}, {Musoke}, {Mej{\'\i}as}, {Nagai}, {Nagar},
  {Nakamura}, {Narayanan}, {Natarajan}, {Nathanail}, {Neilsen}, {Neri}, {Ni},
  {Noutsos}, {Nowak}, {Okino}, {Olivares}, {Ortiz-Le{\'o}n}, {Oyama},
  {{\"O}zel}, {Park}, {Patel}, {Pen}, {Pesce}, {Pi{\'e}tu}, {Plambeck},
  {PopStefanija}, {Porth}, {P{\"o}tzl}, {Prather}, {Preciado-L{\'o}pez},
  {Psaltis}, {Pu}, {Ramakrishnan}, {Rao}, {Rawlings}, {Raymond}, {Rezzolla},
  {Ripperda}, {Roelofs}, {Rogers}, {Ros}, {Rose}, {Roshanineshat}, {Rottmann},
  {Roy}, {Ruszczyk}, {Rygl}, {S{\'a}nchez}, {S{\'a}nchez-Arguelles}, {Sasada},
  {Savolainen}, {Schloerb}, {Schuster}, {Shao}, {Shen}, {Small}, {Sohn},
  {SooHoo}, {Sun}, {Tazaki}, {Tetarenko}, {Tiede}, {Tilanus}, {Titus}, {Toma},
  {Torne}, {Trent}, {Traianou}, {Trippe}, {van Bemmel}, {van Langevelde}, {van
  Rossum}, {Wagner}, {Ward-Thompson}, {Wardle}, {Weintroub}, {Wex}, {Wharton},
  {Wielgus}, {Wong}, {Wu}, {Yoon}, {Young}, {Young}, {Younsi}, {Yuan}, {Yuan},
  {Zensus}, {Zhao}, \& {Zhao}}]{Narayan2021}
{Narayan}, R., {Palumbo}, D. C.~M., {Johnson}, M.~D., {et~al.} 2021, \apj, 912,
  35, \dodoi{10.3847/1538-4357/abf117}

\bibitem[{Oliphant(2007)}]{oliphant2007}
Oliphant, T.~E. 2007, Computing in Science \& Engineering, 9, 10,
  \dodoi{10.1109/MCSE.2007.58}

\bibitem[{Olivares {et~al.}(2019)Olivares, Porth, Davelaar, Most, Fromm,
  Mizuno, Younsi, \& Rezzolla}]{olivares2019}
Olivares, H., Porth, O., Davelaar, J., {et~al.} 2019, Astronomy \&amp;
  Astrophysics, Volume 629, id.A61, 21 pp., 629, A61,
  \dodoi{10.1051/0004-6361/201935559}

\bibitem[{{Pandya} {et~al.}(2016){Pandya}, {Zhang}, {Chandra}, \&
  {Gammie}}]{pandya2016}
{Pandya}, A., {Zhang}, Z., {Chandra}, M., \& {Gammie}, C.~F. 2016, \apj, 822,
  34, \dodoi{10.3847/0004-637X/822/1/34}

\bibitem[{Park {et~al.}(2019)Park, Hada, Kino, Nakamura, Ro, \&
  Trippe}]{park2019}
Park, J., Hada, K., Kino, M., {et~al.} 2019, The Astrophysical Journal, 871,
  257, \dodoi{10.3847/1538-4357/aaf9a9}

\bibitem[{{Pasetto} {et~al.}(2021){Pasetto}, {Carrasco-Gonz{\'a}lez},
  {G{\'o}mez}, {Mart{\'\i}}, {Perucho}, {O'Sullivan}, {Anderson},
  {D{\'\i}az-Gonz{\'a}lez}, {Fuentes}, \& {Wardle}}]{Pasetto2021}
{Pasetto}, A., {Carrasco-Gonz{\'a}lez}, C., {G{\'o}mez}, J.~L., {et~al.} 2021,
  \apjl, 923, L5, \dodoi{10.3847/2041-8213/ac3a88}

\bibitem[{{Perucho} \& {Lobanov}(2007)}]{Perucho2007}
{Perucho}, M., \& {Lobanov}, A.~P. 2007, \aap, 469, L23,
  \dodoi{10.1051/0004-6361:20077610}

\bibitem[{{Perucho} {et~al.}(2010){Perucho}, {Mart{\'\i}}, {Cela}, {Hanasz},
  {de La Cruz}, \& {Rubio}}]{Perucho2010}
{Perucho}, M., {Mart{\'\i}}, J.~M., {Cela}, J.~M., {et~al.} 2010, \aap, 519,
  A41, \dodoi{10.1051/0004-6361/200913012}

\bibitem[{Porth {et~al.}(2021)Porth, Mizuno, Younsi, \& Fromm}]{porth2021}
Porth, O., Mizuno, Y., Younsi, Z., \& Fromm, C.~M. 2021, Monthly Notices of the
  Royal Astronomical Society, 502, 2023, \dodoi{10.1093/mnras/stab163}

\bibitem[{Porth {et~al.}(2017)Porth, Olivares, Mizuno, Younsi, Rezzolla,
  Moscibrodzka, Falcke, \& Kramer}]{porth2017}
Porth, O., Olivares, H., Mizuno, Y., {et~al.} 2017, Computational Astrophysics
  and Cosmology, 4, 1, \dodoi{10.1186/s40668-017-0020-2}

\bibitem[{{Prieto} {et~al.}(2016){Prieto}, {Fern{\'a}ndez-Ontiveros},
  {Markoff}, {Espada}, \& {Gonz{\'a}lez-Mart{\'\i}n}}]{Prieto2016}
{Prieto}, M.~A., {Fern{\'a}ndez-Ontiveros}, J.~A., {Markoff}, S., {Espada}, D.,
  \& {Gonz{\'a}lez-Mart{\'\i}n}, O. 2016, \mnras, 457, 3801,
  \dodoi{10.1093/mnras/stw166}

\bibitem[{{Ricarte} {et~al.}(2023){Ricarte}, {Johnson}, {Kovalev}, {Palumbo},
  \& {Emami}}]{Ricarte2023}
{Ricarte}, A., {Johnson}, M.~D., {Kovalev}, Y.~Y., {Palumbo}, D. C.~M., \&
  {Emami}, R. 2023, Galaxies, 11, 5, \dodoi{10.3390/galaxies11010005}

\bibitem[{Rieger(2019)}]{rieger2019}
Rieger, F.~M. 2019, Galaxies, 7, 78, \dodoi{10.3390/galaxies7030078}

\bibitem[{{Ripperda} {et~al.}(2020){Ripperda}, {Bacchini}, \&
  {Philippov}}]{ripperda2020}
{Ripperda}, B., {Bacchini}, F., \& {Philippov}, A.~A. 2020, \apj, 900, 100,
  \dodoi{10.3847/1538-4357/ababab}

\bibitem[{Ripperda {et~al.}(2022)Ripperda, Liska, Chatterjee, Musoke,
  Philippov, Markoff, Tchekhovskoy, \& Younsi}]{ripperda2022}
Ripperda, B., Liska, M., Chatterjee, K., {et~al.} 2022, The Astrophysical
  Journal, 924, L32, \dodoi{10.3847/2041-8213/ac46a1}

\bibitem[{Rybicki \& Lightman(1979)}]{rybicki1979}
Rybicki, G.~B., \& Lightman, A.~P. 1979, Radiative Processes in Astrophysics
  ({Wiley})

\bibitem[{{Shcherbakov}(2008)}]{Shcherbakov2008}
{Shcherbakov}, R.~V. 2008, \apj, 688, 695, \dodoi{10.1086/592326}

\bibitem[{Sironi {et~al.}(2021)Sironi, Rowan, \& Narayan}]{sironi2021}
Sironi, L., Rowan, M.~E., \& Narayan, R. 2021, The Astrophysical Journal, 907,
  L44, \dodoi{10.3847/2041-8213/abd9bc}

\bibitem[{{Sironi} \& {Spitkovsky}(2014)}]{Sironi2014}
{Sironi}, L., \& {Spitkovsky}, A. 2014, \apjl, 783, L21,
  \dodoi{10.1088/2041-8205/783/1/L21}

\bibitem[{Sobacchi \& Lyubarsky(2018)}]{sobacchi2018a}
Sobacchi, E., \& Lyubarsky, Y.~E. 2018, Monthly Notices of the Royal
  Astronomical Society, 473, 2813, \dodoi{10.1093/mnras/stx2592}

\bibitem[{Stanzione {et~al.}(2020)Stanzione, West, Evans, Minyard, Ghattas, \&
  Panda}]{Frontera}
Stanzione, D., West, J., Evans, R.~T., {et~al.} 2020, in Practice and
  Experience in Advanced Research Computing, PEARC '20 (New York, NY, USA:
  Association for Computing Machinery), 106–111

\bibitem[{Tchekhovskoy {et~al.}(2011)Tchekhovskoy, Narayan, \&
  McKinney}]{tchekhovskoy2011}
Tchekhovskoy, A., Narayan, R., \& McKinney, J.~C. 2011, Monthly Notices of the
  Royal Astronomical Society, 418, L79,
  \dodoi{10.1111/j.1745-3933.2011.01147.x}

\bibitem[{{The GRAVITY Collaboration} {et~al.}(2023){The GRAVITY
  Collaboration}, {Abuter}, {Aimar}, {Amaro Seoane}, {Amorim}, {Baub{\"o}ck},
  {Berger}, {Bonnet}, {Bourdarot}, {Brandner}, {Cardoso}, {Cl{\'e}net},
  {Davies}, {de Zeeuw}, {Dexter}, {Drescher}, {Eckart}, {Eisenhauer},
  {Feuchtgruber}, {Finger}, {F{\"o}rster Schreiber}, {Foschi}, {Garcia}, {Gao},
  {Gelles}, {Gendron}, {Genzel}, {Gillessen}, {Hartl}, {Haubois}, {Haussmann},
  {Hei{\ss}el}, {Henning}, {Hippler}, {Horrobin}, {Jochum}, {Jocou}, {Kaufer},
  {Kervella}, {Lacour}, {Lapeyr{\`e}re}, {Le Bouquin}, {L{\'e}na}, {Lutz},
  {Mang}, {More}, {Ott}, {Paumard}, {Perraut}, {Perrin}, {Pfuhl}, {Rabien},
  {Ribeiro}, {Sadun Bordoni}, {Scheithauer}, {Shangguan}, {Shimizu}, {Stadler},
  {Straub}, {Straubmeier}, {Sturm}, {Tacconi}, {Vincent}, {von Fellenberg},
  {Widmann}, {Wielgus}, {Wieprecht}, {Wiezorrek}, \& {Woillez}}]{Gravity2023}
{The GRAVITY Collaboration}, {Abuter}, R., {Aimar}, N., {et~al.} 2023, arXiv
  e-prints, arXiv:2307.11821, \dodoi{10.48550/arXiv.2307.11821}

\bibitem[{{van der Walt} {et~al.}(2011){van der Walt}, Colbert, \&
  Varoquaux}]{vanderwalt2011}
{van der Walt}, S., Colbert, S.~C., \& Varoquaux, G. 2011, Computing in Science
  and Engineering, 13, 22, \dodoi{10.1109/MCSE.2011.37}

\bibitem[{Vos {et~al.}(2022)Vos, Mo{\'s}cibrodzka, \& Wielgus}]{Vos2022}
Vos, J., Mo{\'s}cibrodzka, M.~A., \& Wielgus, M. 2022, A\&A, 668, A185,
  \dodoi{10.1051/0004-6361/202244840}

\bibitem[{Walker {et~al.}(2018)Walker, Hardee, Davies, Ly, \&
  Junor}]{walker2018a}
Walker, R.~C., Hardee, P.~E., Davies, F.~B., Ly, C., \& Junor, W. 2018, The
  Astrophysical Journal, 855, 128, \dodoi{10.3847/1538-4357/aaafcc}

\bibitem[{{Wielgus} {et~al.}(2022){Wielgus}, {Moscibrodzka}, {Vos}, {Gelles},
  {Mart{\'\i}-Vidal}, {Farah}, {Marchili}, {Goddi}, \& {Messias}}]{wielgus22}
{Wielgus}, M., {Moscibrodzka}, M., {Vos}, J., {et~al.} 2022, \aap, 665, L6,
  \dodoi{10.1051/0004-6361/202244493}

\bibitem[{Wong {et~al.}(2021)Wong, Du, Prather, \& Gammie}]{wong2021}
Wong, G.~N., Du, Y., Prather, B.~S., \& Gammie, C.~F. 2021, The Astrophysical
  Journal, 914, 55, \dodoi{10.3847/1538-4357/abf8b8}

\bibitem[{Xiao(2006)}]{xiao2006}
Xiao, F. 2006, Plasma Physics and Controlled Fusion, 48, 203

\bibitem[{Zavala \& Taylor(2003)}]{zavala2003}
Zavala, R.~T., \& Taylor, G.~B. 2003, The Astrophysical Journal, 589, 126,
  \dodoi{10.1086/374619}

\bibitem[{Zhdankin {et~al.}(2023)Zhdankin, Ripperda, \&
  Philippov}]{zhdankin2023}
Zhdankin, V., Ripperda, B., \& Philippov, A.~A. 2023, Particle Acceleration by
  Magnetic {{Rayleigh-Taylor}} Instability: Mechanism for Flares in Black-Hole
  Accretion Flows, \dodoi{10.48550/arXiv.2302.05276}

\end{thebibliography}
\end{document}